\documentclass[a4paper,fleqn,usenatbib]{mnras}

\usepackage{eso-pic}

\AddToShipoutPictureBG*{%
  \AtPageUpperLeft{%
    \hspace{0.75\paperwidth}%
    \raisebox{-3.5\baselineskip}{%
      \makebox[0pt][l]{\textnormal{DES 2019-0324}}
}}}%

\AddToShipoutPictureBG*{%
  \AtPageUpperLeft{%
    \hspace{0.75\paperwidth}%
    \raisebox{-4.5\baselineskip}{%
      \makebox[0pt][l]{\textnormal{FERMILAB PUB-19-204-AE}}
}}}%

\usepackage{lineno}

\usepackage{graphicx}	
\usepackage{amsmath,amsfonts,amssymb}
\usepackage{makeidx}
\usepackage{epstopdf} 
\usepackage{booktabs,caption,fixltx2e}
\usepackage[flushleft]{threeparttable}
\usepackage{hyperref}
\usepackage{multirow}
\usepackage{rotating}
\usepackage{color}
\usepackage[T1]{fontenc}
\usepackage{ae,aecompl}
\usepackage{soul}

\def \BE{\begin{equation}}
\def \EE{\end{equation}}	
\def \BC{\begin{center}}
\def \EC{\end{center}}
\def \BEA{\begin{eqnarray}}
\def \EEA{\end{eqnarray}}
\def\XMM{XMM-{\it Newton}}

\defcitealias{gupta17}{G17}

\title[Cluster Radio AGN to z$\sim$1]{Constraining Radio Mode Feedback in Galaxy Clusters with the Cluster Radio 
AGN Properties to z$\sim$1}

\newcommand{\Melbourne}{$^{1}$}
\newcommand{\Munich}{$^{2}$}
\newcommand{\ExcellenceCluster}{$^{3}$}
\newcommand{\MPE}{$^{4}$}
\newcommand{\STANFORDkavli}{$^{5}$}
\newcommand{\SLAC}{$^{6}$}
\newcommand{\Fermilab}{$^{7}$}
\newcommand{\Madrid}{$^{8}$}
\newcommand{\UCL}{$^{9}$}
\newcommand{\HARVARD}{$^{10}$}
\newcommand{\MIT}{$^{11}$}
\newcommand{\CIEMAT}{$^{12}$}
\newcommand{\RIOLAB}{$^{13}$}
\newcommand{\ILLINOIS}{$^{14}$}
\newcommand{\Urbana}{$^{15}$}
\newcommand{\BARCELONA}{$^{16}$}
\newcommand{\ASIAA}{$^{17}$}
\newcommand{\RIOOBS}{$^{18}$}
\newcommand{\INDIA}{$^{19}$}
\newcommand{\SANTACRUZ}{$^{20}$}
\newcommand{\MICHIGANA}{$^{21}$}
\newcommand{\MICHIGANP}{$^{22}$}
\newcommand{\IEEC}{$^{23}$}
\newcommand{\ICE}{$^{24}$}
\newcommand{\STANFORD}{$^{25}$}
\newcommand{\Ohio}{$^{26}$}
\newcommand{\Macquarie}{$^{27}$}
\newcommand{\ANU}{$^{28}$}
\newcommand{\PAULO}{$^{29}$}
\newcommand{\AandM}{$^{30}$}
\newcommand{\CATALAN}{$^{31}$}
\newcommand{\PRINCETON}{$^{32}$}
\newcommand{\PORTO}{$^{33}$}
\newcommand{\TriesteA}{$^{34}$}
\newcommand{\TriesteB}{$^{35}$}
\newcommand{\TriesteC}{$^{36}$}
\newcommand{\SOUTHAMPTON}{$^{37}$}
\newcommand{\Lancaster}{$^{38}$}
\newcommand{\OAK}{$^{39}$}
\newcommand{\ARGONNE}{$^{40}$}
\newcommand{\LaSerena}{$^{41}$}

\author[N.~Gupta, M. Pannella, J. J. Mohr, et al.] {
\parbox{\textwidth}{
\Large
N.~Gupta\thanks{nikhel.gupta@unimelb.edu.au}\Melbourne$^,$\Munich$^,$\ExcellenceCluster$^,$\MPE,
M. Pannella\Munich,
J. J. Mohr\Munich$^,$\ExcellenceCluster$^,$\MPE,
M.~Klein\Munich$^,$\MPE,
E.~S.~Rykoff\STANFORDkavli$^,$\SLAC,
J.~Annis\Fermilab,
S.~Avila\Madrid,
F.~Bianchini\Melbourne,
D.~Brooks\UCL,
E.~Buckley-Geer\Fermilab,
E.~Bulbul\HARVARD$^,$\MIT,
A.~Carnero~Rosell\CIEMAT$^,$\RIOLAB,
M.~Carrasco~Kind\ILLINOIS$^,$\Urbana,
J.~Carretero\BARCELONA,
I.~Chiu\ASIAA,
M.~Costanzi\Munich,
L.~N.~da Costa\RIOLAB$^,$\RIOOBS,
J.~De~Vicente\Madrid,
S.~Desai\INDIA,
J.~P.~Dietrich\Munich$^,$\ExcellenceCluster,
P.~Doel\UCL,
S.~Everett\SANTACRUZ,
A.~E.~Evrard\MICHIGANA$^,$\MICHIGANP,
J.~Garc\'ia-Bellido\Madrid,
E.~Gaztanaga\IEEC$^,$\ICE,
D.~Gruen\STANFORD$^,$\STANFORDkavli$^,$\SLAC,
R.~A.~Gruendl\ILLINOIS$^,$\Urbana,
J.~Gschwend\RIOLAB$^,$\RIOOBS,
G.~Gutierrez\Fermilab,
D.~L.~Hollowood\SANTACRUZ,
K.~Honscheid\Ohio,
D.~J.~James\HARVARD,
T.~Jeltema\SANTACRUZ,
K.~Kuehn\Macquarie,
C.~Lidman\ANU,
M.~Lima\PAULO$^,$\RIOLAB,
M.~A.~G.~Maia\RIOLAB$^,$\RIOOBS,
J.~L.~Marshall\AandM,
M.~McDonald\MIT,
F.~Menanteau\ILLINOIS$^,$\Urbana,
R.~Miquel\CATALAN$^,$\BARCELONA,
R.~L.~C.~Ogando\RIOLAB$^,$\RIOOBS,
A.~Palmese,\Fermilab,
F.~Paz-Chinch\'{o}n\ILLINOIS$^,$\Urbana,
A.~A.~Plazas\PRINCETON,
C.~L.~Reichardt\Melbourne,
E.~Sanchez\Madrid,
B.~Santiago\PORTO$^,$\RIOLAB,
A.~Saro\TriesteA$^,$\TriesteB$^,$\TriesteC,
V.~Scarpine\Fermilab,
R.~Schindler\SLAC,
M.~Schubnell\MICHIGANP,
S.~Serrano\IEEC$^,$\ICE,
I.~Sevilla-Noarbe\Madrid,
X.~Shao\Munich,
M.~Smith\SOUTHAMPTON,
J.~P.~Stott\Lancaster,
V.~Strazzullo\Munich,
E.~Suchyta\OAK,
M.~E.~C.~Swanson\Urbana,
V.~Vikram\ARGONNE,
and A.~Zenteno\LaSerena
}
\vspace{0.4cm}
\\
\parbox{\textwidth}{
The authors' affiliations are shown in the end of manuscript.
}
}

\begin{document}
\date{Accepted ???. Received ???; in original form ???} 

\maketitle

\begin{abstract}
We study the properties of the Sydney University Molonglo Sky Survey (SUMSS) 843~MHz radio AGN population in galaxy clusters from two large catalogs created using the Dark Energy Survey (DES): $\sim$11,800 optically selected RM-Y3 and $\sim$1,000 X-ray selected MARD-Y3 clusters. 
We show that cluster radio loud AGN are highly concentrated around cluster centers to $z\sim1$. 
We measure the halo occupation number for cluster radio AGN above a threshold luminosity, finding that the number of radio AGN per cluster increases with cluster halo mass as $N\propto M^{1.2\pm0.1}$ ($N\propto M^{0.68\pm0.34}$) for the RM-Y3 (MARD-Y3) sample.   
Together, these results indicate that radio mode feedback is favoured in more massive galaxy clusters.  
Using optical counterparts for these sources, we demonstrate weak redshift evolution in the host broad 
band colors and the radio luminosity at fixed host galaxy stellar mass. 
We use the redshift evolution in radio luminosity to break the degeneracy 
between density and luminosity evolution scenarios in the redshift trend of the radio AGN luminosity function (LF).  
The LF exhibits a redshift trend of the form $(1+z)^\gamma$ in density and luminosity, respectively, of  $\gamma_{\rm D}=3.0\pm0.4$ and $\gamma_{\rm P}=0.21\pm0.15$ in the RM-Y3 sample, and $\gamma_{\rm D}=2.6\pm0.7$ and $\gamma_{\rm P}=0.31\pm0.15$ in MARD-Y3.  
We discuss the physical drivers of radio mode feedback in cluster AGN, and we use the cluster radio galaxy LF to estimate the average radio-mode feedback energy as a function of cluster mass and redshift and compare it to the core ($<0.1R_{500}$) X-ray radiative losses for clusters at $z<1$.  
\end{abstract}


\section{Introduction} 
\label{sec:Introduction}

A number of studies of low redshift clusters have shown that local instabilities due to active galactic nuclei (AGN) outbursts reheat the 
intracluster medium (ICM) and regulate the cooling in the cluster centre through radio-mode AGN feedback \citep[e.g.][]
{mcnamara05, rafferty06, blanton10, ogrean10, ehlert11, birzan12, myriam12, bharadwaj14, voit16, gaspari20}. There is some 
evidence of the evolution of AGN feedback in massive field galaxies up to $z\sim1.3$ \citep{simpson13}, suggesting 
that the balance between radiative cooling and AGN feedback was achieved in the early universe. However, the 
evolution of AGN feedback in galaxy clusters with redshift and mass is not well studied.
\cite{larrondo13} studied X-ray AGN emission in brightest cluster galaxies (BCGs) up to $z$=0.6, showing that the typical nuclear AGN X-ray luminosity increases by a factor of 10, mostly due to an increase in the fraction of BCGs hosting X-ray AGN between $z=0.1$ and $z=0.6$.
\cite{birzan17} studied 21 Sunyaev-Zel'dovich Effect \citep[SZE;][]{sunyaev72} selected clusters with $0.3<z<1.3$ and found higher radio luminosity as compared to X-ray luminosity at higher redshifts ($z>0.6$), presumably due to increased merging activity of galaxy clusters that may trigger the radio mode feedback.
\cite{yang18} compiled an X-ray luminous BCG sample in the redshift ranges $0.2 < z < 0.3$ and $0.55 < z < 0.75$, and contrastingly found no evidence for evolution. These results should however be considered as preliminary, because the BCG sample is simply collected from all publicly available observations of clusters and groups, and so the sample selection is not well understood.

Previous studies of cluster radio sources  have either not considered the redshift trends in the cluster radio source 
properties \citep[e.g.][]{lin07,lin09,sehgal10,linhenry15} or have shown contrasting trends \citep[e.g.][hereafter G17]
{sommer11, birzan17, lin17,gupta17} due to statistical limitations of galaxy cluster samples extending to higher 
redshift. 
Moreover, given the  relatively featureless shape of the radio galaxy luminosity function (LF, for e.g. \citetalias{gupta17}), it is typically not possible to differentiate a 
change in number density from a change in the typical AGN luminosity (so-called luminosity evolution).  Typically,  the 
reported redshift trends in the 
properties of cluster radio sources have been presented as either pure density or pure luminosity evolution (e.g. \cite{sommer11}).

In this paper, we present the redshift and mass trends of the properties of cluster radio sources. The sources we 
consider come from the 843~MHz Sydney University Molonglo Sky Survey \citep[SUMSS;][]{bock99, mauch03, 
murphy07} radio source catalog, and at the distances where we study these sources they correspond to high 
radio luminosity objects that are dominated by AGN synchrotron emission.  We study the cluster radio AGN to measure 
their spatial distribution in clusters, the halo occupation number (HON) or characteristic number within the cluster virial 
region, the radio luminosity evolution at fixed stellar mass and the luminosity functions.  This study uses a 
large ensemble of optically and X-ray selected galaxy clusters that have been constructed using
the first three years of data from the Dark Energy 
Survey \citep[DES-Y3; see][for discussion of the data release]{abbott18} 
and the ROSAT All Sky Survey 2RXS faint source catalog \citep[RASS;][]{boller16}. 

Our study primarily employs statistical background subtraction to correct for those radio AGN that are not physically 
associated with the clusters.  However, we also use DES-Y3 survey data to identify the optical counterparts of cluster 
radio AGN and to directly select those AGN whose redshifts are consistent with the cluster redshifts.  This also allows us 
to perform Spectral Energy Distribution (SED) analyses of these radio source counterparts to obtain constraints on the 
stellar mass and spectral type of these sources.  This information then allows us to probe the relationship between stellar 
mass and radio luminosity for radio AGN, which provides direct constraints on the redshift evolution of cluster radio AGN. We use 
this information together with the LF to jointly constrain number density and luminosity evolution for the cluster radio 
AGN population.  With measurements of the radio AGN LF and its dependence on cluster mass and redshift, we then 
estimate mass and redshift trends in the radio mode feedback and its relationship to the X-ray radiative losses within 
cluster cores.

The plan of the paper is as follows: In Section~\ref{sec:Data}, we discuss the observations and the data used in this 
work. We describe the SED analysis and the cross-matching of radio to optical sources  in Section~\ref{sec:Crossmatching}.  
Section~\ref{sec:Results} is dedicated to the studies of surface density profiles, luminosity evolution, halo occupation 
numbers and LFs.  Section~\ref{sec:Discussion} contains a discussion of the implications of our measured mass and 
redshift trends for the physical drivers of radio mode feedback and the balance of AGN feedback and X-ray 
radiative losses in the cluster cores.  We summarize our results in Section~\ref{sec:Conclusions}. 
Throughout this paper we assume a flat $\Lambda$CDM cosmology with matter density parameter $\Omega_{\rm M} = 
0.3$ and Hubble constant $H_0$ = 70~$\rm km$ $\rm s^{-1}$ $\rm Mpc^{-1}$.  We take the normalization of the matter 
power spectrum to be $\sigma_8= 0.83$.
\section{Data}
\label{sec:Data}

We study the overdensity of radio point sources toward X-ray and optically 
selected galaxy clusters selected using 
RASS and DES-Y3 observations. We focus on high luminosity radio AGN 
$\log{[\mathrm{P/(W\,Hz^{-1})}]}$\,>\,23 that are selected from the SUMSS 
catalog observed at 843~MHz.  A part of our analysis 
is supported through the identification of the optical counterparts of these sources using the 
DES-Y3 observations (see Section~\ref{sec:HostProperties}), but the rest 
of our results employ statistical background subtraction.

\subsection{SUMSS catalog}
\label{sec:SUMSS}
The Sydney University Molonglo Sky Survey \citep[SUMSS;][]{bock99, mauch03, murphy07} imaged the southern radio 
sky at 843~MHz with a characteristic angular resolution of $\sim$\,$45^{\prime\prime}$ using the Molonglo Observatory Synthesis 
Telescope \citep[MOST,][]{mills81, robertson91}. The survey was completed in early 2007 and covers 8,100 $\rm 
deg^2$ of sky with $\delta$$\leq$$-30^\circ$ and $\vert b \vert$$\geq$$10^\circ$. The catalog contains 210,412 radio 
sources to a limiting peak brightness of 6 mJy beam$^{-1}$ at $\delta$$\leq$$-50^\circ$ and 10 mJy beam$^{-1}$ at $
\delta$$>$$-50^\circ$.  At the SUMSS selection frequency, we expect nearly all sources above the flux selection 
threshold to be synchrotron dominated \citep{dezotti05}. 
The positional uncertainties given in the catalogue are a combination of fitting uncertainties and calibration uncertainties of the MOST and for sources with peak brightness $A_{843}\geq$20 mJy beam$^{-1}$, the accuracy is in the range 1$^{\prime\prime}$ to 2$^{\prime\prime}$ \citep[see section 5.1 of][]{murphy07}. The flux measurements are accurate to within 3~percent.  The catalog is complete to 8 mJy at $\delta$$\leq$$-50^\circ$ and to 18 mJy at $\delta$$>$$-50^\circ$, and we restrict our analysis to these complete subsamples. At the SUMSS frequency, approximately 10~percent of the sources exhibit extent along one axis \citep{mauch03}. 

\begin{figure}
\centering
\includegraphics[width=8.8cm, height=7.5cm]{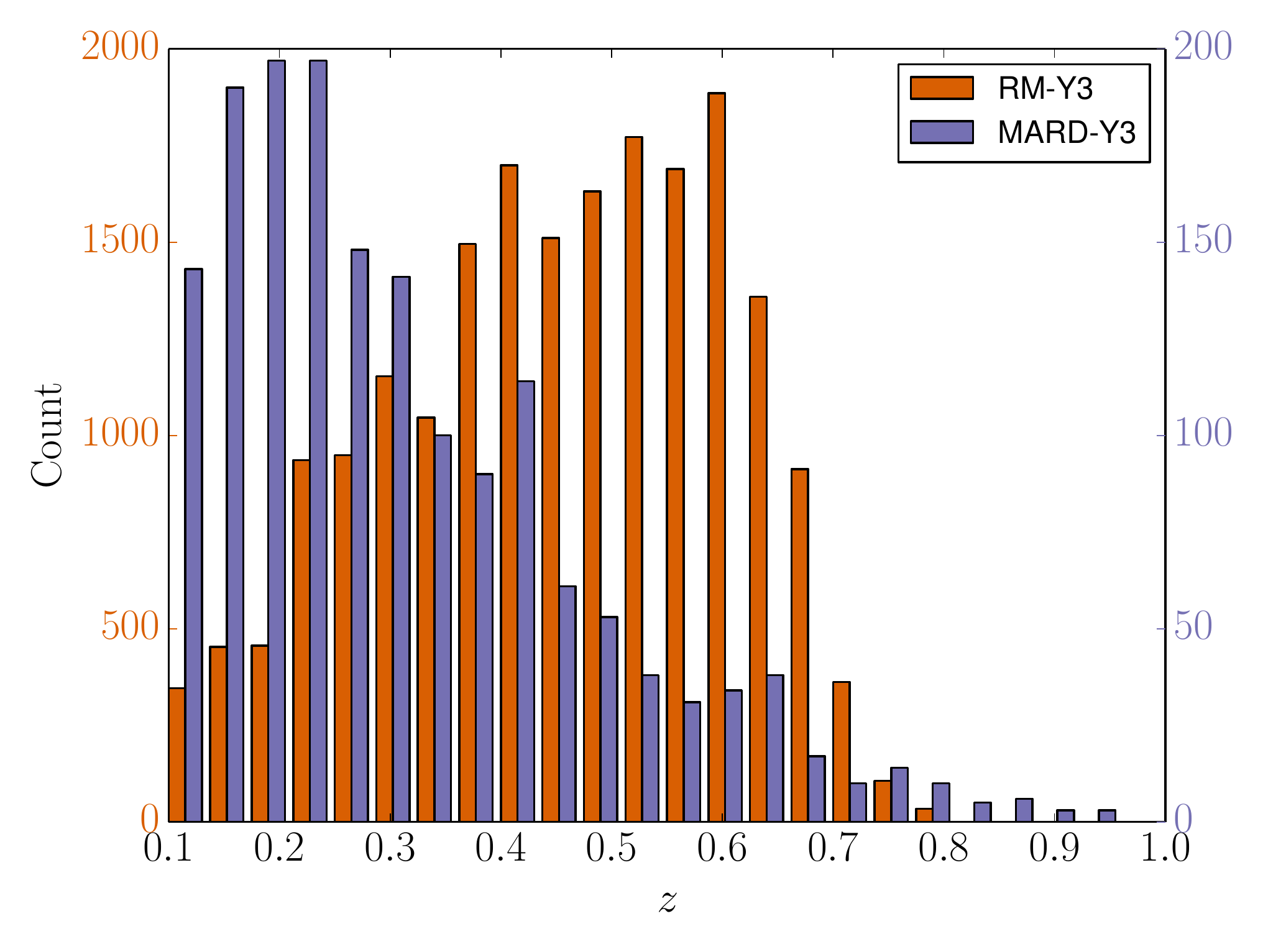}
\includegraphics[width=8.8cm, height=7.5cm]{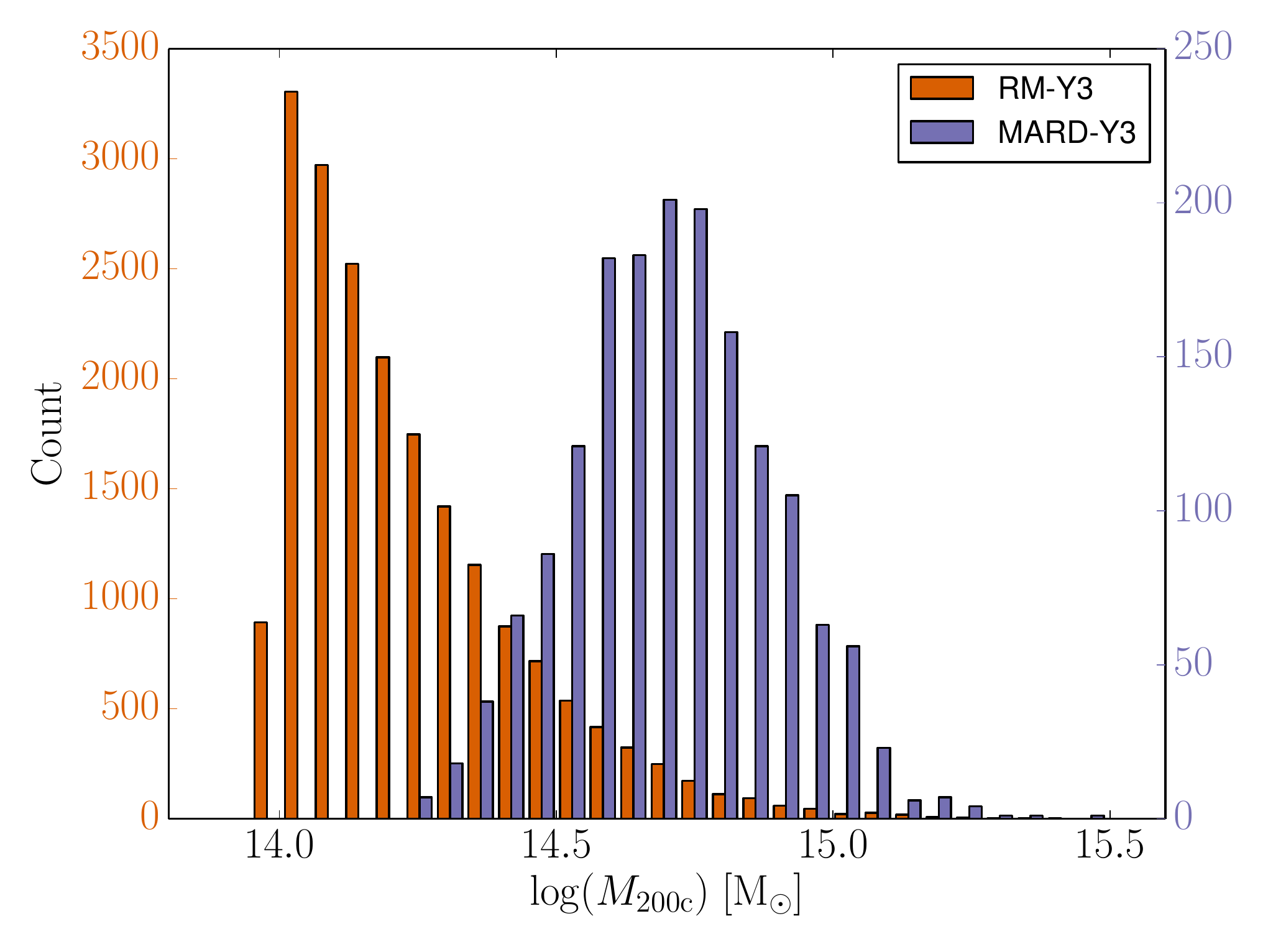}
\vskip-0.1in
\caption{Distribution of RM-Y3 and MARD-Y3 galaxy clusters as a function of redshift (upper panel) and mass (lower 
panel). Red bars show optically selected RM-Y3 clusters with median mass and redshift of $1.52 \times 10^{14}$ $\rm 
M_{\odot}$ and 0.47, respectively.  Blue bars show X-ray selected MARD-Y3 clusters with median mass and redshift of 
$5.06 \times 10^{14}$ $\rm M_{\odot}$ and 0.28, respectively. The color-coded vertical axes labels on the left and right 
sides of the plots represent the number of clusters in the RM-Y3 and MARD-Y3 catalogs, respectively.}
\label{fig:redmapper}
\end{figure}

\subsection{DES-Y3 data}
\label{sec:DESData}
We make use of the DES Y3 GOLD-release multiwalength catalog in its version 2.2, which covers  $\sim 5000$~$\deg^2$ 
of the southern celestial hemisphere and has about 60~percent overlap ($\sim 3000$~$\deg^2$) with the SUMSS survey.
The GOLD catalog is intended to be the basis for cosmology analyses with the Dark Energy Survey 
data. The GOLD catalog consists of a validated object catalog with a set of quality control flags and additional auxiliary 
data such as photometric redshift information based both on neural networks and Bayesian template fitting. 
Specifically, for each object in the DES catalog we make use of total magnitudes with relative uncertainties 
\citep[MAG\_AUTO from {\tt SExtractor};][]{bertin96} in the {\it grizy} bands and DES derived photometric redshifts 
(DNF\_MEAN) alongside further information present in the catalog such as that derived from SPREAD\_MODEL, which 
allows for object star--galaxy classification \citep{desai12}, flagging of bad areas or band specific 
photometric measurement issues \citep[for additional details on the catalogs, see discussion in][]
{morganson18,abbott18}.

\subsection{Galaxy Cluster Catalogs}
\label{sec:ClusterCatalogs}

For this analysis we adopt the RM-Y3 catalog of 19,795 optically selected clusters and the MARD-Y3 catalog of 2,171 
X-ray selected clusters.  The two cluster samples are described briefly in the following subsections.

\subsubsection{Optically selected RM-Y3 catalog}
\label{sec:Redmapper}

We use optically selected galaxy clusters identified with the red-sequence Matched-filter Probabilistic Percolation 
algorithm \citep[redMaPPer;][]{rykoff14} from the Dark Energy Survey first three years of data (RM-Y3).  As the name 
suggests, redMaPPer (RM) can be used to identify clusters as over-densities of red-sequence galaxies. Precisely, the 
algorithm estimates the probability ($P_{\rm mem}$) of a red galaxy to be a cluster member given as
\BE
P_{\rm mem} = \frac{\lambda {\it u}(x)}{ \lambda {\it u}(x) + {\it b}_{gal}},
\EE
where $x$ is a vector that describes the observable properties of a galaxy (e.g., multiple galaxy colors, i-band magnitude, and position), ${\it u}(x)$ is the density profile of cluster normalized to unity, ${\it b}_{gal}$ is the density of background galaxies and the richness ($\lambda$) is estimated by summing up the membership probabilities of galaxies in the cluster region satisfying the constraint equation. RM has been shown to provide excellent photometric redshifts as well as richness  estimates.  For instance, the catalogs have been shown to have high completeness and purity \citep{rozo_rykoff14, rozo14a, rozo14b} when the RM algorithm is applied to the Sloan Digital Sky Survey (SDSS) Stripe 82 data \citep{annis14}, to the eighth SDSS data release \citep[DR8][]{aihara11} and to the DES-Y1 and science verification (DES-SV) data \citep{rykoff16, soergel16}.

The RM-Y3 catalog (Rykoff et al., in prep) adopts a brighter luminosity threshold of 0.4 $\rm L^*$ rather than the 
minimum scatter luminosity threshold of 0.2 $\rm L^*$ \citep{rykoff12} to get a clean sample of clusters above a 
richness threshold. The center of the cluster is taken to be the position of the brightest cluster galaxy (BCG).
There are 19,795 galaxy clusters with $\lambda \geq 20$ in the redshift range of 0.1$\leq z \leq$0.8 with a median 
redshift of 0.47. 

To estimate masses from the measured richnesses, we adopt an externally calibrated $\lambda$--mass relation.  As 
discussed elsewhere \citep[e.g.][]{capasso19b}, the $\lambda$--mass relation for SZE  \citep{saro15} and X-ray 
\citep{capasso19b} selected cluster samples differs from that of optically selected RM clusters  \citep{mcclintock19}. 
The SZE analysis relies on SPT derived cluster masses, which have been shown to be consistent with those derived through weak 
lensing analyses \citep{dietrich19,stern19,bocquet19}. The X-ray analysis relies on masses derived through a dynamical 
analysis whose systematics have been studied with simulations \citep{mamon13} and through cross-comparisons of 
weak lensing and dynamical masses \citep{capasso19a}.  One clear difference is that the RM sample analysis relies on stacked weak lensing information, measuring the mean mass within bins of richness and redshift, whereas the other studies mentioned infer the underlying mass-observable relation using scatter in the observable at fixed mass and applying corrections for the Eddington bias. 
However, the scale of the Eddington bias correction is too small to explain the differences, and thus it may be that the differences are driven by residual contamination in the optically selected RM sample that is not present in the SZE and X-ray selected samples. 
 
Further study is warranted, but because we are employing 
an optically selected RM sample to study the radio galaxy population, we adopt the \citet{mcclintock19} calibration here 
when computing cluster masses,
\BE
M_\mathrm{200m} = A_{\lambda} \left(\frac{\lambda}{\lambda_{\rm P}}\right)^{B_{\lambda}} \left(\frac{1+z}{1+z_{\rm P}}
\right)^{\gamma_{\lambda}},
\label{eqn:lambda_mass}
\EE
with $A_{\lambda}=(3.08\pm0.15) \times 10^{14}$, $B_{\lambda} = 1.356\pm0.052$ and $\gamma_{\lambda} = 
-0.30\pm0.31$. The pivot richness ($\lambda_{\rm P}$) and redshift ($z_{\rm P}$) are given as 40 and 0.35, 
respectively.  We further correct from $M_\mathrm{200m}$ to $M_\mathrm{200c}$ using a model of the concentration-mass 
relation \citep{diemer15}, where $M_{200 \rm m}$ and $M_{200 \rm c}$ are defined as the masses of the cluster within a sphere where the mean 
density is 200 times the mean and the critical density of the Universe, respectively.

For the RM-Y3 sample, we find that $M_\mathrm{200c}$ is in the range of $9.61 \times 10^{13}$ to $2.62 \times 
10^{15}$ $\rm M_{\odot}$ with a median value of $1.52 \times 10^{14}$ $\rm M_{\odot}$.  There are $\sim$11,800 
clusters within $0.1<z<0.8$ in the SUMSS region.
Fig.~\ref{fig:redmapper} shows the mass and redshift distributions of the clusters.

\subsubsection{X-ray selected MARD-Y3 catalog}
\label{sec:MARD}
We use the X-ray selected RASS cluster catalog confirmed and de-contaminated with the multi-component matched filter 
(MCMF) applied to DES-Y3 data to compare with radio source properties in RM-Y3 clusters. This X-ray selected 
cluster catalog \citep[MARD-Y3;][]{klein19} contains galaxy clusters confirmed using a multi-component matched filter 
\citep[][MCMF]{klein18} follow up in 5,000\,deg$^2$ of the DES-Y3 optical data of the $\sim20,000$ sources from the second ROSAT All-Sky Survey 
source catalog (2RXS) presented in \cite{boller16}. The MCMF tool is designed for use on large scale imaging surveys 
such as the DES to do automated confirmation, redshift estimation and suppression of random superpositions of X-ray/
SZE cluster candidates and physically unassociated optical systems.  MCMF is used to identify optical counterparts as 
peaks in galaxy richness as a function of redshift along the line of sight toward each 2RXS source within a search 
region informed by an X-ray prior.  All peaks are assigned a probability $f_{\rm cont}$ of being a random superposition 
and $f_{\rm cont}$ is extracted from the galaxy richness distributions along large numbers of random lines of sight.
In this work, we present radio properties using a catalog with $f_{\rm cont}<0.2$ containing 2,171 galaxy clusters. The catalog 
covers a redshift range of $0.02<z<1.1$ with more than 100 clusters at $z>0.5$, and it has residual contamination of 
9.6~percent. 

The X-ray luminosity of clusters in the catalog is estimated from the source count rate in the 0.1-2.4\,keV band 
within a $5^{\prime}$ aperture radius around each 2RXS source. This simplified luminosity $L_{\rm X}$ has been shown to be
simply related to $L_{\rm 500c}$, the luminosity within a radius within which the mean density is 500 times the critical density of the universe 
at the assumed cluster redshift \citep{klein19}.  
The mass in the catalog is derived using the estimated luminosity at that redshift and an $L_{\rm X}$--mass scaling 
relation from the analysis of \cite{bulbul19}, which uses SZE based mass constraints from the cosmological analysis of 
the SPT-SZ cluster sample \citep{dehaan16} together with deep \XMM\ observations of a subset of those 
clusters to consistently derive multiple observable to mass relations.  These SPT-SZ masses have since been shown to be
consistent with weak lensing \citep{dietrich19,stern19,bocquet19} and dynamical masses \citep{capasso19a}.
The scaling relation has the following form
\BE
\label{EQ:MARDY3-scaling}
L_{\rm X} = A_{\rm X} \left(\frac{M_{500\rm c}}{M_{\rm piv}}\right)^{B_{\rm X}} \left(\frac{E(z)}{E(z_{\rm piv})}\right) 
\left(\frac{1+z}{1+z_{\rm piv}}\right)^{\gamma_{\rm X}}.
\EE
Here, $A_{\rm X}$, $B_{\rm X}$ and $\gamma_{\rm X}$ are the parameters with best fit values of $4.15^{+1.10}_{-0.81}\times 
10^{44}$~erg s$^{-1}$, 1.91$^{+0.18}_{-0.15}$ and $0.20^{+0.41}_{-0.43}$, respectively 
\citep[][table 5 solution II for $L_{\rm X,cin}$]{bulbul19}. 
The pivot mass $M_{\rm piv}$ and redshift $z_{\rm piv}$ are 
$6.35\times 10^{14}$~$\rm M_{\odot}$ and 0.45, respectively. The function
$E(z)\equiv H(z)/H_0$ which gives the ratio between the Hubble parameter and its present-day value, and $M_{500\rm 
c}$ is defined as the mass of the cluster within a sphere where the mean density is 500 times the critical density of the 
Universe. We account for Eddington bias following the description in \cite{mortonson11} and using log-normal scatter in 
the $L_{\rm X}$-mass relation $\sigma_{\ln \rm X}=0.25^{+0.08}_{-0.13}$ \citep{bulbul19}. Further, we correct from $M_\mathrm{500c}$ to 
$M_\mathrm{200c}$ again using a model of the concentration-mass relation \citep{diemer15}.

The median mass and redshift  of the MARD-Y3 sample are $5.06 \times 10^{14}$ $\rm M_{\odot}$ and 0.28, 
respectively, and the distributions are shown in Fig~\ref{fig:redmapper}.  A mass versus redshift plot for the MARD-Y3 sample is presented in \citet[][figure 9 grey and black circles]{klein19}. 
Approximately 1,000 MARD-Y3 clusters between
$0.1<z<0.8$ lie within the SUMSS region and are used for this analysis. This sample is significantly smaller than the RM-Y3 sample, however, the unique mass and redshift range as well as the selection of MARD-Y3 sample in X-ray band provides an interesting complement to the RM-Y3 sample.

The cluster centers in the MARD-Y3 catalog are identified from optical data \citep[see section 3.9;][]{klein19} as the position of the rBCG, the brightest galaxy 
within 1.5\,Mpc of the X--ray centroid that has all colors within 3-$\sigma$ of red sequence at the redshift of cluster, as long as that position is within 
1$'$ of the peak of the red galaxy distribution of the cluster.  
If the rBCG is too offset from the red galaxy population, the cluster center is defined to be that of the red galaxy population.

\begin{figure*}
\centering
\includegraphics[trim={0 5cm 0 1.5cm},clip,width=8.5cm, scale=0.5]{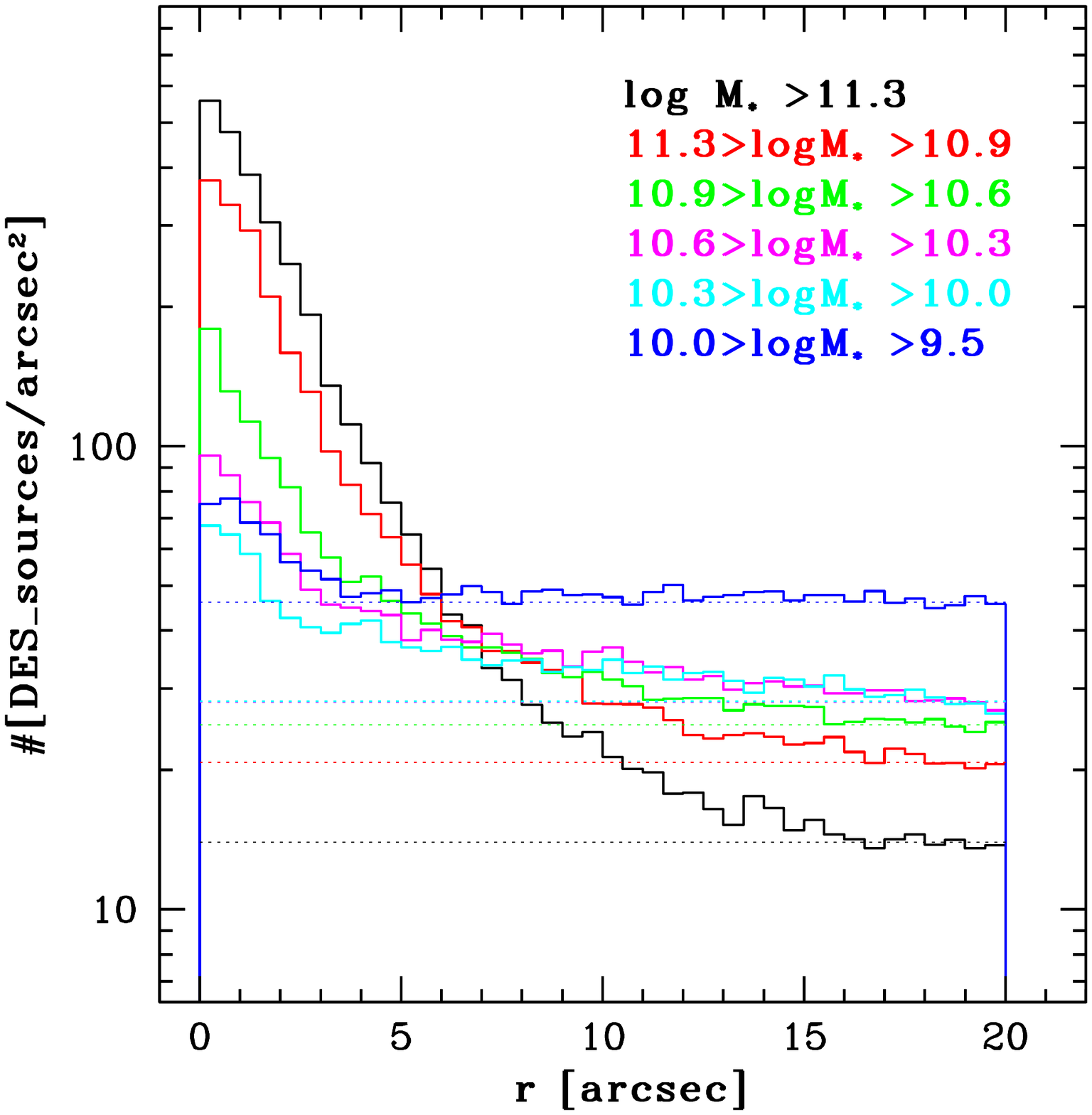}
\includegraphics[width=8.5cm, height=8cm]{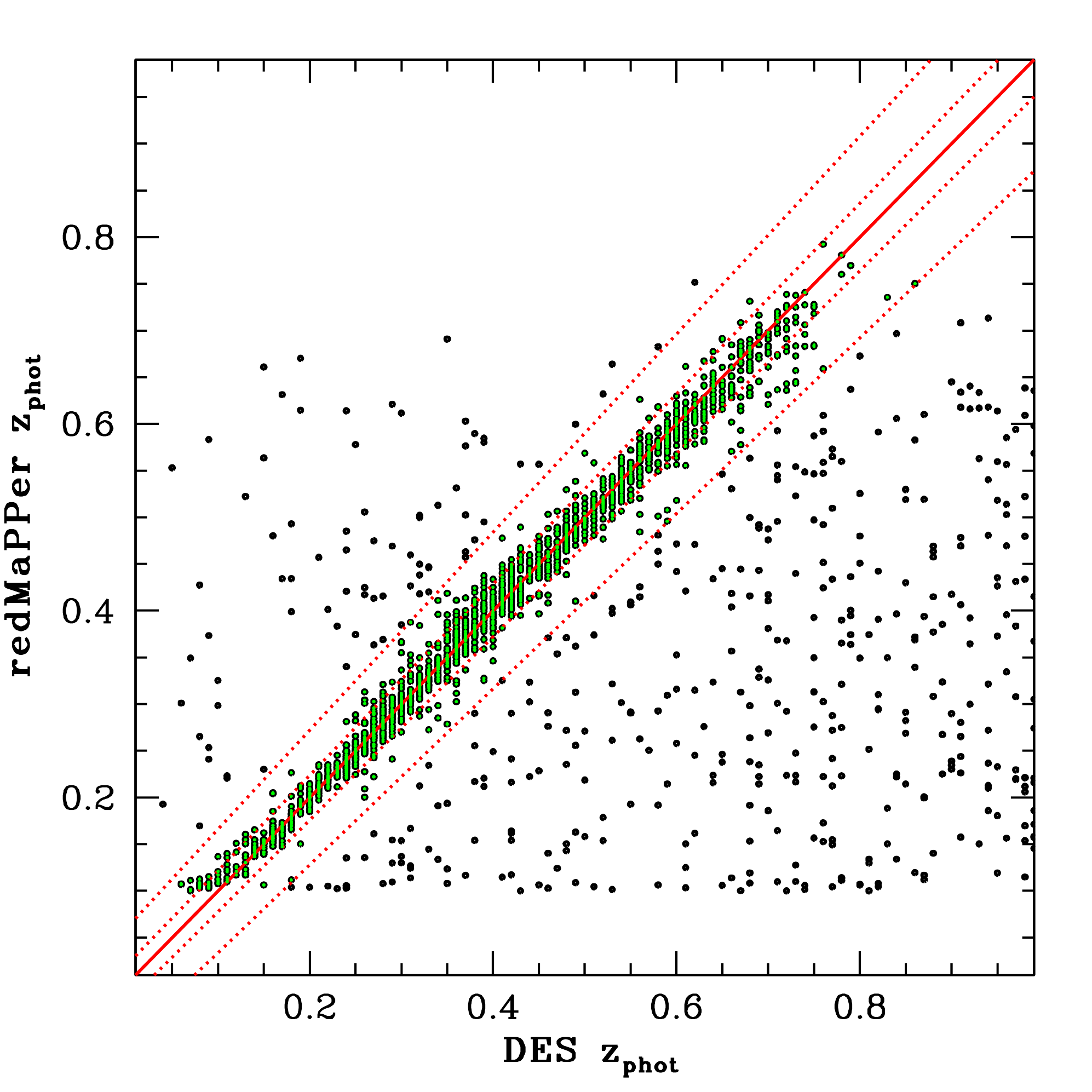}
\vskip-0.1in
\caption{The surface number density of DES galaxies (left) as a function of the offset from SUMSS radio AGN (solid lines) for different stellar mass ranges.  The peak at small offset is produced by physically associated sources, 
and the decline to a flatter distribution at larger radius is caused by physically unassociated sources.  Dotted lines show 
the level of the flat surface density from non--physical cross-matches for the different stellar mass bins.  These surface densities are used to estimate contamination and completeness as a function of stellar mass for the cross--matched sample.  
High mass galaxies are much more likely to 
be physical counterparts of the radio AGN (see Section~\ref{sec:Crossmatching}).  Comparison between redMaPPer cluster redshifts and the photometric redshifts of SUMSS counterparts of DES galaxies used in this work (right). The solid red line is the bisector, while the dotted lines show the 2\% ($\pm1\sigma$) and 6\% ($\pm3\sigma$) bands. These latter values are estimated as the normalized median absolute deviation (NMAD) of the $|\Delta z|/(1+z)$ distribution. Optical counterparts (green points) are assigned to the cluster if their DES photo-z deviates from that of the cluster by less than $\Delta z_{\mathrm{phot}}=0.06 (1+z)$, i.e., three times the NMAD.}
\label{fig:Counterparts}
\end{figure*}

\subsection{Optical counterpart identification}
\label{sec:Crossmatching}

Our core analysis focuses on the statistical overdensity of radio AGN toward galaxy clusters to constrain the properties 
of cluster radio AGN.  However, this analysis is complemented by an analysis of a subset of radio AGN whose optical 
counterparts we have identified in the DES-Y3 dataset.  Here we describe the counterpart identification and stellar mass 
estimation needed to undertake that analysis.

A complete description of the derivation and accuracy of galaxy stellar masses as well as the cross-matching technique used to uniquely associate a DES counterpart to each SUMSS radio source will be provided in a forthcoming work (Pannella et al., in prep.). Here, we briefly summarize the main steps of our procedure:
1) for each radio source above the nominal SUMSS survey completeness limit in radio flux density, we select all DES 
catalog entries lying within 20$^{\prime\prime}$; 2) the catalog of plausible DES counterparts   
is trimmed down by applying a cut in {\it i}$_{mag}<23$ AB magnitudes, 
and by removing unresolved sources from the catalog by imposing that the EXTEND\_CLASS flag is greater than 0, 
and, finally, by removing all galaxies for which the DES photometric redshift 
was not constrained. 

We derive stellar masses for all the objects in our sample with {\it fastpp}\footnote{Publicly available at https://
github.com/cschreib/fastpp} \citep[a C++ version of the SED fitting code {\it FAST};][]{fast} on the
$grizY$ total magnitudes. We use \citet[]{bc03} with
delayed exponentially declining star formation histories (SFHs,
$\psi(t) \propto \frac{t}{\tau^2} \exp(-t/\tau)$) with 0.01$<\tau<$10
Gyr, solar metallicities (Z$_\odot$ = 0.02), the Salpeter initial mass function \citep[][]{salpeter55}, and the \citet{Calzetti00}
reddening law with a range of extinction A$_V$ up to 4 magnitudes. 

We then examine the surface density on the sky of DES sources as a function of their distance from 
the SUMSS radio AGN for different bins of stellar mass. This is shown in the left panel of Fig.~\ref{fig:Counterparts} with 
solid lines in six different colors, corresponding to different ranges of stellar mass.  These surface density profiles exhibit 
a peak at small separation, which corresponds to actual physical counterparts dropping away to a flat surface density 
that corresponds to a region dominated by physically unassociated galaxies projected at random near the radio AGN.  It 
is evident that for higher stellar mass the central peak has a higher contrast with respect to the flat background, which is an 
indication that for these stellar mass ranges a larger fraction of galaxies correspond to physical counterparts of the 
radio AGN.  This is not surprising, because AGN powered radio sources are typically hosted by passive, massive and
bulge-dominated galaxies \citep{best07,lin07,kauffmann08,lin10,best12}. 


By fitting for the level of the flat surface density from random (i.e., non--physical) cross-matches (see dotted lines in the left panel of Fig.~2), we can then estimate
for any given matching radius the resulting contamination of the sample.  In addition, we can integrate over the portion of the central
peak extending beyond the matching radius to estimate the incompleteness within each mass bin for a given matching radius.
For the purposes of this analysis, we adopt a simple matching criteria that produces a sample with no more than 15~percent contamination.  
Practically, this means that we define the cross--matched host for each radio AGN to be 
the nearest DES galaxy with $i<23.0$ that has stellar mass in one of the top two
bins ($\log{M_*}>11.3$ and $11.3>\log{M_*}>10.9$), within a maximum offset distance of 10$''$ or 6$''$, respectively.  
Within the lower mass bins it is not possible to define a sensible maximum offset distance where the contamination is as low as 15~percent.
The estimated completeness of this cross--matched sample is $\sim$60~percent.  

This host identification procedure produces a catalog of 24,998 unique associations of SUMSS detected radio AGN to a DES galaxy host. In this work we specifically concentrate on the
2,264 candidate cluster radio AGN that lie within the radius $\theta_{200}$ of a 
RM-Y3 cluster.  Among this latter sub-sample, only 1,643 (73~percent) have photometric redshifts that place them within a cluster (see the right panel of Fig.~\ref{fig:Counterparts}).
To assign cluster membership to the radio AGN, we compare the DES  photometric redshift (DNF\_MEAN) of each source to the 
redshift of the RM-Y3 cluster.  If a source deviates from the cluster redshift by less than 3 times the normalized median absolute deviation (NMAD) ($\delta z = \Delta z/
(1+z) = $~0.02), we consider it to be a cluster member and therefore a confirmed cluster radio AGN.
We do not carry out separate cross-matching in the MARD-Y3 sample,
because the sample is an order of magnitude smaller than the RM-Y3 sample.


\section{Cluster radio AGN properties}
\label{sec:Results}

We present measurements of the radial distribution of AGN around clusters in Section~\ref{sec:RadialProfile}, and 
present new constraints on the radio luminosity and color evolution of radio AGN using the sample of matched optical counterparts 
in Section~\ref{sec:HostProperties}.   Leveraging these results, we then present the cluster radio AGN luminosity function 
in Section~\ref{sec:LF} and the halo occupation number in Section~\ref{sec:HON}.  

%
\begin{figure}
\centering
\includegraphics[width=8.5cm, height=8cm]{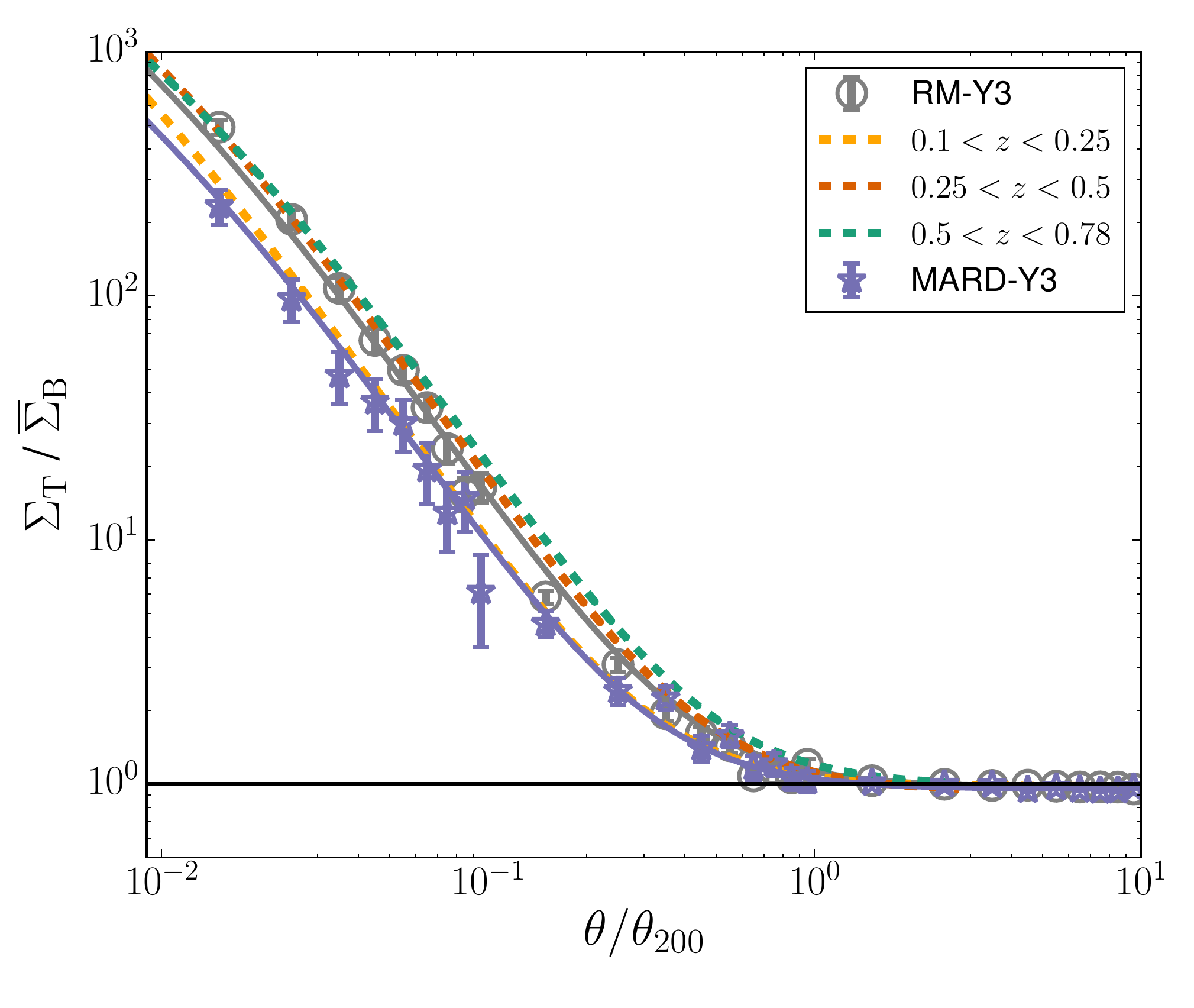}
\vskip-0.1in
\caption{The projected radial distribution of radio AGN observed at 843~MHz around the centers of the optically 
selected RM-Y3 (gray) and X-ray selected MARD-Y3 (blue) galaxy clusters.  The best fit projected NFW models are 
also shown as solid lines in gray and blue with the model parameters listed in Table~\ref{Table:NFWparameters}.  The fits 
are evaluated using much smaller bins ($\sim$ 0.001 $\theta/\theta_{200}$) in comparison to the data points shown 
here, which use larger bins to reduce Poisson noise for the figure. We also show the best fit NFW models in three 
redshift bins using the RM-Y3 catalog represented by dashed lines.  We see a decrease in concentration of radio 
sources in clusters with increasing redshift for both RM-Y3 and MARD-Y3 selected cluster samples (see Table~\ref{Table:NFWparameters}).}
\label{fig:SD_profiles}
\end{figure}

\subsection{Radial distribution}
\label{sec:RadialProfile}

We study the radial distribution of radio AGN in the cluster $\theta_{\rm 200c}$  region by stacking the flux limited and complete samples of 
radio AGN overlapping the RM-Y3 and MARD-Y3 samples.  To do this we adopt the optical centers for both samples, described in detail in separate references \citep{rykoff14,klein19}.

Following \citetalias{gupta17}, we use the projected NFW 
profile $\Sigma(x)$ \citep{navarro97, bartelmann96} as a fitting function for the radial distribution. Here $x=r/r_{\rm s}$ 
and $r_{\rm s}=R_{\rm 200c}/c$, where $c$ is the concentration parameter and $x$ is equivalent to $c$ for $r=R_{\rm 
200c}$.  The total surface density of the radio AGN ($\Sigma_{\rm T}$) as a function of $x$ around the cluster has both 
cluster and background components ($\Sigma_{\rm B}$) and is written as
\BE
\Sigma_{\rm T(x)}=\Sigma(x)+\Sigma_{\rm B},
\EE 
and to reduce the covariance between the central amplitude and concentration, we write this in terms of the total number of 
galaxies in the cluster sample as 
\BE
\Sigma N_{\rm T}(x)=\Sigma N(x)+\Sigma_{\rm B}A,
\label{eqn:no_gal}
\EE 
where $A$ is the solid angle of the annulus or bin and the total number of background subtracted galaxies $\Sigma N(x)=\Sigma N_{200}$ for 
$r=R_{\rm 200c}$. We fit our stacked distribution of radio AGN to a model with three parameters: $c$, $\Sigma N_{200}$ and $
\Sigma_{\rm B}$ \citepalias[see][]{gupta17}. We stack radio AGN out to $10\times\theta_{\rm 200c}$ to allow for a good 
constraint on the effective background density $\Sigma_{\rm B}$.  We refer to $\Sigma_{\rm B}$ as the effective background density,
because our stacks are constructed using sources down to the flux limit of the SUMSS survey, and as described in Section~\ref{sec:SUMSS} this
flux limit has two different values, depending on the declination of the cluster.

In the fit we employ the \citet{cash79} statistic 
\BE
\begin{split}
\label{eqn:cash_lik}
\it{C} = \sum_{i} & \left(
\Sigma N_{{\rm T},i}^{\rm d} \ln(\Sigma N_{{\rm T},i}^{\rm m}) - \Sigma N_{{\rm T},i}^{\rm m}\right. 
 - \Sigma N_{{\rm T},i}^{\rm d} \ln(\Sigma N_{{\rm T},i}^{\rm d}) \\
& \left.+ \Sigma N_{{\rm T},i}^{\rm d},
\right),
\end{split}
\EE
where $\Sigma N_{\rm T, i}^{\rm m}$ is the total number of galaxies from the model as in equation~(\ref{eqn:no_gal}) and 
$\Sigma N_{\rm T, i}^{\rm d}$ is the total number of galaxies in the observed data in the $i^{\rm th}$ angular bin.
We use the Markov Chain Monte Carlo (MCMC) code, {\tt emcee} \citep[a Python implementation of an affine invariant 
ensemble sampler;][]{mackey13} to fit the model to the data throughout this work. In the fitting we adopt a bin size 
corresponding to $\theta_{\rm 200c}/1000$ and fit over the region extending to $10\theta_{\rm 200c}$. 
The concentration parameter is sampled in $\rm log$ space during the fit and the profile is centrally concentrated with 
$c=143 ^{+10}_{-9}$ and $144 ^{+30}_{-25}$ for the RM-Y3 and MARD-Y3 cluster samples, respectively.

We use this model of the radial distribution of radio AGN in the next section to correct the projected LF to the 
LF within the cluster virial region defined by $R_{\rm 200c}$ \citep[following][]{lin04a}.  
We measure the surface density profiles by dividing RM-Y3 and MARD-Y3 catalogs in three redshift bins and see a 
clear tendency for higher concentration at lower redshifts in both catalogs. The best fit NFW parameters are presented 
in Table~\ref{Table:NFWparameters}.  
We use the flux limited sample rather than a luminosity limited sample for this measurement because of the larger number of available systems. This does not affect the trends in concentration seen here.

We test the reliability of trends in concentration for mis-centring in the RM-Y3 sample. We select clusters for which the probability that each of the alternate cluster centres is the correct centre $P_{\rm cen}>0.95$ \citep{rykoff14}. We find that this sub-sample of clusters show similar concentration trends with $c=250^{+40}_{-35}$, $141 ^{+21}_{-18}$ and  $107 ^{+18}_{-16}$ for low to high redshift bin in Table~\ref{Table:NFWparameters}.

High central concentrations of radio sources in galaxy clusters have been observed in previous studies. \citet{lin07} studied 
radial distribution of cluster radio AGN observed at 1.4~GHz with $P>10^{23}$~W Hz$^{-1}$ in a sample of X-ray 
selected galaxy clusters with $z<0.2$, finding $c=52^{+24}_{-14}$. \citetalias{gupta17} studied radial profiles of cluster 
radio AGN at 150~GHz in the Meta-Catalog of X-ray detected Clusters of galaxies (MCXC) with median 
$z=0.1$ and found $c \sim 100$.  The results presented here indicate that centrally concentrated cluster radio AGN are 
present over a broad range of cluster mass $M_{200c}>1\times10^{14}$ and out to redshift $z\sim0.8$.

In Fig.~\ref{fig:SD_profiles} we show the best fit surface density profiles and data. To create these plots we combine 
many bins to reduce the noise in the measured radial profiles. Following \citetalias{gupta17}, we normalize the vertical axes 
of this plot with the mean number density of background sources ($\overline{\Sigma}_{\rm B}$). Fig.~\ref{fig:SD_profiles} shows that $\overline{\Sigma}_{\rm B}$ is a good estimation of the background number density of the clusters, as $\Sigma_{\rm T}$/$\overline{\Sigma}_{\rm B}$ is 
consistent with 1 outside the cluster.

\begin{table}
\centering
\caption{Best fit projected NFW model parameters for the radial distribution of a complete sample of radio AGN 
observed at 843~MHz in a stack of optically selected RM-Y3 and X-ray selected MARD-Y3 galaxy clusters. The best fit 
parameters are also shown for subsets of these clusters in three redshift bins where for each we present concentration 
$c$, the estimate of the total number of radio AGN within $R_{200c}$ in our sample $\Sigma N_{200}$ and the effective  background 
surface density of radio AGN $\Sigma_{\rm B}$.  
}
\label{Table:NFWparameters}
\begin{center}
\begin{tabular}{lcccc}
\hline\hline
\multicolumn{1}{l}{Cluster sample} & \multicolumn{1}{c}{$c$} & \multicolumn{1}{c}{$\Sigma N_{200}$} & \multicolumn{1}{c}{$
\Sigma_{\rm B}$ [$\rm deg^{-2}$]} \\ 
\hline
RM-Y3
(all)                    & $143 ^{+11}_{-10}$	       & $2198^{+55}_{-53.}$      & $20.23\pm0.05$ \\[3pt]
$0.1<z<0.25$  & $203^{+35}_{-29}$	   & $553^{+27}_{-26}$         & $20.33\pm0.11$ \\[3pt]
$0.25<z<0.5$  & $140^{+16}_{-15}$        & $1068^{+36}_{-37}$       & $20.00\pm0.12$ \\[3pt]
$0.5<z<0.8$    & $105^{+16}_{-13}$        & $583^{+27}_{-26}$         & $20.51\pm0.20$ \\[3pt]
\hline
MARD-Y3  
(all)                    & $144^{+30}_{-25}$        & $425^{+30}_{-29}$         & $19.89\pm0.12$ \\
$0.1<z<0.25$  & $193^{+60}_{-28}$	        & $221^{+18}_{-17}$            & $19.91\pm0.15$ \\[3pt]
$0.25<z<0.5$  & $120^{+40}_{-28}$        & $160^{+15}_{-14}$           & $19.62\pm0.24$ \\[3pt]
$0.5<z<0.8$    & $100^{+58}_{-40}$        & $48^{+08}_{-07}$            & $21.20\pm0.68$ \\[3pt]
\hline\hline
\end{tabular}
\end{center}
\end{table} 
\begin{figure*}
\centering
\includegraphics[width=8.5cm, height=8cm]{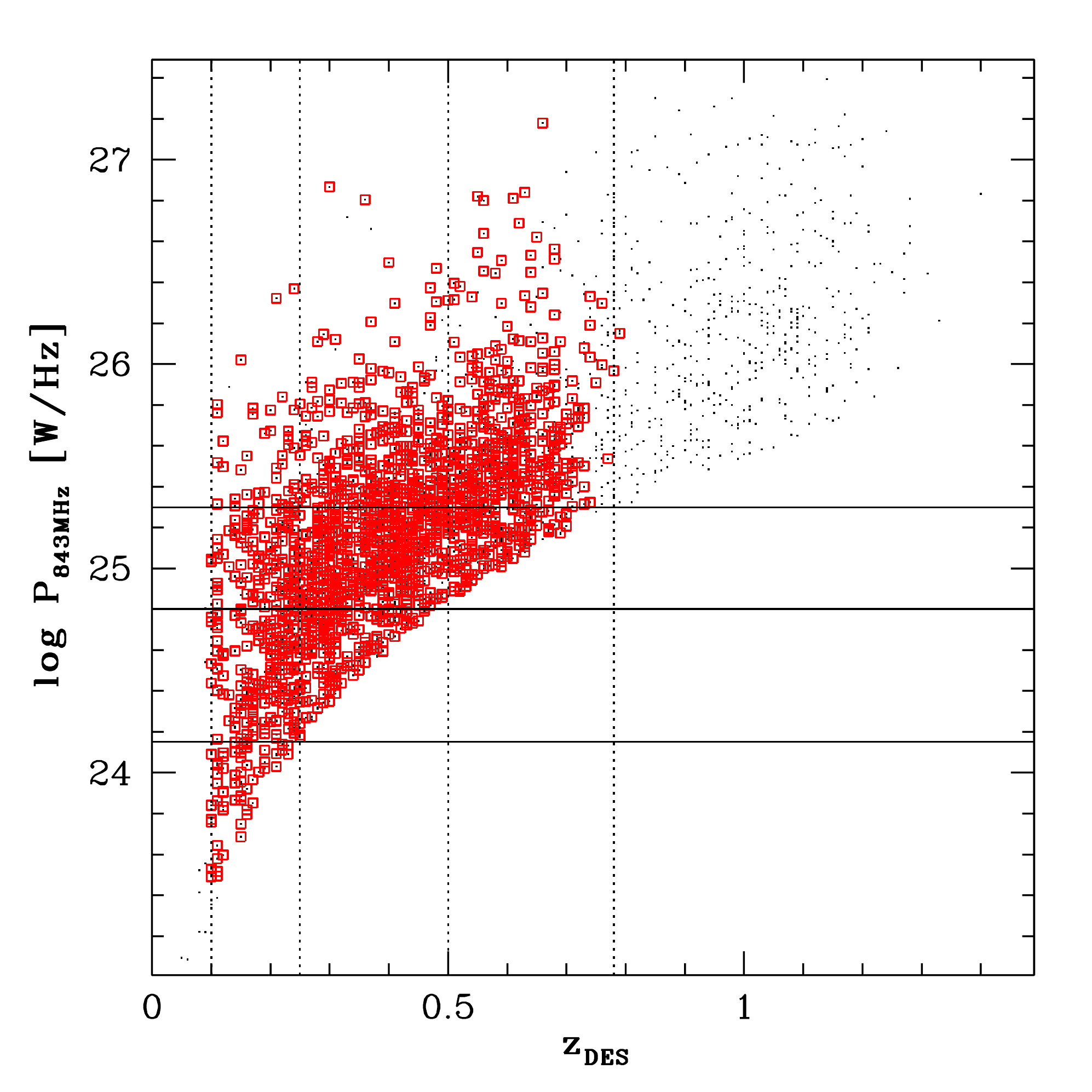}%
\includegraphics[width=8.5cm, height=8cm]{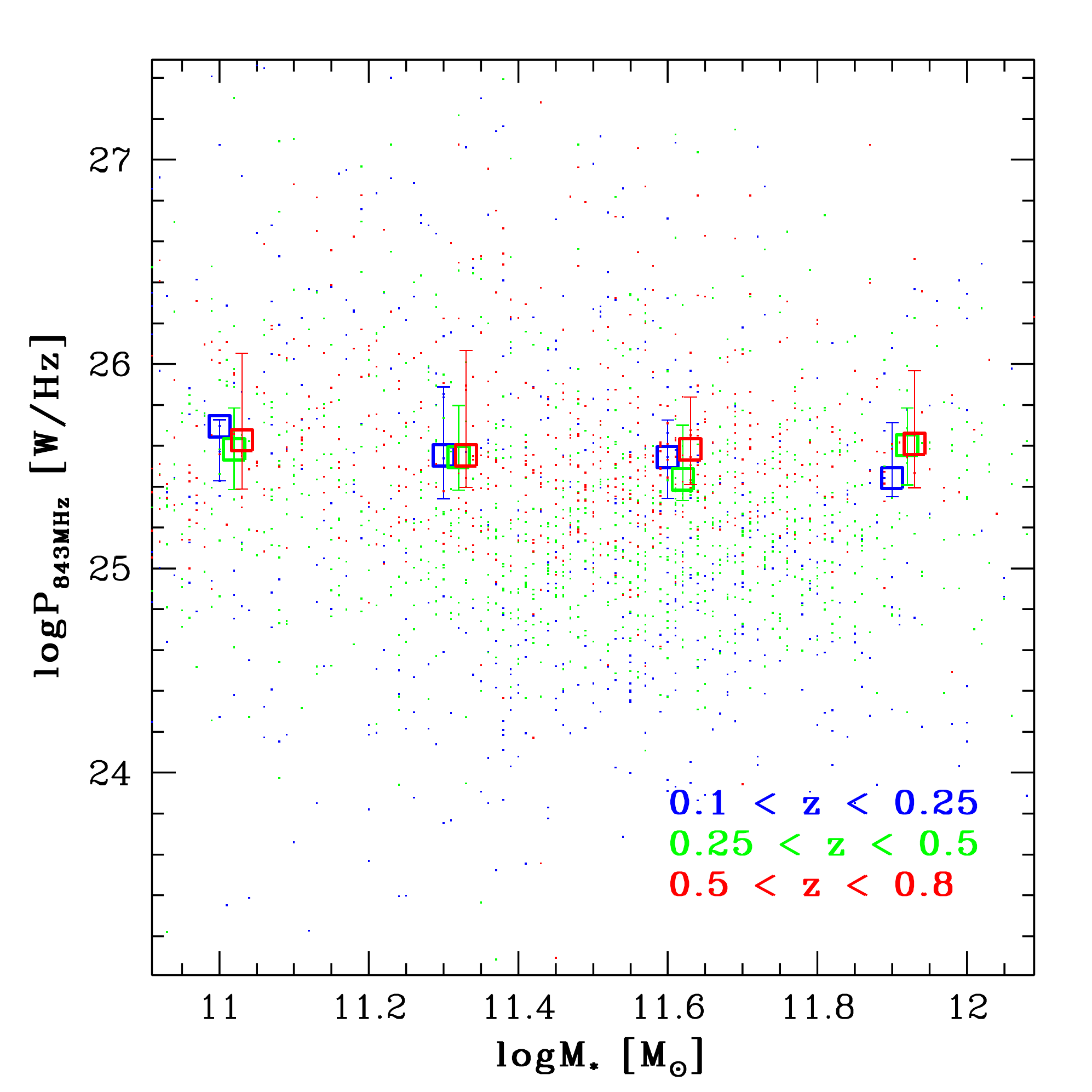}
\vskip-0.1in
\caption{Luminosity distribution of SUMSS sources (left) observed at 843~MHz as a function of redshift. The redshifts 
are obtained by cross matching SUMSS with the DES-Y3 catalogs. The black points show radio sources inside $\theta_{200}$ of
RM clusters, the red squares show instead the radio sources identified as cluster members.  The vertical dotted lines represent the three redshift bins used in this work, while the solid horizontal lines show the luminosity completeness levels at different redshifts. Distribution of SUMSS detected cluster members (right) in radio luminosity and stellar mass, 
color coded by redshift (redshift ranges labeled). Empty squares show median luminosities in bins of stellar mass and 
redshift over a luminosity range where completeness is assured at all redshifts, log P$_{\rm 843MHz}$ > 25.3, see left panel. Our results 
are consistent with a modest increase in the luminosity of cluster radio AGN at fixed stellar mass over the explored 
redshift range (see Section~\ref{sec:LuminosityEvolution}).}
\label{fig:LMRelation}
\end{figure*}

\subsection{Host properties of the cross--matched sample}
\label{sec:HostProperties}

In the RM-Y3 cluster sample there are 1,643 cluster radio AGN with identified optical counterparts.  This cross--matched sample 
has an estimated contamination of 15~percent and an estimated completeness of 60~percent (see Section~\ref{sec:Crossmatching}).  
Of the 11,800 RM-Y3 clusters lying within the SUMSS region, 1,579 (13~percent) host at least one confirmed cluster radio AGN associated with the cluster by its redshift.  The vast majority of these (95~percent) 
contain a single cluster radio AGN, 87 contain two radio AGN, one cluster contains three 
and one other contains four radio AGN all consistent with their cluster redshift.  Finally, 1,044 (63~percent)
of the total sample of 1,643 cross--matched AGN are projected to 
lie within $0.1r_{200}$, consistent with the centrally concentrated radial profile for the 
statistically defined complete sample in Section~\ref{sec:RadialProfile}. 

In the subsections below we describe the radio power to stellar mass relation of this sample (Section~\ref{sec:LuminosityEvolution}) 
and then examine the typical broad band colors as a function of redshift (Section~\ref{sec:SEDEvolution}).

\subsubsection{Radio luminosity evolution at fixed host mass}
\label{sec:LuminosityEvolution}

To constrain the evolution of the radio luminosity with redshift at fixed stellar mass, we use the ensemble of SUMSS 
sources with DES-Y3 counterparts.  In particular, we take the high luminosity sample of radio sources with $
\log{[\mathrm{P/(W\,Hz^{-1})}]}>25.3$ in the whole redshift range of DES-Y3 RM clusters, corresponding to 510 cluster radio AGN. 
This choice of luminosity cut is justified in 
Fig.~\ref{fig:LMRelation}, where we show the relation between the 843~MHz luminosity and the redshift of radio AGN. 
Above this luminosity cut, we have a sample of sources that can be studied over the full redshift range of our cluster 
samples, whereas if we were to push to lower radio luminosities we would only expect to find those sources in our flux 
limited sample over a narrower range of lower redshifts.

We fit a power law relation of the following form to constrain a relation between 843~MHz radio power
($P_\mathrm{843MHz}$, calculated using the redshift of each radio source) and the stellar mass ($M_{*}$) of radio 
sources obtained from the SED fitting analysis described in Section~\ref{sec:Crossmatching}
\BE
\label{eq:LM_rel}
P_\mathrm{843MHz}  = 10^{A_{\rm PM}} \left(\frac{M_{*}}{M_\mathrm{*,piv}}\right)^{B_{\rm PM}}  \left(\frac{1+z}
{1+z_\mathrm{piv}}\right)^{\gamma_{\rm PM}},
\EE
where $M_\mathrm{*,piv}=3.7\times 10^{11}$\,$\rm M_{\odot}$ and $z_\mathrm{piv}=0.51$ are the median stellar mass and redshift of the subsample.
 The best fit parameters are $A_{\rm PM}=25.51\pm0.006$, $B_{\rm PM}=0.01\pm0.02$ and $\gamma_{\rm PM}
=0.30\pm0.15$. Along with these parameters we also vary the log normal scatter in radio luminosity at fixed stellar 
mass in the relation, finding a best fit value $\sigma_{\rm log{P}}=0.31\pm0.01$. 

These results make it clear that, unlike X-ray AGN, cluster high power radio AGN exhibit no correlation ($B_{\rm PM}=0.01\pm0.02$) 
between their radio luminosity and their stellar mass
(which is a predictor of the underlying supermassive black hole mass) at all redshifts probed by our sample.  
This  characteristic of radio mode feedback has previously been noted \citep{best07,lin07}.

The relation has a scatter of $\Delta\log_{10}{P}=0.31\pm0.01$, and the 
population we are studying exhibits a weak trend for the radio luminosity at fixed stellar mass to increase with 
redshift $\gamma_{\rm PM}=0.30\pm0.15$.  Along with this weak luminosity evolution with redshift, it has been noted
that the mechanical feedback in comparison to the radio luminosity is also approximately constant with redshift
to $z\sim0.6$ \citep{larrondo12}.  This provides an indication that individual radio galaxies are
only moderately more radio luminous at high redshift and that the feedback events driving radio emission and radio mode
feedback are not changing dramatically with redshift.
We will use this new constraint on cluster radio AGN evolution 
to disentangle density evolution and luminosity evolution 
in the cluster radio AGN luminosity function in Section~\ref{sec:LF}.

\begin{figure}
\centering
\vskip-0.35in
\includegraphics[width=8.5cm, height=8cm]{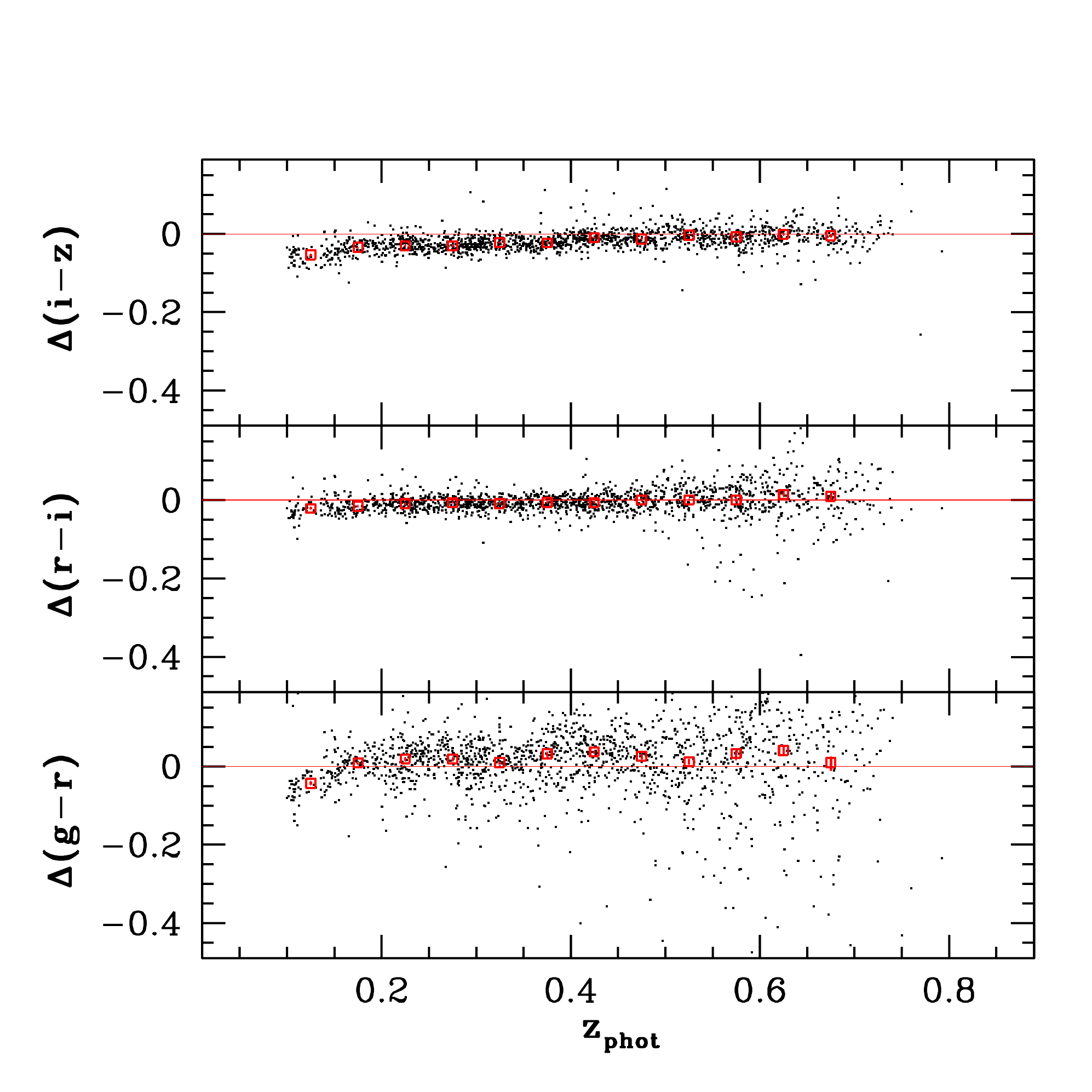}%
\vskip-0.1in
\caption{Broad band color offsets from the red sequence as a function of redshift 
for the cross--matched hosts of the SUMSS cluster radio AGN.
At all redshifts the host population has colors very similar to typical red sequence galaxies.  Furthermore, 
there is  no obvious evidence for an evolution
of median rest-frame colors with redshift (see discussion in Section~\ref{sec:SEDEvolution}).  
Overplotted as boxes are the median color offsets of the hosts within redshift bins.}
\label{fig:HostColors}
\end{figure}

\subsubsection{SED evolution}
\label{sec:SEDEvolution}

Broad band colors constrain the SED of the cluster radio AGN hosts and provide insights into the presence 
of ongoing star formation or optical AGN emission.
In Fig.~\ref{fig:HostColors} we plot the broad band color offsets of the radio AGN hosts from the mean 
red sequence (RS) color as a function of redshift.  The RS colors are defined using a sample of $\sim$10$^3$ 
clusters with spectroscopic redshifts studied within 
 DES as part of the calibration of an optical counterpart and redshift estimation code called 
 MCMF \citep{klein18,klein19}.  Because the RS color model makes use of single object fitting (SOF) magnitudes 
 present in the DES catalog (see \citealt{klein19} and references therein for more details), for this purpose we 
 use the SOF magnitudes for the SUMSS source counterparts. 
 
 It is apparent that the hosts of radio AGN tend to have integrated colors very close to the RS at all redshifts.  
 This is consistent with the established picture that radio AGN are mostly hosted in massive galaxies 
 with very little or no ongoing star formation over the explored redshift range.   
   
 The redshift evolution of observed colors does not suggest any clear change in the stellar population properties
 of the host galaxies over time. When comparing rest-frame properties, e.g., the $g$-$r$ at z$\sim$0.4 and 
 the $r$-$i$ at z$\sim$0.7--- both approximately tracing the rest-frame $U$-$B$ color--- one finds only a weak 
 ($\approx$-0.03 mag) signature for color evolution that is very likely 
 driven by the intrinsic uncertainties and scatter of the RS model itself.  The notably higher scatter in $g$-$r$ 
 beyond $z\sim0.4$ and in $r$-$i$ beyond $z\sim0.6$, while reflecting both higher measurement uncertainties and that these 
 bands are tracing rest frame portions of the SED that are most sensitive to recent star formation, could be an indication that a minority population of ``bluer" hosts are actually entering the cross-matched sample.
 
In summary, our analysis does not suggest any major shift happening as a function of redshift in the stellar population 
properties of the radio loud AGN host galaxies out to $z\sim0.8$.  This places limits on the suggested transition in the 
cluster radio AGN from a Low Excitation Radio Galaxy (LERG) dominated to a High Excitation Radio Galaxy 
(HERG) dominated population \citep{birzan17}.  A direct classification as HERG 
or LERG would require spectra, which are currently not available for significant numbers of the AGN hosts we study here.

\begin{figure*}
\centering
\vskip-0.2in
\includegraphics[width=8.5cm, height=8cm]{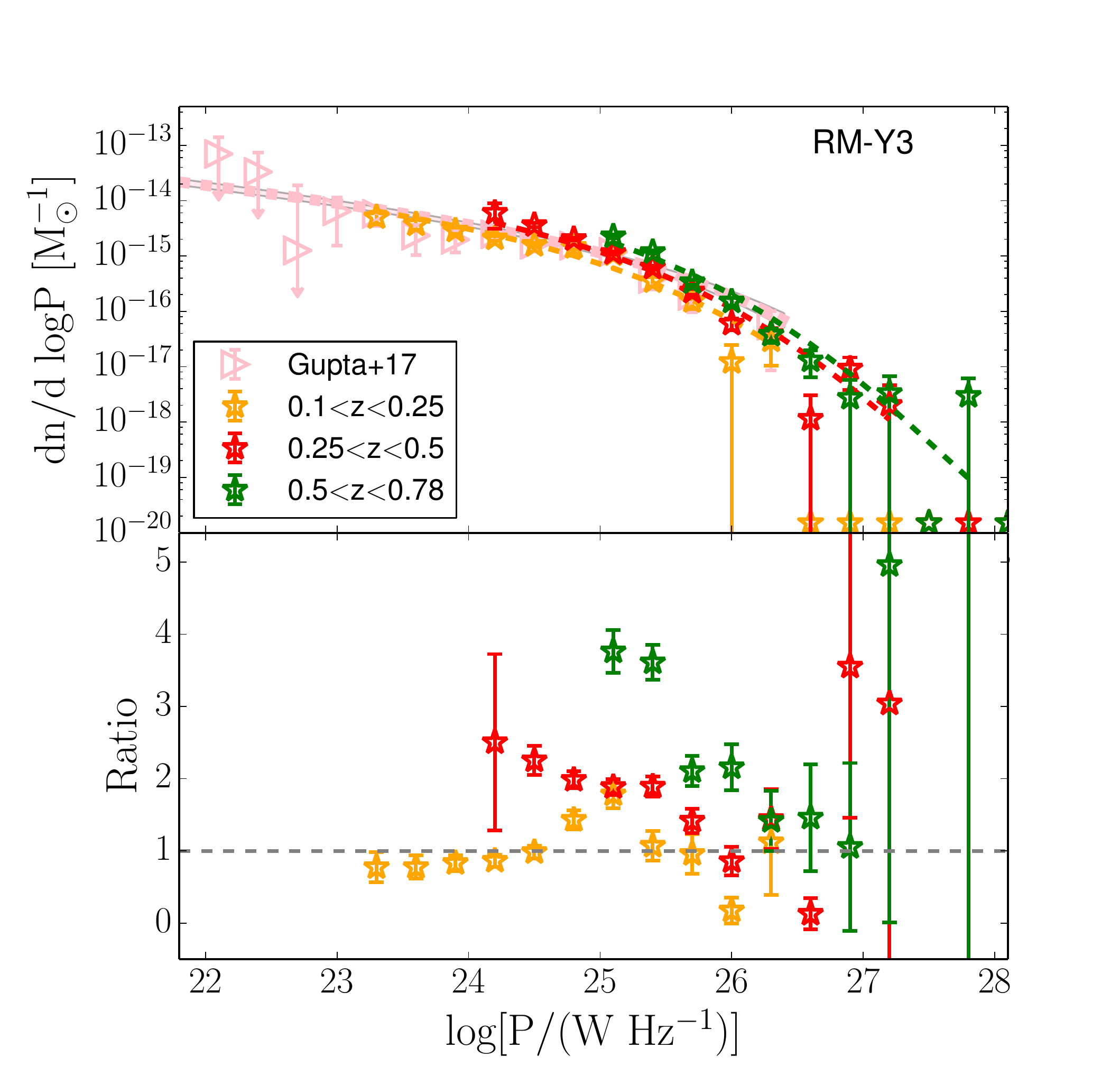}
\includegraphics[width=8.5cm, height=8cm]{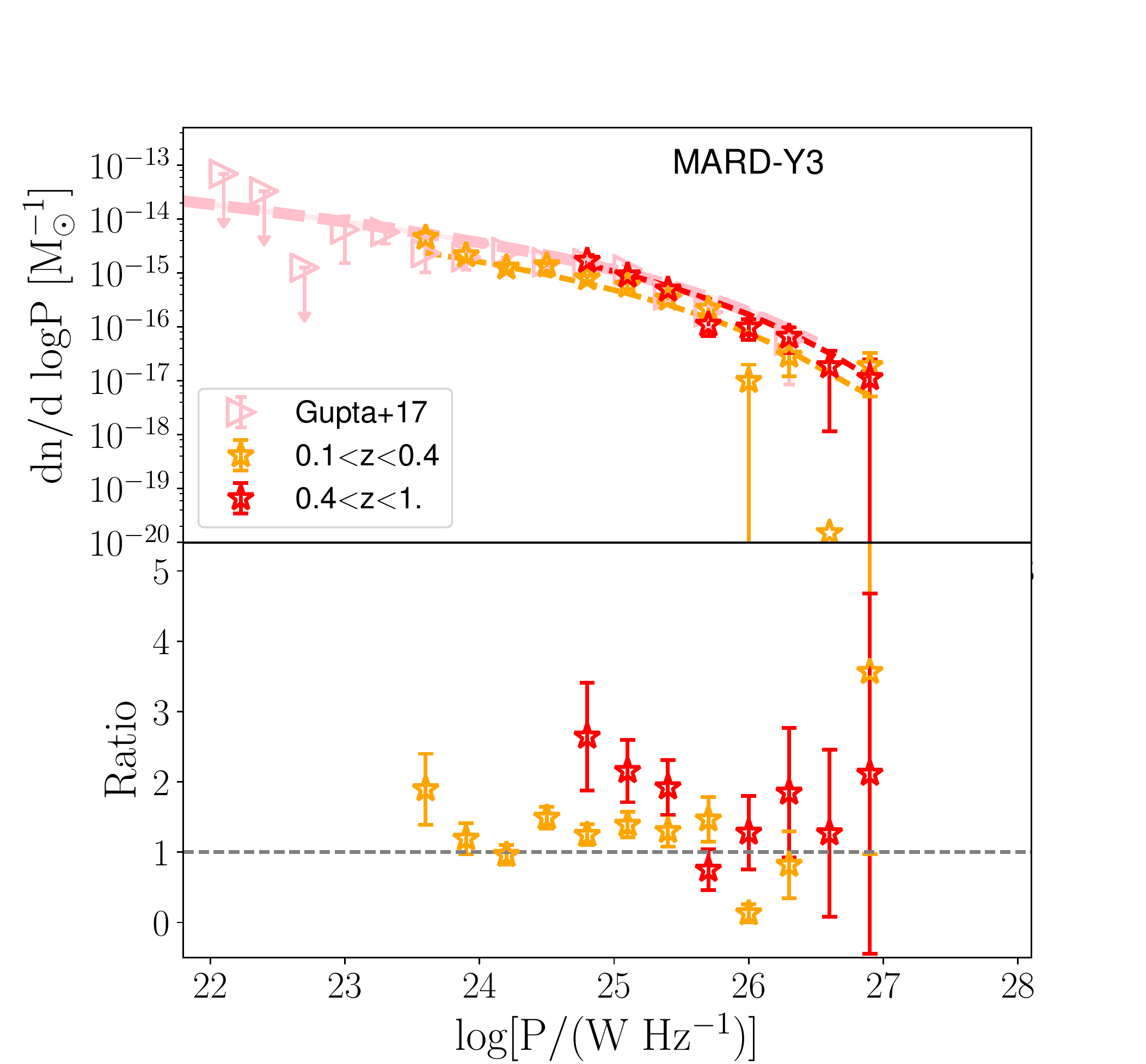}
\vskip-0.1in
\caption{SUMSS based 843~MHz cluster radio AGN LFs observed from RM-Y3 (left) and MARD-Y3 (right) catalogs.  Upper 
panels show the LFs obtained using the statistical background subtraction algorithm described in 
Section~\ref{sec:LFMethod}. The Poisson 
uncertainties are represented by error bars. The datasets are fitted with the LF model by varying $y$, $x$ and $
\gamma$ (density and luminosity evolution) parameters as discussed in Section~\ref{sec:LFModeling}. Different lines 
indicate the best fit model LFs (see Table~\ref{tab:LF}). For convenience, the luminosity bins containing negative 
values in the background subtracted counts are represented as points at the bottom of the figure. Pink points and line represent the best fit model and model uncertainties from \citetalias{gupta17}. Lower panels show 
the ratio between the measurements and the best fit model for the lowest redshift bin in this work.
While fitting we divide the cluster samples into 15 redshift bins, but for this figure we show the results in only two or 
three redshift bins to reduce the Poisson noise.}
\label{fig:LF843}
\end{figure*}
\begin{table}
\caption{The best fit LF parameters for different samples of cluster radio AGN. $\gamma_{\rm D}$ and $\gamma_{\rm P}
$ are defined as the density and luminosity redshift evolution parameters, respectively. The joint constraints on $
\gamma_{\rm D}$ and $\gamma_{\rm P}$ are obtained when a Gaussian prior on $\gamma_{\rm P}$ is adopted from 
the best fit redshift evolution of the stellar mass - luminosity relation.}
\label{tab:LF}
\begin{center}
\begin{tabular}{cccc}
\hline\hline
$y$ & $x$ & $\gamma_{\rm D}$ & $\gamma_{\rm P}$ \\
\hline
\multicolumn{4}{c}{RM-Y3} \\[1pt]
$25.82\pm0.04$   & $26.56\pm0.06$ & $3.16\pm0.43$ & $-$  \\[1pt] 
$25.82\pm0.04$   & $26.56\pm0.06$ & $-$ & $3.16\pm0.42$  \\[1pt] 
$25.75\pm0.04$   & $26.40\pm0.06$ & $3.00\pm0.42$ & $0.21\pm0.15$  \\[1pt] 
\hline
\multicolumn{4}{c}{MARD-Y3} \\[1pt]
$25.83\pm0.12$   & $26.86\pm0.14$ & $2.60\pm0.71$ & $-$  \\[1pt] 
$25.82\pm0.12$   & $26.84\pm0.14$ & $-$ & $2.62\pm0.70$  \\[1pt] 
$25.74\pm0.11$   & $26.70\pm0.15$ & $2.05\pm0.66$ & $0.31\pm0.15$  \\[1pt] 
\hline\hline
\end{tabular}
\end{center}
\end{table}
\subsection{Luminosity function}
\label{sec:LF}
We construct radio LFs by counting the excess of radio AGN toward RM-Y3 and MARD-Y3 galaxy clusters to study their evolution
to $z\sim1$.  In addition, we construct the LF from the cross--matched sample for the purpose of cross-comparison with the LF of the 
complete sample.  In all cases we 
apply the redshift dependent $k$-correction with a spectral index of -0.7 to estimate the luminosity at the same 
rest frame frequency for all redshifts.   

In the following subsections we describe the LF construction in detail 
(Sections~\ref{sec:LFMethod} and \ref{sec:LFCrossmatched}), present the modeling (Section~\ref{sec:LFModeling}) and best 
fit parameters (Section~\ref{sec:LFMeasurements}), compare the LF of the complete sample with that of the 
cross--matched sample (Section~\ref{sec:LFComparison}) and then finally 
discuss how our results compare to those from previous studies (Section~\ref{sec:LFLiterature}).

\subsubsection{Statistical LF construction}
\label{sec:LFMethod}

This adopted method to construct cluster galaxy LFs using statistical background subtraction is described in 
detail in previous works \citep[][G17]{lin04a,lin07}.  Briefly, we adopt the 
cluster redshift to estimate the radio source luminosities for individual sources that lie within an angular distance $\theta_{\rm 200c}$ of the cluster center. We combine the point sources lying within $\theta_{\rm 200c}$ of all clusters in logarithmic luminosity 
bins.  This then produces a combination of the luminosity function of the true cluster radio AGN together with the 
contamination from foreground or background radio AGN that are randomly superposed on the cluster.  We then build a model of the 
contamination using the $\log N - \log S$ extracted for the full SUMSS population, cycling through the cluster list using 
the cluster redshift to transform from radio source flux density to luminosity and scaling by the associated solid angle of 
the virial region for each cluster.  For the reasons discussed in \citetalias{gupta17}, 
we then subtract the contamination model and divide the resulting LFs by the sum of the virial 
masses of the clusters that contribute to each of the luminosity bins.  In doing this we are normalizing the LF in units
of $M_\odot^{-1}$, which is a convenient proxy for volume in the case of collapsed objects like clusters and groups.
In Section~\ref{sec:HON} we study the halo occupation number or HON of radio AGN using the LF redshift evolution derived here and
show that the number of radio galaxies scales approximately linearly with the cluster halo mass.

For the RM-Y3 catalog there are 3601 (1449, 2152) total (background, background subtracted) radio AGN observed 
above the previously discussed flux limits. In the MARD-Y3 case the corresponding numbers are 910 (455, 455).  
The total number of AGN in the RM-Y3 cluster sample allow us to estimate the incompleteness in the cross--matched sample, 
and this indicates that the $\sim60$~percent completeness
estimated from the surface density profiles for our matching criteria is reasonably accurate (see Section~\ref{sec:Crossmatching}).

We validate our LF construction code and our method by analyzing simulated samples that are 10 times the size 
of our radio source catalogs and that 
are created using the best fit LFs reported in Table~\ref{tab:LF} (for density evolution).  We recover the input parameters 
to within the statistical uncertainties.

\subsubsection{Construction of LF for cross--matched sample}
\label{sec:LFCrossmatched}

Although the cross--matched sample is estimated to be only $\le60$~percent complete and to have contamination of 15~percent, we 
nevertheless construct the LF to allow for comparison to the statistically derived LF.
For this purpose, we use the subset of SUMSS sources with identified optical counterparts whose redshifts and sky 
locations place them within a RM-Y3 cluster (as noted earlier, we do not do cross--matching for the radio AGN in the 
much smaller MARD-Y3 cluster sample).
In this case we adopt the cluster redshift for all 
SUMSS sources from that cluster, convert from flux to luminosity and build up a vector in logarithmic luminosity space 
that contains the sum of all identified sources.  We divide this vector by the sum of the virial masses of all the clusters 
that could have contributed to each luminosity bin (whether they actually contain a radio AGN or not), using the 
minimum luminosity probed by the cluster given the SUMSS flux limit and the cluster redshift.

This method is attractive in that there is no need to subtract off contamination from randomly superposed radio AGN.  
This reduces the Poisson noise in the final LF and is helpful in better defining the behavior of the rare, most 
luminous radio AGN.  However, given the incompleteness and contamination in the cross--matched sample, we do not 
expect perfect agreement with the estimates of the LF extracted statistically using the method described in the 
previous section.  Comparisons are presented in Section~\ref{sec:LFComparison} below.

\subsubsection{LF modeling}
\label{sec:LFModeling}

We fit our LFs using the functional form from \citet{condon02}, given as
\BE
\label{eqn:Condon_fit}
\log \left( \frac{{\rm d}n}{{\rm d} \log P} \right) = y - \left[ b^2 + \left( \frac{\log P-x}{w} \right)^2 \right]^{1/2} - 1.5 \log P,
\EE
where the parameters $b$, $x$ and $w$, control the shape of the LF and $y$ is its amplitude.

Assuming that the overall shape of the LFs remains constant, the only changes can be in the density and luminosity of the 
sources \citep{malchalski00}. The density evolution corresponds to a vertical shift in the LFs and can be quantified as
\BE
\label{eqn:density_evolution}
\frac{{\rm d}n (z)}{{\rm d} \log P} = \frac{{\rm d}n (z=z_\mathrm{C})}{{\rm d} \log P} \times \left(\frac{1+z}{1+z_{\rm C}}
\right)^{\gamma_{\rm D}},
\EE
similarly, the luminosity evolution corresponds to a horizontal shift in the LFs because of the evolving luminosities of the 
sources
\BE
\label{eqn:lum_evolution}
P(z) = P(z=z_\mathrm{C}) \times \left(\frac{1+z}{1+z_{\rm C}}\right)^{\gamma_{\rm P}},
\EE
where $z_{\rm C}=0.47$ corresponds to the median redshift of the RM-Y3 cluster sample and for comparison we take 
the same $z_{\rm C}$ for the MARD-Y3 cluster sample. $\gamma_{\rm D}$ and $\gamma_{\rm P}$ correspond to the power 
law index for density and luminosity evolution, respectively, of the LFs.

We again perform an MCMC analysis with the Cash statistic to fit the LFs. Following \citetalias{gupta17}, we fit for the AGN 
part of the LF, fix the values of the two shape parameters $b$ and $w$ to those determined in \cite{condon02} and vary  
$x$ and $y$ along with the density or luminosity evolution power law index. As reported in a previous study using X-ray 
selected clusters \citepalias{gupta17}, we find consistent results when $b$ and $w$ are fixed to either \cite{condon02} 
or \cite{best12} best fit values. We evaluate the likelihood of a given model by scaling the LF model with the total cluster 
mass contributing to each luminosity bin and then adding the statistically determined background number of galaxies to 
the corresponding luminosity bin. 
\begin{figure}
\centering
\vskip-0.2in
\includegraphics[width=8.7cm, height=9cm]{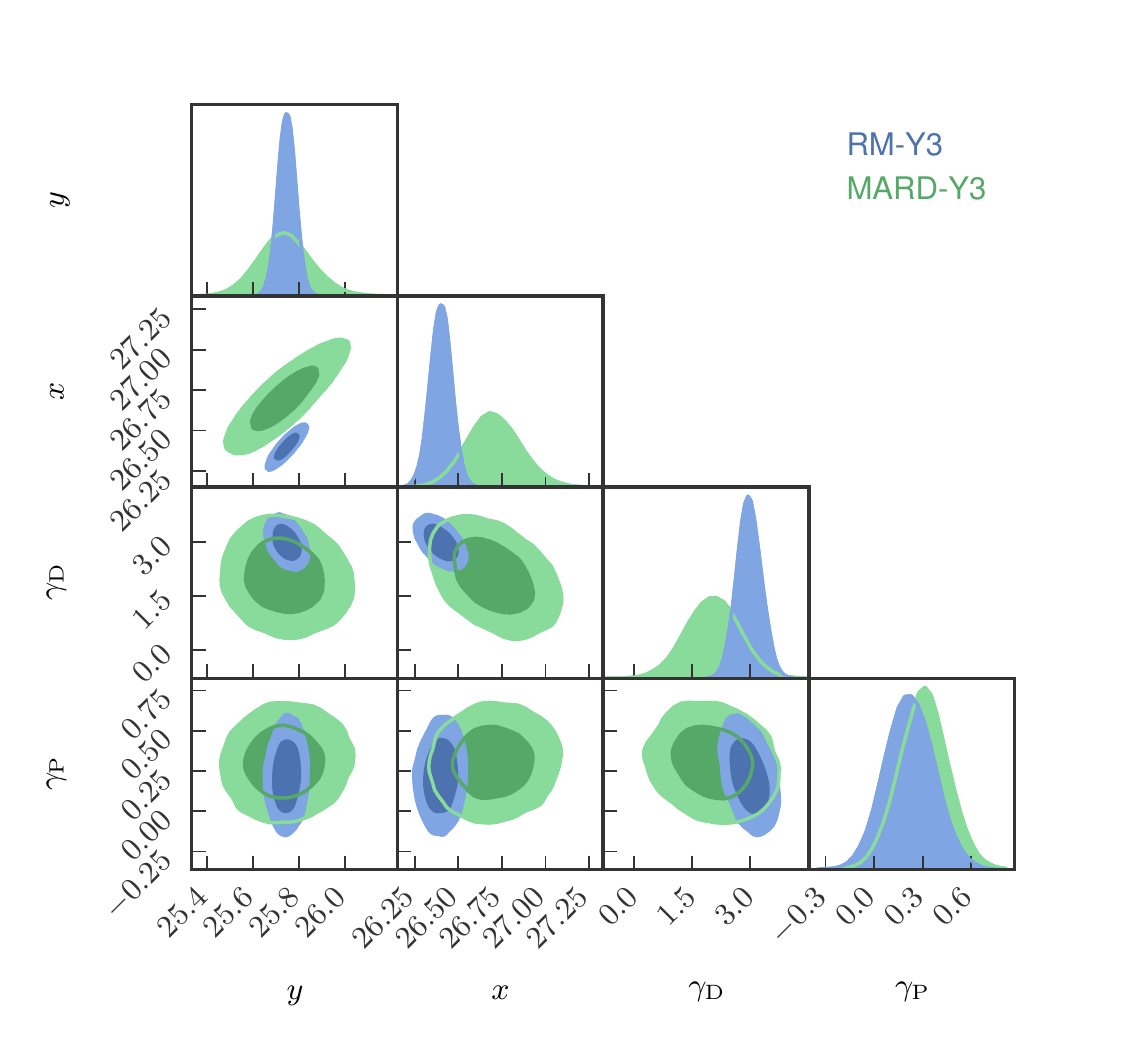}
\vskip-0.1in
\caption{Joint LF parameter posterior distributions for a combined density ($\gamma_\mathrm{D}$) and luminosity ($
\gamma_\mathrm{P}$) evolution fit to the radio AGN LFs (RM-Y3 in blue; MARD-Y3 in green).   Parameter values are 
reported in Table~\ref{tab:LF}.  For this fit, we adopted priors on the luminosity evolution parameter $
\gamma_\mathrm{P}=0.30\pm0.15$, determined through a direct analysis of the radio luminosity to stellar mass relation 
as described in Section~\ref{sec:LuminosityEvolution}.  The smaller MARD-Y3 sample prefers a fit with fewer extreme, 
high luminosity AGN.}
\label{fig:triangle}
\end{figure}

We correct our LF amplitudes by scaling them with a deprojection factor ($D_{\rm prj}$) that accounts for the cylindrical 
to spherical projection bias of radio AGN. This correction is very small because the radio AGN in clusters have a high 
NFW concentration; we find $D_{\rm prj}\sim0.92$.

\subsubsection{Best fit LF parameters and uncertainties}
\label{sec:LFMeasurements}

LFs for the full sample constructed statistically (Section~\ref{sec:LFMethod}) are shown in the upper 
panels of Fig.~\ref{fig:LF843} for RM-
Y3 (left) and MARD-Y3 (right) catalogs. 
We choose all sources to construct the 
LF with a flux limit at the 100~percent completeness of the SUMSS catalog as described in Section~\ref{sec:SUMSS}.  
In this figure, we plot the background subtracted observed counts in larger 
luminosity bins and in three redshift bins.  However, this figure does not represent the fitting method, where the 
observed counts are divided into much finer luminosity bins and in 15 redshift bins (with similar numbers of clusters in 
each bin) to get the model parameters. We also show the best fit model and model uncertainties from 
\citetalias{gupta17} in pink, where the median redshift of that sample is $z=0.1$. 

The bottom panels of Fig.~\ref{fig:LF843} show the ratios of the measured LFs for both cluster samples and at different redshift bins
 to the best fit model in the lowest redshift bin in this work.  It is visually apparent that the optically selected RM-Y3 and X-ray 
 selected MARD-Y3 samples provide similar radio AGN LFs over this redshift range. Because the MARD-Y3 cluster sample
 is smaller, we show only two redshift ranges.
 
In the lower panels one can see some evidence of a change in LF shape with increasing redshift.  For instance, there is a 
larger increase in the LF amplitude at lower luminosities but for luminosity $\log{[\mathrm{P/(W\,Hz^{-1})}]}> 10^{26.5}$, no evolution with 
redshift is evident.  We have not attempted independent fits within each redshift range where all parameters 
are free to vary, simply because of the limited sizes of our AGN samples.  However, with future samples it should be possible
to further examine whether there is more evolution at lower radio power than at higher.

Table~\ref{tab:LF} contains the best fit parameters for the LFs for the scenarios where either pure density evolution or 
pure luminosity evolution is taken into account. As expected, we find that the measured LFs alone are insufficient to simultaneously 
constrain density and luminosity evolution in the MCMC analysis. 

The parameter uncertainties shown in Table~\ref{tab:LF} are marginalized over uncertainties in the $\lambda$-mass 
relation when RM-Y3 catalog is used and the $L_\mathrm{X}$-mass relation when the MARD-Y3 sample is used.   
As an example, we marginalize over the uncertainties in the $\lambda$-mass relation described 
in Section~\ref{sec:Redmapper} by first measuring the luminosity functions under single parameter excursions of $\pm$ 
2-$\sigma$ for the $\lambda$-mass parameters $A_{\lambda}$, $B_{\lambda}$ and $\gamma_{\lambda}$.  We then extract 
the derivatives of the LF parameters $y$, $x$ and $\gamma$ with respect to each of the $\lambda$-mass parameters 
and finally propagate the uncertainties in the $\lambda$-mass relation to the luminosity function parameters using 
Gaussian error propagation, further assuming no parameter covariance in the $\lambda$-mass relation parameters. 
This is a good approximation as there are negligible degeneracies between the parameters of the $\lambda$-mass 
relation \citep{mcclintock19}.
We find that the contributions to the LF parameter uncertainties from the remaining uncertainties in the $\lambda$-mass 
relation are smaller than the statistical or sample size contributions to the uncertainties. Nevertheless, we present 
overall combined constraints (including the systematics from uncertainties in $\lambda$-mass) in Table~\ref{tab:LF}.

We also use the constraints on the radio luminosity evolution with redshift at fixed stellar mass 
(Section~\ref{sec:LuminosityEvolution}), to simultaneously constrain density and 
luminosity evolution.  These results are shown in the third set of results for each sample in Table~\ref{tab:LF}. 
Adopting the best fit value and uncertainty in $\gamma_{\rm PM}=0.30\pm0.15$ as a prior on the 
luminosity evolution, we estimate joint parameter constraints on luminosity and density evolution parameters of the LF.  
Fig.~\ref{fig:triangle} shows the 2-D marginalized 1-$\sigma$ MCMC constraints for luminosity function parameters and 
it reveals the significant parameter covariance between the shape and amplitude parameters $x$ and $y$, 
which is boosting the fully marginalized uncertainties on each parameter. 

There is a statistically significant shift in the joint space of shape and amplitude parameters 
between the RM-Y3 and MARD-Y3 samples.  This shift is 
apparent as a $\sim$2$\sigma$ offset in the fully marginalized posterior of the shape parameter
$x$ in Table~\ref{tab:LF}.  We expect that this difference is driven by a steeper fall--off of the RM-Y3 
sample at highest radio powers, which then 
drives corresponding changes in the amplitude due to the correlation between the shape and amplitude parameters.
The MARD-Y3 sample is smaller, and the highest luminosity portion of the 
LF with $\log{[\mathrm{P/(W\,Hz^{-1})}]}> 10^{27}$ is more poorly probed than in the RM-Y3 sample, and for that 
reason we restrict the fit for the smaller sample to lower power AGN.  Thus, the portion of the LF causing this apparent
difference between the radio AGN LF of these two samples is only probed in one of the two samples.  With future, larger
X-ray selected samples from eROSITA \citep{predehl10,merloni12} it should be possible to probe these highest luminosity,
rarest radio AGN more precisely and better understand any possible differences between 
X-ray and optically selected cluster samples.

The best fit density evolution parameters for the X-ray and optically selected cluster samples are statistically consistent, but the RM-Y3 sample suggests a more rapid evolution.  In the combined density and luminosity evolution models, the density evolution parameter for the MARD-Y3 is $\gamma_\mathrm{D}=2.05\pm0.66$ and for the optically selected RM-Y3 is  $\gamma_\mathrm{D}=3.00\pm0.42$.  Again, these uncertainties include the underlying uncertainties in the observable mass relations for each cluster sample.

\begin{figure}
\centering
\vskip-0.40in
\includegraphics[width=9cm, height=10cm]{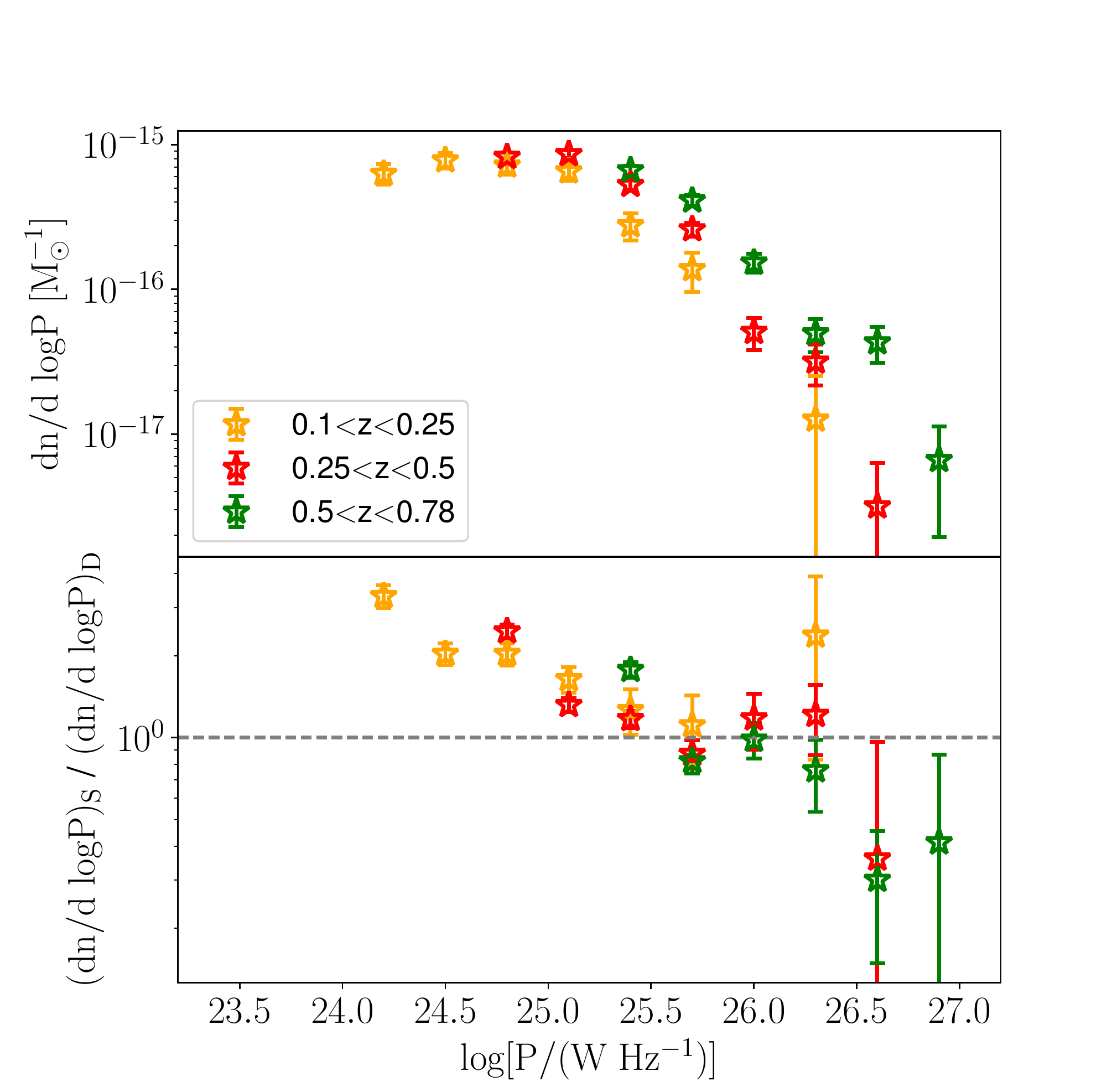}
\vskip-0.1in
\caption{The cluster radio AGN luminosity functions obtained using the incomplete optically cross--matched AGN sample 
(Section~\ref{sec:LFCrossmatched}) for the RM-Y3 clusters is in the upper panel. The ratio 
(lower panel) between the statistical (subscript S) (see Section~\ref{sec:LFMethod} and Fig.~\ref{fig:LF843}) and 
cross--matched (subscript D) LFs is shown for three redshift bins.  There is consistency for the most luminous radio AGN 
($\log{[\mathrm{P/(W\,Hz^{-1})}]}>25.5$) where the completeness of the cross--matched sample is highest, 
but for lower luminosity AGN the incompleteness increases (see Section~\ref{sec:LFComparison}).}
\label{fig:LFcomparison}
\end{figure}

\subsubsection{Comparison with the cross--matched sample}
\label{sec:LFComparison}

In Fig.~\ref{fig:LFcomparison}, we present the 843 MHz LFs (top panel) in different redshift bins, constructed 
using the sample of optically cross--matched radio AGN as described in Section~\ref{sec:LFCrossmatched}. 
We compare these LFs with those produced using statistical background subtraction and for which the sample is complete, as
described in Section~\ref{sec:LFMethod} and shown in Fig.~\ref{fig:LF843}.

On the luminous end,  the LFs are in good agreement, but at radio power $\log{[\mathrm{P/(W\,Hz^{-1})}]}<25.5$, the 
LFs from the cross--matched sample are systematically lower in amplitude than the statistically reconstructed LFs.  
Given the estimated completeness ($\sim 60$~percent) and contamination ($\sim15$~percent) in the cross--matched sample (discussed in 
Section~\ref{sec:Crossmatching}), some differences are expected.  As described in Section~\ref{sec:Crossmatching}, only the two most
massive sub-classes of galaxies ($\log(M_*)>10.9$) are used in this cross matching so that the sample contamination can 
be kept at or below the 15~percent level.

The fact that the LF of the cross--matched sample underestimates
the statistical LF mostly at the lowest radio luminosities in each bin may well be an indication of higher contamination and incompleteness in the
lowest flux radio bins, where sources have the largest positional uncertainties and are therefore more likely to be cross--matched to the
wrong counterpart.   Another effect that may be playing a role is that double tail radio sources may be resolved more effectively at lower redshift.  Resolving such sources would create two radio sources and increase the chances that both would be cross--matched 
to the wrong counterpart.

The differences between the LFs calculated using the cross--matched and the complete sample 
pose challenges for field LF studies with the SUMSS sample where
cross-matching is required, but as described in Section~\ref{sec:LFMethod} above, for the cluster radio AGN LF 
one need not use a cross--matched sample.

\subsubsection{Comparison of LF redshift trends to previous results}
\label{sec:LFLiterature}

Similar trends with redshift have been seen in previous studies of field radio AGN LFs. LFs of optically selected Quasi-
Stellar Objects (QSOs) at $z\leq2.2$ showed a luminosity evolution with $\gamma_{\rm P}= 3.2\pm0.1$ \citep{boyle88}. 
\citet{malchalski00} and \cite{brown01} studied a sample of 1.4~GHz radio sources at low and intermediate redshifts 
and suggested a luminosity evolution of $\gamma_{\rm P}= 3\pm1$ and $4\pm1$, respectively. In a recent study, 
\citet{pracy16}  derived 1.4~GHz LFs for radio AGN separated into LERGs and HERGs.  
They found that the LERG population displays little or no evolution, while the 
HERG population evolves more rapidly as $\gamma_{\rm P}= 7.41^{+0.79}_{-1.33}$ or $\gamma_{\rm D}= 2.93^{+0.46}
_{-0.47}$.  HERGs have bluer color and a weaker 4000 \AA{} break, which are indications of ongoing star formation 
activity. LERGs, however, appear to be preferentially located at the centers of groups or clusters and are fueled by 
feedback from their hot gas haloes \citep{lin07,kauffmann08, lin10, best12}.  As presented in Section~\ref{sec:SEDEvolution} above,
our sample of cross--matched hosts of cluster radio AGN show no significant shifts in color out to $z\sim0.8$, and
so presumably the population we are studying with our cluster radio AGN LF is a LERG population. 

\cite{smolcic09}  explored the cosmic evolution of AGN with low radio powers ($\log{[\mathrm{P/(W\,Hz^{-1})}]}<25.7$) out to $z = 1.3$ and found pure density evolution $\gamma_{\rm D}= 1.1\pm0.1$ and pure luminosity evolution $\gamma_{\rm P}= 0.8\pm0.1$.
\cite{strazzullo10} carried out a multi-wavelength analysis of Deep {\it Spitzer} Wide-area InfraRed Extragalactic Legacy 
Surveys' Very Large Array field (SWIRE VLA) and found $\gamma_{\rm P}= 2.7\pm0.3$ and $3.7^{+0.3}_{-0.4}$ for AGN 
and starforming populations, respectively. Similarly, \cite{mcalpine13} studied pure density and luminosity evolution for a 
combined dataset of $\sim$900 VLA observed galaxies in the field, finding $\gamma_{\rm P}=1.18\pm0.21$ and 
$2.47\pm0.12$ for AGN and star forming galaxies, respectively. \cite{janssen12} demonstrated that in the local Universe 
a sub-population of LERGs are hosted in blue star forming galaxies, with these blue LERGs becoming increasingly 
important at higher radio power. Thus, it is possible that the contribution of such blue LERGs increases towards higher 
redshifts, rendering the initial assumption that all AGN are hosted by red passive galaxies invalid.  \cite{smolcic17} studied 
a COSMOS sample of radio AGN at 3\,GHz out to $z=5$ and constrained pure density evolution $\gamma_{\rm D}= 2.00\pm0.18$ 
and pure luminosity evolution $\gamma_{\rm P}= 2.88\pm0.34$.

Also for clusters, \cite{green16} have shown that at least 14~percent of BCGs show a significant color offset from 
passivity in a population of ~980 X-ray detected clusters ($0.03 < z <0.5$). In their table~2 and figure~16, they show the 
offset to passivity as a function of X-ray luminosity of host clusters, and they find larger fractions of galaxies with offset 
from passivity in high luminosity clusters, which in their sample are preferentially at higher redshift. For samples of X-ray 
and optically selected galaxy clusters, \cite{sommer11}, show $\gamma_{\rm P}=8.19\pm2.66$ and $\gamma_{\rm D}=3.99\pm1.24$, 
respectively using 1.4~GHz detected radio AGN from the FIRST survey in a redshift range of 0.1 to 0.3. They also find a 
steep pure density evolution with $\gamma_{\rm D}=9.40\pm1.85$ for an X-ray selected sample of galaxy clusters. In a 
recent work, \cite{birzan17} investigated AGN feedback in a large sample of SZE selected clusters from SPT and ACT 
surveys and found $\sim7$ times more SUMSS sources in $z>0.6$ clusters than in the $z<0.6$ sample, which they suggest may be 
due to the differences in the feedback mechanisms onto the super massive black holes (SMBHs) in the low and high 
luminosity sources. 
In another recent work, \cite{lin17} constructed the radio LFs for 1.4~GHz sources in clusters out to $z\sim1$ and found 
an over-abundance of radio AGN in clusters compared to the field population \citep[see also][]{lin07,best07}. They find that 
cluster galaxies at $z > 0.77$ are about 1.5 to 2 times more likely to be active in the radio compared to those in lower redshift clusters. 
In comparison to these previous studies, the current work is based on a large cluster sample that allows us to push to higher redshifts and more importantly we present a first analysis to constrain both density and luminosity evolution of LFs simultaneously.
\begin{figure}
\centering
\vskip-0.3in
\includegraphics[width=9cm, height=8cm]{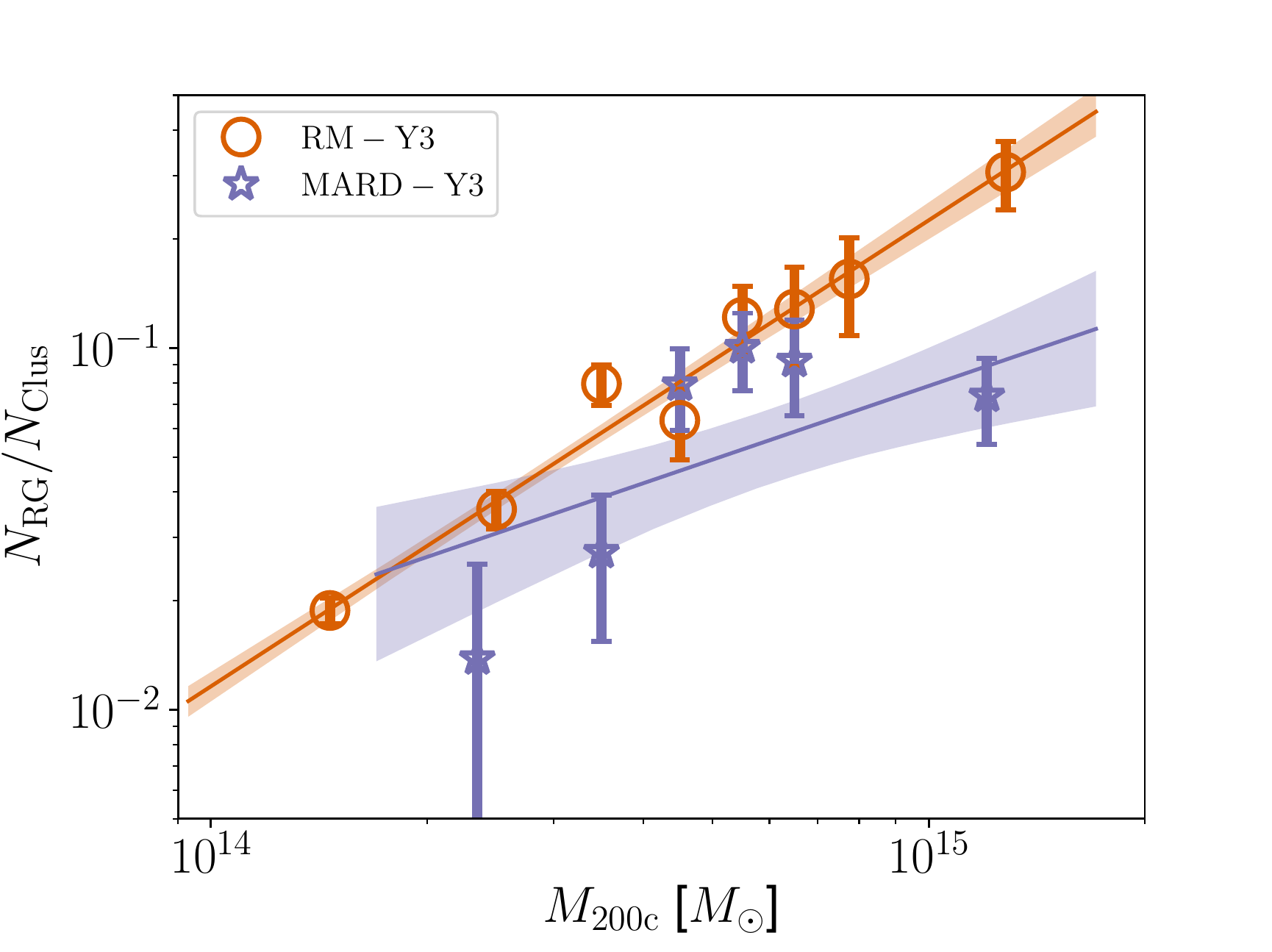}
\vskip-0.1in
\caption{HONs: The mean number of radio sources with $\log{[\mathrm{P/(W\,Hz^{-1})}]}>25.5$ per galaxy cluster as a 
function of cluster mass at $z=0.47$ using optically selected RM-Y3 (red) and X-ray selected MARD-Y3 (blue) clusters. The solid 
line is the best fit power law model and the shaded region shows the 1-$\sigma$ model uncertainty. The extent of the shaded region shows the range of mass used in the analysis. The plot is presented at the 
pivot redshift $z_\mathrm{C}=0.47$ for both cluster samples, and the redshift dependence is corrected using the 
measured redshift trend $(1+z)^{\gamma_{\rm D}}$ from the LF analysis, where $\gamma_{\rm D}$ is the density evolution only result 
for each sample as presented in Table~\ref{tab:LF}.}
\label{fig:HON}
\end{figure}

\subsection{Halo occupation number (HON)}
\label{sec:HON}

We define the halo occupation number (HON) as the average number of background subtracted radio AGN per cluster 
with $\log{[\mathrm{P/(W\,Hz^{-1})}]}>25.5$. We estimate the HON in a stack of RM-Y3 and MARD-Y3 clusters in various 
mass bins to study 
the mass trends for observed cluster radio sources. We account for the redshift trends estimated for combined density and luminosity
evolution in the previous section and solve for the mass trends with the pivot redshift $z_\mathrm{C}=0.47$.

Fig.~\ref{fig:HON} shows the HON of radio sources in RM-Y3 and MARD-Y3 clusters, where we also show the best fit 
power law of the form
\BE
\label{eqn:MF_PowerLaw}
{N_{\rm RG}} = A_{\rm H}  \left(\frac{M_{\rm 200c}}{1.5\times 10^{14}}\right)^{B_{\rm H}}\left(\frac{1+z}
{1+z_{\rm C}}\right)^{\gamma_{\rm D}},
\EE
where $N_{\rm RG}$ describes the average number of radio AGN in a cluster with $\log{[\mathrm{P/(W\,Hz^{-1})}]}>25.5$ 
and $A_{\rm H}$ and $B_{\rm H}$ are the normalization and mass trend of the power law best constrained to 
$0.019\pm0.001$ and $1.2\pm0.1$, respectively for RM-Y3 sample.  For the MARD-Y3 sample, we find $A_{\rm H} = 0.021^{+0.013}_{-0.010}$ and $B_{\rm H} = 0.68\pm0.34$, which is consistent with the mass trend (at 1.5$\sigma$) for the RM-Y3 sample, but also only $\sim$2$\sigma$ away from $B_\mathrm{H}=0$.  The two datasets together suggest a trend $B_\mathrm{H}\sim1$.
In fitting the relation, we adopt the mean mass $<M_{\rm 200c}>$ of the clusters within 
each mass bin, and we have adopted the best fit density evolution presented for the 
combined density and luminosity evolution model in 
Table~\ref{tab:LF} for each cluster sample.

 In Fig.~\ref{fig:HON}, we show the data points in finer mass bins for the RM-Y3 sample and broader mass bins for the MARD-
Y3 sample along with best fit model and 1-$\sigma$ model uncertainties as the shaded region around the models.  
The HON for the MARD-Y3 sample tends to lie below that of the RM-Y3 sample, but it is the number of radio AGN in the highest
mass bin that most strongly suggests differences between the samples.  We have examined this and see that if we choose a higher luminosity 
cut $\log{[\mathrm{P/(W\,Hz^{-1})}]}>26.0$ the uncertainties increase but the two samples also show more similar behavior.  The two mass
trends are statistically consistent, suggesting a mass trend $B_\mathrm{H}\sim1$.  This
motivates our choice of normalizing the LFs by the mass of the galaxy 
clusters.  Such a scaling has been suggested before \citep{lin07}, but never with such a large sample of clusters where the
sample alone could demonstrate that the probability of a cluster containing a radio AGN above a threshold luminosity scales
roughly linearly with the cluster halo mass.  We discuss this 
result further within the context of our other measurements in the following section.

\begin{figure}
\centering
\vskip-0.3in
\includegraphics[width=8.5cm, height=8cm]{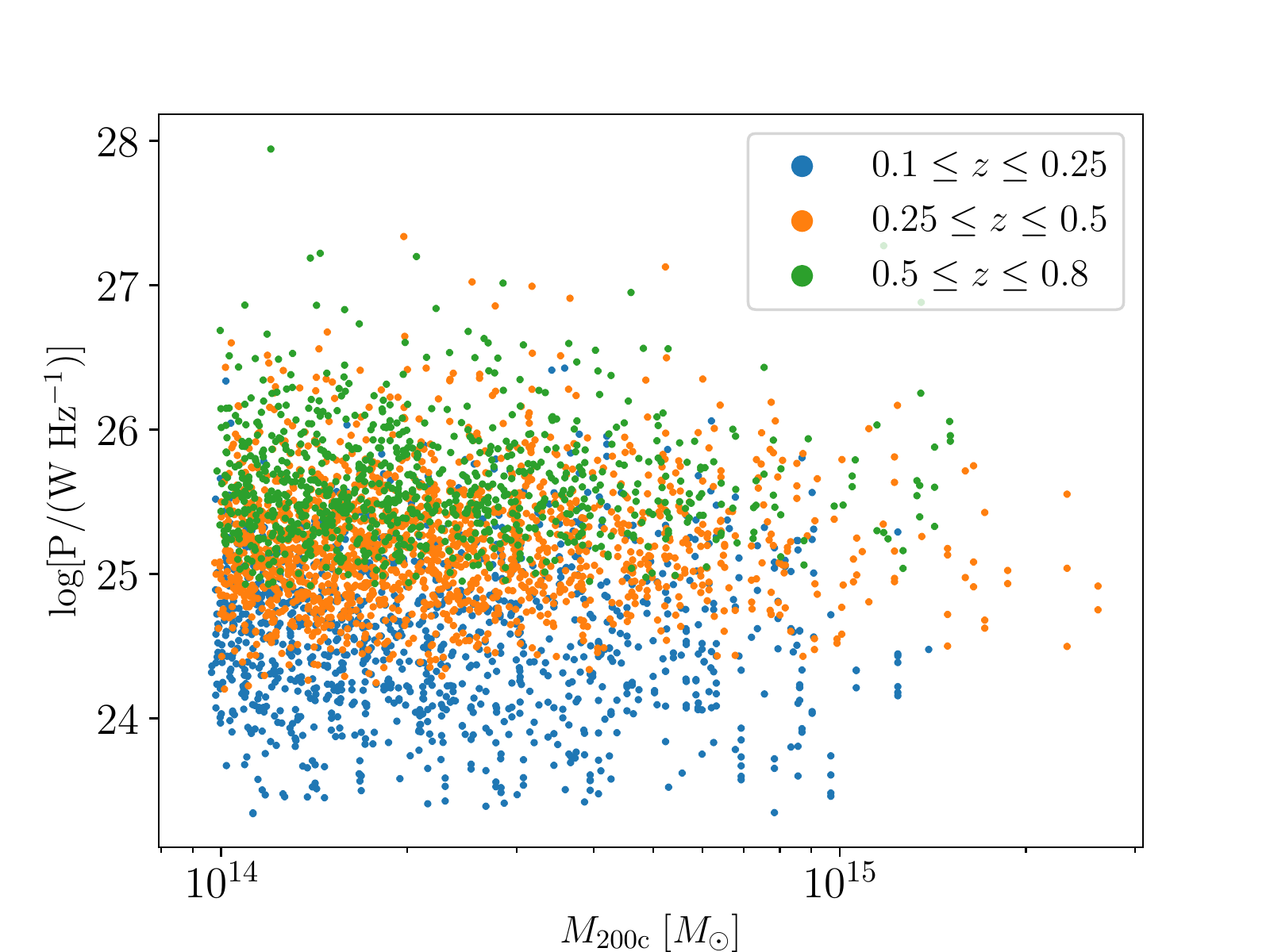}
\vskip-0.1in
\caption{The radio luminosities for the cross-matched cluster radio AGN sample are plotted versus cluster halo mass.  Given that the SUMSS catalog is flux limited, we separate the samples into different redshift ranges (color coded).  There is no clear evidence for a cluster halo mass dependence on the radio AGN luminosity.
}
\label{fig:mass-lum-cor}
\end{figure}

We have further examined whether there is a trend in radio AGN luminosity as a function of cluster mass.  In Fig.~\ref{fig:mass-lum-cor} we plot the radio AGN sample separated into three different redshift ranges in the space of radio luminosity versus cluster halo mass.  The data provide no clear evidence for a cluster halo mass dependence of the radio AGN luminosity.  Thus, the approximately linear cluster halo mass dependence of the HON of radio AGN  with $\log{[\mathrm{P/(W\,Hz^{-1})}]}>25.5$ (see Fig.~\ref{fig:HON}) cannot be explained by radio AGN being more luminous in higher mass clusters.

 \section{Discussion}
 \label{sec:Discussion}
 
 In this section we use the measurements of the cluster radio AGN properties and their evolution to 
 $z\sim0.8$ presented in the previous section 
 to discuss environmental influences on radio AGN 
 and also to examine the radio mode feedback as a function of mass and 
redshift within galaxy clusters.
 
 \subsection{Environmental influences on radio AGN}
 \label{sec:Environment}
 
 We discuss the key results of our study within the context of past studies and then turn to possible scenarios that 
 could explain the trends in the cluster radio AGN we have presented.
 
 \subsubsection{Cluster Radio AGN Properties in Context}

The mass trend in the radio AGN HON presented in Section~\ref{sec:HON} ($B_{\rm H}=1.2\pm0.1$ for RM-Y3 and $B_{\rm H}=0.7\pm0.3$ for MARD-Y3) is similar
to that reported for cluster galaxies as a whole. 
In an analysis of the near-infrared (NIR) K-band 
properties of galaxies within 93 galaxy clusters and groups using data from the Two Micron All Sky Survey (2MASS), \citet{lin04a} reported a 
mass trend parameter $B_{\rm H}=0.84\pm0.04$, indicating that in high mass galaxy clusters there are fewer galaxies 
per unit mass as compared to low mass clusters.  This trend is seen to remain largely unchanged to $z\sim1$ in a 
sample of SPT SZE selected galaxy clusters \citep{hennig17}.  
The number of radio AGN per unit cluster mass, on the other hand, increases strongly with redshift for both of the
samples studied here ($\gamma_\mathrm{D}=3.0\pm0.4$ and $2.1\pm0.7$ for RM-Y3 and MARD-Y3, respectively).

The HON mass scaling and normalization taken together with the centrally concentrated radial distribution presented in 
Section~\ref{sec:RadialProfile}, provides a picture of a cluster radio AGN population within a cluster that is dominated by a 
single, radio luminous AGN lying in or near the cluster core.  These galaxies are preferentially giant, passive ellipticals, and their 
probability of being radio loud increases with their stellar mass \citep{lin07,best07}.  In Fig.~\ref{fig:HON} we can see that, 
as a population, these galaxies exhibit a $\sim$10~percent probability of exceeding our adopted, nominal radio luminosity of 
$\log{[\mathrm{P/(W\,Hz^{-1})}]}>25.5$ if they lie in a $\sim6\times10^{14}M_\odot$ cluster at $z\sim0.5$.  
The fact that the probability of exceeding this threshold luminosity
increases with cluster mass suggests a relationship between the conditions required for radio mode 
feedback and the cluster halo mass.  

One could argue that given the number of potential galaxy hosts for a radio loud AGN scales approximately with cluster 
mass \citep[$B_{\rm H}=0.84\pm0.04$;][]{lin04a}, it may simply be that the probability of a radio AGN scales 
approximately with the number of potential hosts.  However, given that the radio AGN lie preferentially in centrally 
located giant ellipticals (e.g., the BCG), this argument is not fully satisfying.  There is only one BCG in each cluster, 
regardless of the cluster mass.  Post-merger scenarios where there are multiple, similar mass BCG candidates that are on 
their way to merging through the action of dynamical friction occur both in low and high mass clusters and would therefore
not seem to explain the cluster halo mass trend we observe.

It has been previously shown that the most massive BCGs lie in the most massive galaxy clusters, although the mass 
trend is very weak as $M_\mathrm{BCG}\sim M_\mathrm{200c}^{0.26\pm0.04}$ 
\citep[][note that these BCG stellar masses do not include the intracluster light]{lin04b}, $M_\mathrm{BCG}\sim M_\mathrm{200c}^{0.24\pm0.08}$ \citep{zhang16} and $M_\mathrm{BCG}\sim M_\mathrm{500c}^{0.4\pm0.1}$ \citep{kravtsov18}.  The halo 
mass trend in the HON is much steeper, suggesting that either (1) the galaxy mass dependence of a galaxy being radio 
loud is extremely steep, e.g., $\propto M_\mathrm{BCG}^B$ with $B\sim4$, which would be inconsistent with the 
findings in \citet[figure 8]{linden07} and \citet[][figure 16]{lin17}, or (2) that the radio AGN HON mass trend is not 
driven by the mass of the central galaxy alone.

The redshift trend in the HON and the density evolution of the LF provides another interesting clue to 
environmental influences on radio AGN.  
The host galaxy described above in a $6\times10^{14}M_\odot$ cluster at $z\sim0.5$
with a 10~percent chance of hosting a radio AGN that exceeds our 
adopted radio luminosity threshold would have a corresponding probability of $\sim$25~percent at $z\sim1$ (equation~\ref{eqn:MF_PowerLaw}).  
Because we know that the HON of the full galaxy population within the cluster virial region does not evolve with 
redshift (or evolves very weakly) to $z\sim1$ \citep{hennig17} and we know that BCG mass growth is 
rather slow over this redshift range 
\citep[1.8 to 1.2 times mass increase from $z=1$ to the 
present; e.g,.][]{lidman12,burke15}, it would be 
difficult to argue that the radio AGN population trends we are seeing 
are related to either galaxy number or galaxy 
host mass. A strong possibility is that there is an environmentally driven effect that is connected to the cluster halo 
mass and that is more efficient at higher redshift.

Another important clue is that the typical radio luminosity at fixed host stellar mass increases only modestly with redshift 
as $(1+z)^{0.30\pm0.15}$ (Section~\ref{sec:LuminosityEvolution}).  This suggests that the radio mode events at high 
and low redshift are similar \citep[see also][and references therein]{galametz09, larrondo12}, but that they are simply 
more frequent within cluster halos at higher redshift.  Moreover, our samples provides no evidence that radio AGN luminosity depends on the cluster halo mass.

\subsubsection{Confining Pressure of the ICM}

Of course, the cooling gas from the ICM could affect the radio activity.  The radio-loud fraction 
increases for BCGs closest to the peak in the ICM emission. 
This suggests that the availability of a fuel supply of cooling gas from the halo 
environment impacts or enhances AGN activity in the most massive galaxies \citep[e.g.][]{stott12}.
The emission lines from BCGs (principally H$\alpha$) indicate the presence of a strong cooling cluster core that 
has been shown to generally host more powerful radio sources. For BCGs in line emitting clusters, the X-ray 
cavity power correlates with both the extended and core radio emission, suggestive of steady fueling of the 
AGN in these clusters \citep{hogan15}.  However, there is no clear mass or redshift dependence of the 
fraction of clusters with 
cool cores \citep{semler12,mcdonald13}, so it seems unlikely that central cooling of the ICM is 
responsible for the strong cluster halo mass and redshift trends that we observe in the cluster radio AGN population.

In a recent work, \cite{lin18} studied a sample of 2300 radio AGN at $z<0.3$ to investigate the 
likely sources of AGN activity in massive galaxies. They found 
that for triggering the radio emission, both the stellar mass and the dark matter halo mass play an important 
role in both central and satellite galaxies. On the other hand, they found no convincing evidence linking the 
elevated radio activity in massive halos to the higher galaxy density therein. They also found that ICM entropy, cooling
time and pressure are playing roles in triggering radio mode feedback.
Their working hypothesis is that stellar mass loss from evolved stars is the source of the material accreted, and that
the ICM pressure plays a role in confining that fuel within the host galaxy, making it more likely to be accreted.

This scenario is interesting within the context of our results, because the ICM pressure increases 
strongly with cluster mass and redshift.  
Moreover, central galaxies would be preferred because the ram pressure stripping by the ICM of the stellar mass lost 
from evolved stars would be much lower or even absent in central giant ellipticals that have small peculiar velocities.
Recent ICM studies of cluster samples spanning a similar redshift range to that of our radio AGN study have examined 
ICM mass fraction, ICM temperature, X-ray luminosity and an integrated ICM pressure $Y_\mathrm{X}$ 
\citep[e.g.][and references therein]{chiu16b,chiu18,bulbul19}.  In these studies the ICM mass fraction is observed to increase 
with cluster mass and, at a fixed mass, to remain roughly constant with redshift. The ICM density profiles 
outside the central core regions have been shown to evolve approximately self-similarly with redshift \citep{mcdonald13}.  
ICM temperature increases with cluster mass
and evolves with redshift roughly self-similarly, implying higher ICM temperatures at higher redshift for a cluster of a given mass.
Taken together, these results imply a confining pressure $P$ around the central giant elliptical that scales with mass as $P\propto M$ and
with redshift as $P\propto [E(z)]^{8/3}$.  Without a model that connects the confining pressure to the radio mode feedback it is not
possible to comment on whether these similarly strong mass and redshift dependences in central confining pressure and 
the probability of a centrally located giant elliptical being radio loud are pure coincidence or indicative of a physical link.

\subsubsection{Mergers of Infalling Gas Rich Galaxies with BCG}

Another possible scenario that appears to be consistent with the observations would be 
the interaction or merger of an infalling galaxy that has residual star formation with a central, giant elliptical.  
Such an infalling galaxy would have to be on an approximately radial orbit, 
and so such events would be expected to be rare.  But the population of 
potential gas rich, infalling galaxies would scale roughly as the mass of the cluster and would also increase with redshift.  
Specifically, clusters with $M_{200}>3\times10^{14}M_\odot$ have blue fractions that are approximately constant as a 
function of cluster mass, but the blue fractions increase with cluster redshift from 20~percent at $z\sim0.1$ to $\sim$50~percent at 
$z\sim1$ \citep{hennig17}.  Thus, the number of potential merging, gas rich, infalling galaxies increases with the cluster virial mass 
and also increases with redshift, providing a potential explanation for the trends we report here in the cluster radio AGN.

Studies of BCG growth and the stellar mass--size relation for quiescent galaxies over 
this same redshift range have suggested that the dominant mode of growth for central giant 
ellipticals is through merging with low mass galaxies \citep{burke15}. 
Another study of BCGs in 14 galaxy clusters between $z=0.84$ and $z=1.46$ indicates that 3 of 
the 14 BCGs are likely to experience a major merger within 600~Myr, suggesting that mergers with 
giant ellipticals are the primary mode for BCG growth; however, the results from this small sample 
do not exclude the possibility that minor mergers could play an important role in shaping how central 
giant ellipticals appear today \citep{lidman13}.
In the process of merging, the gas rich infalling galaxy would deliver its gas to the BCG.  Central 
cluster galaxies are known to have reservoirs of molecular gas that are detected through their 
CO line emission at low redshifts \citep[e.g.][]{edge01, mcnamara14, temi18, vantyghem19, rose19} 
and at redshifts as high as $z=1.7$ \citep{webb17}. These molecular gas clouds are observed to 
have line of sight velocities of the order of $\sim$ 200 to 300 $\rm km~s^{-1}$, which would support 
a chaotic cold accretion type model in BCGs \citep[e.g.,][]{david14, tremblay16}.  However, 
stronger evidence that mergers of infalling, gas rich galaxies with central giant ellipticals are 
responsible for feeding and triggering the AGN accretion that leads eventually to radio mode 
feedback is needed to build confidence that this is a driver of the cluster AGN mass and redshift 
trends observed in the current work.

We note here also that ram pressure stripping will remove gas from any infalling, 
gas rich galaxies, and this low entropy gas will naturally sink toward the cluster center.  The 
supply of this low entropy, sinking gas would increase with cluster mass because there are 
more infalling, gas rich galaxies in higher mass clusters.  Moreover,  it would increase with 
redshift, because cluster galaxy populations within the virial region $R_{200c}$ have higher 
blue fractions-- and therefore higher gas content-- at higher redshift \citep[e.g.,][]{hennig17}.  
As with the previous scenario of merging, this scenario deserves further study.

\begin{figure*}
\centering
\vskip-0.2in
\includegraphics[width=8.5cm, height=8cm]{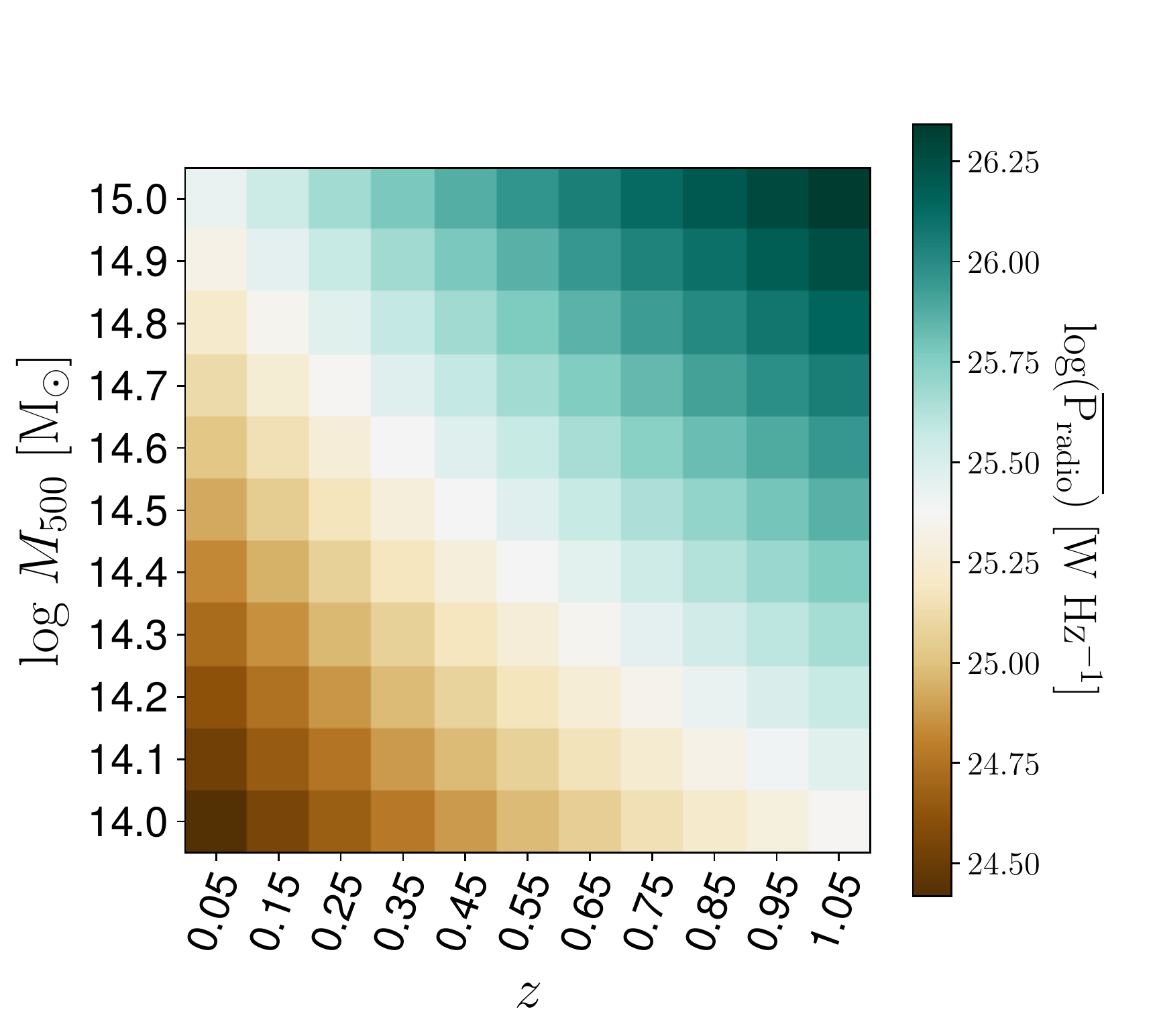}
\includegraphics[width=8.5cm, height=8.3cm]{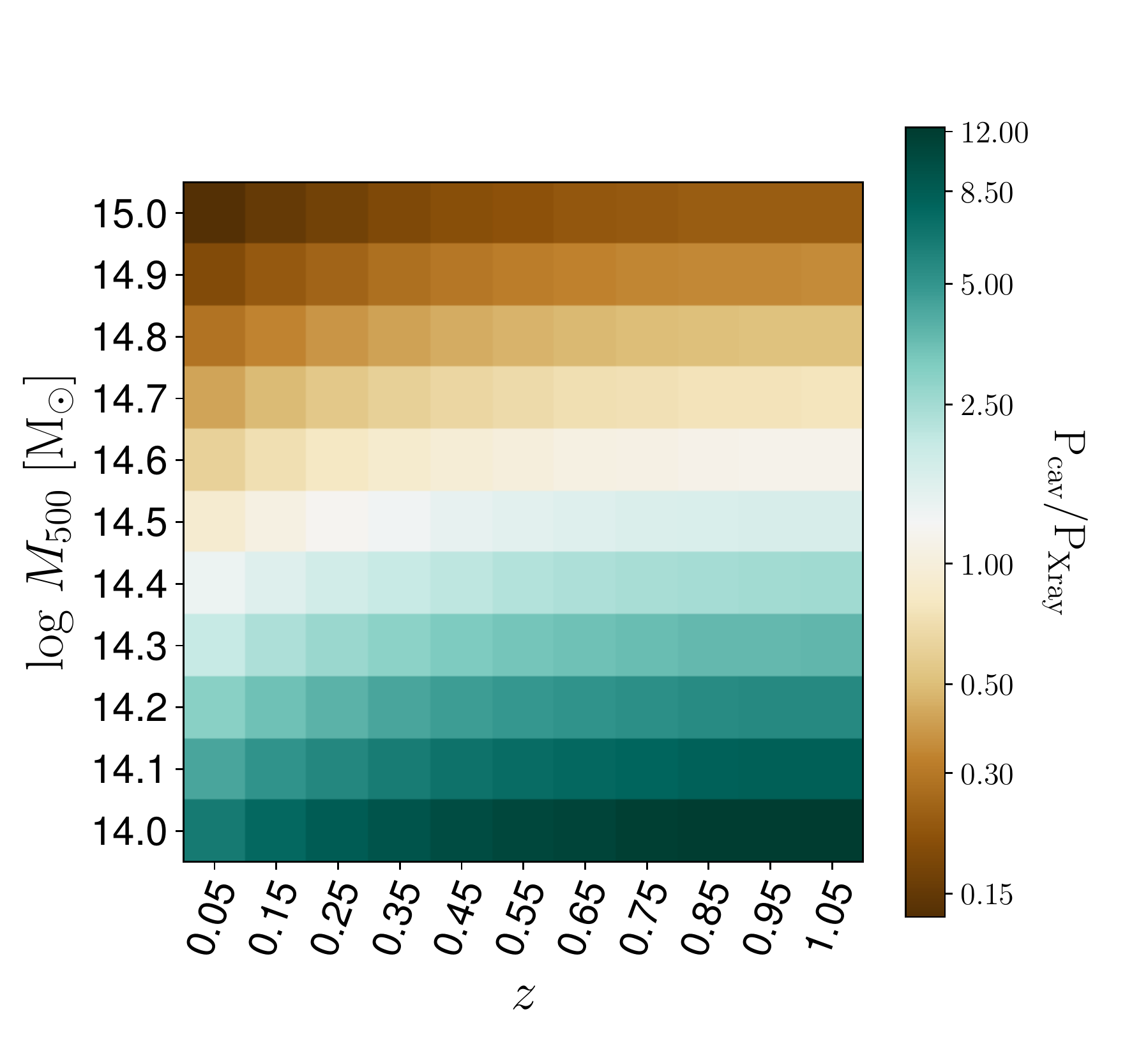}
\vskip-0.1in
\caption{Mean radio power at 843 MHz (left) of cluster radio AGN as a function of cluster mass and redshift calculated 
using the radio LF presented in Section~\ref{sec:LF}.  
On the right, we plot the ratio of the estimated mechanical feedback energy $\mathrm{P_{cav}}$ from the radio AGN to 
the central X-ray luminosities ($r<0.1R_{500}$) from the ICM as a function of cluster mass and redshift (see 
Section~\ref{sec:AGNFeedback}) using the relationship between feedback energy and radio power presented in 
\citet{cavagnolo10}. We note that the X-ray core luminosities are well constrained 
at $M_{500}>3\times10^{14}M_\odot$ \citep{bulbul19}.}

\label{fig:AGNFeedback}
\end{figure*}

\subsection{Radio AGN feedback }
\label{sec:AGNFeedback}

To make an estimate of the radio mode feedback from AGN in clusters, we first calculate the mean total cluster radio AGN 
power $\rm P_{Radio}$ using our measured LF extracted from the analysis of the RM-Y3 cluster sample (the case
with a combination of luminosity and number density evolution presented in Table~\ref{tab:LF}). We 
multiply the LF at a given redshift with the mass of the cluster of interest and integrate the product of the radio power 
and the LF in the radio power range of $10^{23}$ to $10^{28}$ $\rm WHz^{-1}$ to get the expected average radio power at 
the rest frame frequency of 843~MHz.
This range of power is selected, because it is comparable to the minimum and maximum radio powers of the 
radio AGN observed in our sample of clusters.  Moreover, this radio power lower limit is close to that marking 
the transition from a star forming dominated to an AGN dominated radio LF.  

In Fig.~\ref{fig:AGNFeedback}, we show the mean total radio power (in colorbar)  as a function of cluster mass and 
redshift.  There are roughly two orders of magnitude in dynamic range of the mean radio power over the mass 
and redshift range explored here.  
The mean radio power is approximately 10$\times$ higher for 10$^{15}M_\odot$ clusters than for 10$^{14}M_\odot$, 
consistent with what we would expect given the mass dependence of the halo occupation number (Section~\ref{sec:HON}).  
Moreover, there is a strong redshift dependence that leads to a similar $\sim$10$\times$ increase in mean radio power 
for a cluster of a given mass at $z=1$ in comparison to $z=0$.  This behavior follows from the measured redshift trends in 
the density evolution of the radio LF (Table~\ref{tab:LF}).

Studies of ICM cavities in ensembles of clusters containing radio galaxies \citep{cavagnolo10,larrondo12,larrondo15} 
have demonstrated correlations between the radio power and the amount of work done to create the cavities in the ICM.  
Thus, to estimate 
the typical radio mode feedback in clusters as a function of redshift, we use the mean total radio power calculated above 
together with the correlation between radio power and cavity power from  \citet{cavagnolo10}.  
Specifically, we take our estimated
843~MHz radio power and convert it to an expected 1.4~GHz radio power 
using the spectral index $\alpha=-0.7$ at all 
redshifts.  Further, we change units from W to erg s$^{-1}$ and estimate the total radio luminosity 
$L1400$ by scaling the estimated radio power by 1.4~GHz. 
We then use the \citet{cavagnolo10} relation to estimate the cavity power $\rm P_{cav}$
\begin{equation}
\mathrm{P_{cav}}=5\times10^{43} \left( {L1400\over 10^{40}\ \mathrm{erg\ s^{-1}}}\right)^{0.7} \mathrm{erg\ s^{-1}}.
\end{equation}
This cavity power represents an estimate of the minimum mechanical feedback into the ICM.  We calculate this estimate of the 
radio mode feedback as a function of mass and redshift as for the mean total radio power $\mathrm{P_{radio}}$ before.

The bulk of this radio mode feedback is  concentrated in the cluster core 
($2\over3$ of the cluster radio AGN in our sample lie 
within $0.1R_{200}$, corresponding to the central 0.1~percent of the cluster virial volume). 
To place this feedback energy in context, 
we compare it to the X-ray radiative losses from the core region. 
Specifically, we measure the core X-ray luminosities within $0.1R_{500}$ using and ensemble of SZE selected galaxy clusters that have 
been observed with the \XMM\ X-ray observatory.  This sample of 59 clusters with $0.2<z<1.5$ has been studied extensively 
in \cite{bulbul19}, where both core included and core excluded quantities have been presented. 

We use the measurements of the core luminosities from that analysis together with mass and redshift measurements for 
the subset of the galaxy cluster sample where the core region of interest is larger than the \XMM\ point spread function (PSF).  
The best fit core ($\le0.1R_{500}$) X-ray luminosity to mass relation that we find is
\BEA
\label{eq:Lx}
\nonumber
{\rm P}_{\rm Xray} = 16.88^{+1.53}_{-1.40} \times 10^{44} {\rm erg\ s^{-1}} \left(\frac{M_{500}}{M_{\rm piv}}\right)^{2.4 \pm 
0.2} \\
\left(\frac{E(z)}{E(z_{\rm piv})}\right)^{7/3} \left(\frac{(1+z)}{(1+z_{\rm piv})}\right)^{-0.7\pm 0.4},
\EEA
where $M_{\rm piv}$ and $z_{\rm piv}$ are $6.35\times 10^{14}$ $\rm M_\odot$ and 0.45, respectively. In 
addition, we find a log normal intrinsic scatter of $0.60^{+0.06}_{-0.05}$ in this relation, which is considerably larger than 
the scatter of the total cluster X-ray luminosity to mass relations presented for the same sample ($\sim0.3$).  
This is not surprising, because 
it has been long established that variations in the structure of cluster cores are the primary driver of the scatter in the X-ray
luminosity at fixed temperature \citep{fabian94a,mohr97} or mass.  

We note here that this X-ray core luminosity to mass scaling
relation is constrained using clusters in the mass range above $3\times10^{14} M_\odot$ and over the bulk of the 
redshift range of interest.   In the following analysis we extrapolate this relation to $10^{14} M_\odot$ to make estimates over
the full mass range where we study the cluster radio AGN LF.  Moreover, we emphasize that this is the average central 
X-ray luminosity within a region that is a fraction of the virial region.  It is not the luminosity within a cooling radius where 
the ICM is estimated to be radiatively unstable.  Our goal here is to examine typical radio mode feedback in the cluster 
core to the typical X-ray radiative losses in the core.

In the right panel of  Fig.~\ref{fig:AGNFeedback}, we show the ratio between the estimated  mean radio mode 
feedback and the mean X-ray radiative losses. 
We find that the ratio between the radio cavity power and the core X-ray luminosity is correlated with the redshift of the 
clusters and is anti-correlated with its mass. For instance, the ratio of the radio feedback to X-ray radiative 
losses for a $10^{14} \rm M_{\odot}$ galaxy cluster is $\sim 2$ times larger at $z\sim1.05$ as 
compared to the ratio at $z\sim 0.05$. On the other hand the 
ratio is $\sim 50$ times larger for a $10^{14} \rm M_{\odot}$ galaxy cluster as compare to a cluster with mass 
$10^{15} \rm M_{\odot}$.  

While these mass and redshift trends are interesting, 
the absolute value of the ratio suffers from several uncertainties.  For example, it 
could be shifted up or down simply by changing the radius within which 
we estimate the X-ray radiative losses.  Moreover, we have used the $\mathrm{P_{cav}}$
estimated from the entire radio AGN population, whereas only $\sim60$~percent of that 
population lies at $\le 0.1R_{500}$.  Finally, our estimate of the mechanical feedback 
$\mathrm{P_{cav}}$ is derived using
the energy required to evacuate the cavities in the ICM observed around 
a sizeable number of clusters.   
$\mathrm{P_{cav}}$ serves as a kind of lower limit to the actual 
mechanical feedback, with some authors arguing the total feedback could
be a factor of 4 to 16 times higher \citep[e.g.,][]{cavagnolo10}.

Our current estimates indicate that the average radio mode feedback within clusters is comparable to or smaller than the average 
central radiative losses within $0.1R_{500}$ for clusters with 
$M_{500}\gtrsim 5\times 10^{14}$ $\rm M_{\odot}$ at redshifts up to $z\sim1$
that we probe here.  It is within this core region where one can identify cool core structures within the ICM 
and where it is posited that the feedback from radio AGN may be playing a crucial role in replacing the radiative losses from X-ray emission and preventing a cooling instability or cooling flow \citep[e.g.][]{fabian12}.
Thus, in this mass range one might expect it to be more likely for cool core structures to emerge 
and for there to be no strong redshift trends.  Indeed, in SZE selected cluster samples where the selection is unaffected by 
the presence or absence of cool core structures, it has been demonstrated that there is no significant evolution in the cool 
core fraction with redshift for clusters with masses $M_{500}>3\times10^{14}M_\odot$ \citep{semler12,mcdonald12}.

At lower masses the mean mechanical feedback from radio mode feedback is many times larger than the core radiative cooling,
and one might expect it to be less likely for cool core structures to emerge.  We note that cool core 
systems in groups and low mass clusters have been identified in the nearby universe \citep{panagoulia14}, 
but it is not yet clear whether this represents a lower fraction of systems than is observed at higher masses.
Moreover, in this low mass range one might expect the excess of feedback energy to increase the 
entropy within the cluster core and beyond, perhaps even redistributing the ICM mass and leading to a reduction 
in the ICM mass fraction within the virial region $R_{500}$.  
In other words, the mass trend in the ratio of radio mode feedback and X-ray radiative losses implies a mass 
dependent ICM mass fraction for clusters over the mass range investigated here.  Moreover, this is true over the 
entire redshift range studied here.  
In the local universe this trend for lower mass clusters 
to exhibit lower ICM fractions was first reported for a large cluster ensemble by \citet{mohr99} (using emission weighted 
mean X-ray temperatures as a mass proxy).  More recently, this mass trend has been demonstrated in a study from group to cluster mass scales using the HIFLUGCS sample \citep[see figure 2;][]{main17} and in the SPT selected cluster sample, where
there is no clear redshift trend over the full redshift range 
studied here using clusters with masses $M_{500}>3\times10^{14}M_\odot$ \citep{chiu16b,chiu18,bulbul19}.  
Thus, the measured mass and redshift trends in the ICM mass fraction are qualitatively consistent with the expectations, given the
measured mass and redshift trends of the ratio of radio mode feedback and X-ray radiative losses.

Thus, the radio LFs for clusters with $M_{200c}>10^{14}M_\odot$ and extending to redshift $z\sim1$ that we present 
here imply significant mean radio mode mechanical feedback that offsets or dominates the mean core X-ray radiative losses for 
all but the most massive clusters studied here ($M_{500}\geq5\times10^{14}M_\odot$).  If the estimated radio mode feedback 
$\mathrm{P_{cav}}$ underestimates the true radio feedback by a factor of 4 to 16, then at no mass scale would the typical, observed
radiative losses within $0.1R_{500}$ be important.   
We emphasize that there is significant scatter $\sigma_{\ln{\mathrm{P_{Xray}}}}=0.6$ in the 
X-ray radiative losses at fixed mass and redshift, and that the radio galaxy 
population is rare and therefore there is significant stochasticity in the radio mode feedback as well.  We use the measured LFs and trends with mass and redshift to examine the stochasticity, finding that the standard deviation in the log of the mean radio luminosity varies from $\sigma_{\ln{\mathrm{P_{Radio}}}}=0.3$ for the $10^{15}M_\odot$ 
clusters at $z\sim1$ (where radio AGN are most abundant) to
$\sigma_{\ln{\mathrm{P_{Radio}}}}=0.6$ for $10^{14}M_\odot$ groups at $z\sim0$ (where radio AGN are least abundant).
These variations in both feedback and radiative losses would lead to the emergence of systems 
where cooling would dominate over feedback and the cool cores we observe could form.  Once the cool core 
starts to form there is evidence of enhanced radio mode feedback, suggesting that the cooling ICM could
itself drive new radio mode feedback \citep[][see figure 5]{fabian12}.

\section{Conclusions}
\label{sec:Conclusions}

We report measurements of the properties of cluster radio AGN to $z\sim1$ employing the 
largest cluster ensembles extending
to high redshift ever used.
These properties of radio AGN include 
their radial distribution within the cluster, radio luminosity evolution at fixed 
stellar mass, luminosity functions (LFs) and halo occupation number (HON) above a luminosity threshold.  
These measurements are made using
SUMSS selected sources observed at 843~MHz within optically selected RM-Y3 
and X-ray selected MARD-Y3 cluster catalogs  
that have been produced using the first three years of DES observations and the RASS X-ray survey.   

We focus on the excess population of radio 
sources associated with the virial regions of these clusters. The median redshifts of RM-Y3 and MARD-Y3 clusters 
are 0.47 and  0.28, respectively, and the catalogs extend to $z=0.8$ and beyond. The mass of RM-Y3 (Rykoff et al., in prep) 
and MARD-Y3 \citep{klein19} clusters is estimated using the richness-mass relation from \cite{mcclintock19} and the 
$L_X$--mass relation from \cite{bulbul19}, respectively, where the mass ranges correspond to $M_{200c}>1\times10^{14}
M_\odot$ and $M_{200c}>3\times10^{14}M_\odot$, respectively.

We find that the radial profile of the sources in RM-Y3 and MARD-Y3 catalogs is highly concentrated at the center of the
cluster over the whole redshift range and is consistent with an NFW model with concentration $c\sim140$.  
With this concentration,
approximately $2\over3$ of the cluster radio AGN lie within $0.1R_{200}$ of the cluster centers, which are 
defined using the cluster galaxies.  

A study of the cross--matched hosts of the cluster radio AGN indicates that the AGN radio luminosity at fixed stellar mass
is constant or increases only modestly as $(1+z)^{0.30\pm0.15}$ 
with redshift.  The broad band colors of these hosts are consistent with the 
cluster red sequence color and remain so without any major changes 
over the full redshift range (see Section~\ref{sec:HostProperties}).  We use the 
luminosity evolution constraint to break the degeneracy in density 
and luminosity evolution for the cluster radio AGN LF.

We construct the LFs assuming that the overdensity of radio AGN toward a cluster is at the redshift of the cluster, and we 
correct for the non-cluster sources by employing a statistical background correction (see Section~\ref{sec:LF}). 
We find that at fixed cluster halo mass, the cluster radio galaxies
are more numerous at higher redshift.  We find best fit redshift trends $(1+z)^\gamma$ in combined density and 
luminosity evolution of $\gamma_{\rm D}=3.00\pm0.42$ and $\gamma_{\rm P}=0.21\pm0.15$, 
respectively for the RM-Y3 sample.
Evolution is statistically consistent but more gradual for the X-ray selected MARD-Y3 sample: 
$\gamma_{\rm D}=2.05\pm0.66$ and $\gamma_{\rm P}=0.31\pm0.15$.
The uncertainties on the luminosity function parameters are marginalized over the uncertainties 
in the cluster observable-mass relations. Our results provide a clear indication that density evolution with redshift
dominates over luminosity evolution with redshift in cluster radio AGN populations.

Furthermore, we estimate the halo occupation number for radio AGN above a fixed luminosity threshold
in a stack of RM-Y3 and MARD-Y3 clusters using the LF evolution identified above. We show that the number of radio AGN above a luminosity threshold scales with the cluster mass as $N\propto M^{B_{\rm H}}$ with $B_{\rm H}=1.2\pm 0.1$ for the RM-Y3 sample and with $B_{\rm H}=0.68\pm 0.34$ for the MARD-Y3 sample, indicating
a relationship between the probability of a cluster containing a radio loud AGN and the cluster halo mass. Furthermore, we show that there is no compelling evidence for a dependence of radio AGN luminosity on cluster halo mass (see Section~\ref{sec:HON}).

We use these measurements to inform a discussion of the environmental influences on radio mode feedback (see Section~\ref{sec:Environment}).  
We consider two scenarios, one where the confining pressure of the ICM around the central giant ellipticals is responsible for the observed radio feedback mode and another 
where the merger of infalling, gas rich galaxies with centrally located 
giant ellipticals could also lead to the mass and redshift trends we measure in the radio AGN population. We note also that even in the absence of merging, gas rich galaxies will lose their gas through ram pressure stripping, and this low entropy gas will sink to the cluster center, potentially providing fuel for radio mode accretion.

We go on in Section~\ref{sec:AGNFeedback} 
to use the radio AGN LF to
estimate the radio mode feedback as a function of mass and redshift
 out to $z\sim1$ and compare that to the mean core X-ray radiative losses 
from the ICM at each mass and redshift.  The radio mode feedback dominates the core radiative losses in low 
mass systems and is comparable to or smaller than those radiative losses for systems with 
$M_{500}>5\times10^{14}M_\odot$ over the full redshift range.  
We note the imbalance of radio mode feedback and X-ray radiative losses is qualitatively consistent with what 
one would expect to explain the absence of redshift evolution of the cool core population and the trend of increasing 
ICM mass fraction with mass observed within clusters over the same mass and redshift range.

\section*{Acknowledgements}

We thank Jeremy Sanders for helpful discussions.  
The Melbourne group acknowledges support from 
the Australian Research Council's Discovery Projects scheme (DP150103208). 
The Munich group acknowledges the support of the International Max Planck Research School 
on Astrophysics of the Ludwig-Maximilians-Universit\"at, 
the Max-Planck-Gesellschaft Faculty Fellowship program at the Max Planck Institute for Extraterrestrial Physics, 
the DFG Cluster of Excellence ``Origin and Structure of the Universe'', the Verbundforschung 'D-MeerKAT' award 05A2017
and the Ludwig-Maximilians-Universit\"at. 
AS is supported by the ERC-StG 'ClustersXCosmo' grant agreement 71676, 
and by the FARE-MIUR grant 'ClustersXEuclid' R165SBKTMA.

SPT is supported by the National Science Foundation through grant PLR-1248097.  Partial support is also provided by the NSF Physics Frontier Center grant PHY-1125897 to the Kavli Institute of Cosmological Physics at the University of Chicago, the Kavli Foundation and the Gordon and Betty Moore Foundation grant GBMF 947. This research used resources of the National Energy Research Scientific Computing Center (NERSC), a DOE Office of Science User Facility supported by the Office of Science of the U.S. Department of Energy under Contract No. DE-AC02-05CH11231. 

Funding for the DES Projects has been provided by the U.S. Department of Energy, the U.S. National Science Foundation, the Ministry of Science and Education of Spain, 
the Science and Technology Facilities Council of the United Kingdom, the Higher Education Funding Council for England, the National Center for Supercomputing 
Applications at the University of Illinois at Urbana-Champaign, the Kavli Institute of Cosmological Physics at the University of Chicago, 
the Center for Cosmology and Astro-Particle Physics at the Ohio State University,
the Mitchell Institute for Fundamental Physics and Astronomy at Texas A\&M University, Financiadora de Estudos e Projetos, 
Funda{\c c}{\~a}o Carlos Chagas Filho de Amparo {\`a} Pesquisa do Estado do Rio de Janeiro, Conselho Nacional de Desenvolvimento Cient{\'i}fico e Tecnol{\'o}gico and 
the Minist{\'e}rio da Ci{\^e}ncia, Tecnologia e Inova{\c c}{\~a}o, the Deutsche Forschungsgemeinschaft and the Collaborating Institutions in the Dark Energy Survey. 

The Collaborating Institutions are Argonne National Laboratory, the University of California at Santa Cruz, the University of Cambridge, Centro de Investigaciones Energ{\'e}ticas, 
Medioambientales y Tecnol{\'o}gicas-Madrid, the University of Chicago, University College London, the DES-Brazil Consortium, the University of Edinburgh, 
the Eidgen{\"o}ssische Technische Hochschule (ETH) Z{\"u}rich, 
Fermi National Accelerator Laboratory, the University of Illinois at Urbana-Champaign, the Institut de Ci{\`e}ncies de l'Espai (IEEC/CSIC), 
the Institut de F{\'i}sica d'Altes Energies, Lawrence Berkeley National Laboratory, the Ludwig-Maximilians Universit{\"a}t M{\"u}nchen and the associated Excellence Cluster Universe, 
the University of Michigan, the National Optical Astronomy Observatory, the University of Nottingham, The Ohio State University, the University of Pennsylvania, the University of Portsmouth, 
SLAC National Accelerator Laboratory, Stanford University, the University of Sussex, Texas A\&M University, and the OzDES Membership Consortium.

Based in part on observations at Cerro Tololo Inter-American Observatory, National Optical Astronomy Observatory, which is operated by the Association of 
Universities for Research in Astronomy (AURA) under a cooperative agreement with the National Science Foundation.

The DES data management system is supported by the National Science Foundation under Grant Numbers AST-1138766 and AST-1536171.
The DES participants from Spanish institutions are partially supported by MINECO under grants AYA2015-71825, ESP2015-66861, FPA2015-68048, SEV-2016-0588, SEV-2016-0597, and MDM-2015-0509, 
some of which include ERDF funds from the European Union. IFAE is partially funded by the CERCA program of the Generalitat de Catalunya.
Research leading to these results has received funding from the European Research
Council under the European Union's Seventh Framework Program (FP7/2007-2013) including ERC grant agreements 240672, 291329, and 306478.
We  acknowledge support from the Brazilian Instituto Nacional de Ci\^encia e Tecnologia (INCT) e-Universe (CNPq grant 465376/2014-2).

This manuscript has been authored by Fermi Research Alliance, LLC under Contract No. DE-AC02-07CH11359 with the U.S. Department of Energy, Office of Science, Office of High Energy Physics. The United States Government retains and the publisher, by accepting the article for publication, acknowledges that the United States Government retains a non-exclusive, paid-up, irrevocable, world-wide license to publish or reproduce the published form of this manuscript, or allow others to do so, for United States Government purposes.
This manuscript has been authored by Fermi Research Alliance, LLC under Contract No. DE-AC02-07CH11359 with the U.S. Department of Energy, Office of Science, Office of High Energy Physics. The United States Government retains and the publisher, by accepting the article for publication, acknowledges that the United States Government retains a non-exclusive, paid-up, irrevocable, world-wide license to publish or reproduce the published form of this manuscript, or allow others to do so, for United States Government purposes.

\bibliographystyle{mnras}
\bibliography{LF_SPT}

\begin{thebibliography}{}
\makeatletter
\relax
\def\mn@urlcharsother{\let\do\@makeother \do\$\do\&\do\#\do\^\do\_\do\%\do\~}
\def\mn@doi{\begingroup\mn@urlcharsother \@ifnextchar [ {\mn@doi@}
  {\mn@doi@[]}}
\def\mn@doi@[#1]#2{\def\@tempa{#1}\ifx\@tempa\@empty \href
  {http://dx.doi.org/#2} {doi:#2}\else \href {http://dx.doi.org/#2} {#1}\fi
  \endgroup}
\def\mn@eprint#1#2{\mn@eprint@#1:#2::\@nil}
\def\mn@eprint@arXiv#1{\href {http://arxiv.org/abs/#1} {{\tt arXiv:#1}}}
\def\mn@eprint@dblp#1{\href {http://dblp.uni-trier.de/rec/bibtex/#1.xml}
  {dblp:#1}}
\def\mn@eprint@#1:#2:#3:#4\@nil{\def\@tempa {#1}\def\@tempb {#2}\def\@tempc
  {#3}\ifx \@tempc \@empty \let \@tempc \@tempb \let \@tempb \@tempa \fi \ifx
  \@tempb \@empty \def\@tempb {arXiv}\fi \@ifundefined
  {mn@eprint@\@tempb}{\@tempb:\@tempc}{\expandafter \expandafter \csname
  mn@eprint@\@tempb\endcsname \expandafter{\@tempc}}}

\bibitem[\protect\citeauthoryear{{Abbott} et~al.,}{{Abbott}
  et~al.}{2018}]{abbott18}
{Abbott} T.~M.~C.,  et~al., 2018, \mn@doi [\apjs] {10.3847/1538-4365/aae9f0},
  \href {http://adsabs.harvard.edu/abs/2018ApJS..239...18A} {239, 18}

\bibitem[\protect\citeauthoryear{{Aihara} et~al.,}{{Aihara}
  et~al.}{2011}]{aihara11}
{Aihara} H.,  et~al., 2011, \mn@doi [\apjs] {10.1088/0067-0049/193/2/29}, \href
  {http://adsabs.harvard.edu/abs/2011ApJS..193...29A} {193, 29}

\bibitem[\protect\citeauthoryear{{Annis} et~al.,}{{Annis}
  et~al.}{2014}]{annis14}
{Annis} J.,  et~al., 2014, \mn@doi [\apj] {10.1088/0004-637X/794/2/120}, \href
  {http://adsabs.harvard.edu/abs/2014ApJ...794..120A} {794, 120}

\bibitem[\protect\citeauthoryear{{Bartelmann}}{{Bartelmann}}{1996}]{bartelmann96}
{Bartelmann} M.,  1996, \aap, \href
  {http://adsabs.harvard.edu/abs/1996A%26A...313..697B} {313, 697}

\bibitem[\protect\citeauthoryear{{Bertin} \& {Arnouts}}{{Bertin} \&
  {Arnouts}}{1996}]{bertin96}
{Bertin} E.,  {Arnouts} S.,  1996, \aaps, \href
  {http://adsabs.harvard.edu/abs/1996A%26AS..117..393B} {117, 393}

\bibitem[\protect\citeauthoryear{{Best} \& {Heckman}}{{Best} \&
  {Heckman}}{2012}]{best12}
{Best} P.~N.,  {Heckman} T.~M.,  2012, \mn@doi [\mnras]
  {10.1111/j.1365-2966.2012.20414.x}, \href
  {http://adsabs.harvard.edu/abs/2012MNRAS.421.1569B} {421, 1569}

\bibitem[\protect\citeauthoryear{{Best}, {von der Linden}, {Kauffmann},
  {Heckman}  \& {Kaiser}}{{Best} et~al.}{2007}]{best07}
{Best} P.~N.,  {von der Linden} A.,  {Kauffmann} G.,  {Heckman} T.~M.,
  {Kaiser} C.~R.,  2007, \mn@doi [\mnras] {10.1111/j.1365-2966.2007.11937.x},
  \href {http://adsabs.harvard.edu/abs/2007MNRAS.379..894B} {379, 894}

\bibitem[\protect\citeauthoryear{{Bharadwaj}, {Reiprich}, {Schellenberger},
  {Eckmiller}, {Mittal}  \& {Israel}}{{Bharadwaj} et~al.}{2014}]{bharadwaj14}
{Bharadwaj} V.,  {Reiprich} T.~H.,  {Schellenberger} G.,  {Eckmiller} H.~J.,
  {Mittal} R.,   {Israel} H.,  2014, \mn@doi [\aap]
  {10.1051/0004-6361/201322684}, \href
  {https://ui.adsabs.harvard.edu/#abs/2014A&A...572A..46B} {572, A46}

\bibitem[\protect\citeauthoryear{{B{\^i}rzan}, {Rafferty}, {Nulsen},
  {McNamara}, {R{\"o}ttgering}, {Wise}  \& {Mittal}}{{B{\^i}rzan}
  et~al.}{2012}]{birzan12}
{B{\^i}rzan} L.,  {Rafferty} D.~A.,  {Nulsen} P.~E.~J.,  {McNamara} B.~R.,
  {R{\"o}ttgering} H.~J.~A.,  {Wise} M.~W.,   {Mittal} R.,  2012, \mn@doi
  [\mnras] {10.1111/j.1365-2966.2012.22083.x}, \href
  {http://adsabs.harvard.edu/abs/2012MNRAS.427.3468B} {427, 3468}

\bibitem[\protect\citeauthoryear{{B{\^i}rzan}, {Rafferty}, {Br{\"u}ggen}  \&
  {Intema}}{{B{\^i}rzan} et~al.}{2017}]{birzan17}
{B{\^i}rzan} L.,  {Rafferty} D.~A.,  {Br{\"u}ggen} M.,   {Intema} H.~T.,  2017,
  \mn@doi [\mnras] {10.1093/mnras/stx1505}, \href
  {http://adsabs.harvard.edu/abs/2017MNRAS.471.1766B} {471, 1766}

\bibitem[\protect\citeauthoryear{{Blanton}, {Clarke}, {Sarazin}, {Randall}  \&
  {McNamara}}{{Blanton} et~al.}{2010}]{blanton10}
{Blanton} E.~L.,  {Clarke} T.~E.,  {Sarazin} C.~L.,  {Randall} S.~W.,
  {McNamara} B.~R.,  2010, \mn@doi [Proceedings of the National Academy of
  Science] {10.1073/pnas.0913904107}, \href
  {https://ui.adsabs.harvard.edu/#abs/2010PNAS..107.7174B} {107, 7174}

\bibitem[\protect\citeauthoryear{{Bock}, {Large}  \& {Sadler}}{{Bock}
  et~al.}{1999}]{bock99}
{Bock} D.~C.-J.,  {Large} M.~I.,   {Sadler} E.~M.,  1999, \mn@doi [\aj]
  {10.1086/300786}, \href {http://adsabs.harvard.edu/abs/1999AJ....117.1578B}
  {117, 1578}

\bibitem[\protect\citeauthoryear{{Bocquet} et~al.,}{{Bocquet}
  et~al.}{2019}]{bocquet19}
{Bocquet} S.,  et~al., 2019, \mn@doi [\apj] {10.3847/1538-4357/ab1f10}, \href
  {https://ui.adsabs.harvard.edu/abs/2019ApJ...878...55B} {878, 55}

\bibitem[\protect\citeauthoryear{{Boller}, {Freyberg}, {Tr{\"u}mper}, {Haberl},
  {Voges}  \& {Nandra}}{{Boller} et~al.}{2016}]{boller16}
{Boller} T.,  {Freyberg} M.~J.,  {Tr{\"u}mper} J.,  {Haberl} F.,  {Voges} W.,
  {Nandra} K.,  2016, \mn@doi [\aap] {10.1051/0004-6361/201525648}, \href
  {http://adsabs.harvard.edu/abs/2016A%26A...588A.103B} {588, A103}

\bibitem[\protect\citeauthoryear{{Boyle}, {Shanks}  \& {Peterson}}{{Boyle}
  et~al.}{1988}]{boyle88}
{Boyle} B.~J.,  {Shanks} T.,   {Peterson} B.~A.,  1988, \mn@doi [\mnras]
  {10.1093/mnras/235.3.935}, \href
  {http://adsabs.harvard.edu/abs/1988MNRAS.235..935B} {235, 935}

\bibitem[\protect\citeauthoryear{{Brown}, {Webster}  \& {Boyle}}{{Brown}
  et~al.}{2001}]{brown01}
{Brown} M.~J.~I.,  {Webster} R.~L.,   {Boyle} B.~J.,  2001, \mn@doi [\aj]
  {10.1086/320410}, \href {http://adsabs.harvard.edu/abs/2001AJ....121.2381B}
  {121, 2381}

\bibitem[\protect\citeauthoryear{{Bruzual} \& {Charlot}}{{Bruzual} \&
  {Charlot}}{2003}]{bc03}
{Bruzual} G.,  {Charlot} S.,  2003, \mnras, \href
  {http://adsabs.harvard.edu/cgi-bin/nph-bib_query?bibcode=2003MNRAS.344.1000B&db_key=AST}
  {344, 1000}

\bibitem[\protect\citeauthoryear{{Bulbul} et~al.,}{{Bulbul}
  et~al.}{2019}]{bulbul19}
{Bulbul} E.,  et~al., 2019, \mn@doi [\apj] {10.3847/1538-4357/aaf230}, \href
  {http://adsabs.harvard.edu/abs/2019ApJ...871...50B} {871, 50}

\bibitem[\protect\citeauthoryear{{Burke}, {Hilton}  \& {Collins}}{{Burke}
  et~al.}{2015}]{burke15}
{Burke} C.,  {Hilton} M.,   {Collins} C.,  2015, \mn@doi [\mnras]
  {10.1093/mnras/stv450}, \href
  {http://adsabs.harvard.edu/abs/2015MNRAS.449.2353B} {449, 2353}

\bibitem[\protect\citeauthoryear{{Calzetti}, {Armus}, {Bohlin}, {Kinney},
  {Koornneef}  \& {Storchi-Bergmann}}{{Calzetti} et~al.}{2000}]{Calzetti00}
{Calzetti} D.,  {Armus} L.,  {Bohlin} R.~C.,  {Kinney} A.~L.,  {Koornneef} J.,
   {Storchi-Bergmann} T.,  2000, \mn@doi [\apj] {10.1086/308692}, \href
  {http://adsabs.harvard.edu/abs/2000ApJ...533..682C} {533, 682}

\bibitem[\protect\citeauthoryear{{Capasso} et~al.,}{{Capasso}
  et~al.}{2019a}]{capasso19a}
{Capasso} R.,  et~al., 2019a, \mn@doi [\mnras] {10.1093/mnras/sty2645}, \href
  {http://adsabs.harvard.edu/abs/2019MNRAS.482.1043C} {482, 1043}

\bibitem[\protect\citeauthoryear{{Capasso} et~al.,}{{Capasso}
  et~al.}{2019b}]{capasso19b}
{Capasso} R.,  et~al., 2019b, \mn@doi [\mnras] {10.1093/mnras/stz931}, \href
  {https://ui.adsabs.harvard.edu/abs/2019MNRAS.486.1594C} {486, 1594}

\bibitem[\protect\citeauthoryear{{Cash}}{{Cash}}{1979}]{cash79}
{Cash} W.,  1979, \mn@doi [\apj] {10.1086/156922}, \href
  {http://adsabs.harvard.edu/abs/1979ApJ...228..939C} {228, 939}

\bibitem[\protect\citeauthoryear{{Cavagnolo}, {McNamara}, {Nulsen}, {Carilli},
  {Jones}  \& {B{\^i}rzan}}{{Cavagnolo} et~al.}{2010}]{cavagnolo10}
{Cavagnolo} K.~W.,  {McNamara} B.~R.,  {Nulsen} P.~E.~J.,  {Carilli} C.~L.,
  {Jones} C.,   {B{\^i}rzan} L.,  2010, \mn@doi [\apj]
  {10.1088/0004-637X/720/2/1066}, \href
  {http://adsabs.harvard.edu/abs/2010ApJ...720.1066C} {720, 1066}

\bibitem[\protect\citeauthoryear{{Chiu} et~al.,}{{Chiu} et~al.}{2016}]{chiu16b}
{Chiu} I.,  et~al., 2016, \mn@doi [\mnras] {10.1093/mnras/stv2303}, \href
  {http://adsabs.harvard.edu/abs/2016MNRAS.455..258C} {455, 258}

\bibitem[\protect\citeauthoryear{{Chiu} et~al.,}{{Chiu} et~al.}{2018}]{chiu18}
{Chiu} I.,  et~al., 2018, \mn@doi [\mnras] {10.1093/mnras/sty1284}, \href
  {http://adsabs.harvard.edu/abs/2018MNRAS.478.3072C} {478, 3072}

\bibitem[\protect\citeauthoryear{{Condon}, {Cotton}  \& {Broderick}}{{Condon}
  et~al.}{2002}]{condon02}
{Condon} J.~J.,  {Cotton} W.~D.,   {Broderick} J.~J.,  2002, \mn@doi [\aj]
  {10.1086/341650}, \href {http://adsabs.harvard.edu/abs/2002AJ....124..675C}
  {124, 675}

\bibitem[\protect\citeauthoryear{{David} et~al.,}{{David}
  et~al.}{2014}]{david14}
{David} L.~P.,  et~al., 2014, \mn@doi [\apj] {10.1088/0004-637X/792/2/94},
  \href {http://adsabs.harvard.edu/abs/2014ApJ...792...94D} {792, 94}

\bibitem[\protect\citeauthoryear{{Desai} et~al.,}{{Desai}
  et~al.}{2012}]{desai12}
{Desai} S.,  et~al., 2012, \mn@doi [\apj] {10.1088/0004-637X/757/1/83}, \href
  {http://adsabs.harvard.edu/abs/2012ApJ...757...83D} {757, 83}

\bibitem[\protect\citeauthoryear{{Diemer} \& {Kravtsov}}{{Diemer} \&
  {Kravtsov}}{2015}]{diemer15}
{Diemer} B.,  {Kravtsov} A.~V.,  2015, \mn@doi [\apj]
  {10.1088/0004-637X/799/1/108}, \href
  {http://adsabs.harvard.edu/abs/2015ApJ...799..108D} {799, 108}

\bibitem[\protect\citeauthoryear{{Dietrich} et~al.,}{{Dietrich}
  et~al.}{2019}]{dietrich19}
{Dietrich} J.~P.,  et~al., 2019, \mn@doi [\mnras] {10.1093/mnras/sty3088},
  \href {http://adsabs.harvard.edu/abs/2019MNRAS.483.2871D} {483, 2871}

\bibitem[\protect\citeauthoryear{{Edge}}{{Edge}}{2001}]{edge01}
{Edge} A.~C.,  2001, \mn@doi [\mnras] {10.1046/j.1365-8711.2001.04802.x}, \href
  {http://adsabs.harvard.edu/abs/2001MNRAS.328..762E} {328, 762}

\bibitem[\protect\citeauthoryear{{Ehlert} et~al.,}{{Ehlert}
  et~al.}{2011}]{ehlert11}
{Ehlert} S.,  et~al., 2011, \mn@doi [\mnras]
  {10.1111/j.1365-2966.2010.17801.x}, \href
  {https://ui.adsabs.harvard.edu/#abs/2011MNRAS.411.1641E} {411, 1641}

\bibitem[\protect\citeauthoryear{{Fabian}}{{Fabian}}{1994}]{fabian94a}
{Fabian} A.~C.,  1994, \mn@doi [\araa] {10.1146/annurev.aa.32.090194.001425},
  \href {http://adsabs.harvard.edu/abs/1994ARA%26A..32..277F} {32, 277}

\bibitem[\protect\citeauthoryear{{Fabian}}{{Fabian}}{2012}]{fabian12}
{Fabian} A.~C.,  2012, \mn@doi [\araa] {10.1146/annurev-astro-081811-125521},
  \href {http://adsabs.harvard.edu/abs/2012ARA%26A..50..455F} {50, 455}

\bibitem[\protect\citeauthoryear{{Foreman-Mackey}, {Hogg}, {Lang}  \&
  {Goodman}}{{Foreman-Mackey} et~al.}{2013}]{mackey13}
{Foreman-Mackey} D.,  {Hogg} D.~W.,  {Lang} D.,   {Goodman} J.,  2013, \mn@doi
  [\pasp] {10.1086/670067}, \href
  {http://adsabs.harvard.edu/abs/2013PASP..125..306F} {125, 306}

\bibitem[\protect\citeauthoryear{{Galametz} et~al.,}{{Galametz}
  et~al.}{2009}]{galametz09}
{Galametz} A.,  et~al., 2009, \mn@doi [\apj] {10.1088/0004-637X/694/2/1309},
  \href {http://adsabs.harvard.edu/abs/2009ApJ...694.1309G} {694, 1309}

\bibitem[\protect\citeauthoryear{{Gaspari}, {Tombesi}  \& {Cappi}}{{Gaspari}
  et~al.}{2020}]{gaspari20}
{Gaspari} M.,  {Tombesi} F.,   {Cappi} M.,  2020, \mn@doi [Nature Astronomy]
  {10.1038/s41550-019-0970-1}, \href
  {https://ui.adsabs.harvard.edu/abs/2020NatAs...4...10G} {4, 10}

\bibitem[\protect\citeauthoryear{{Gitti}, {Brighenti}  \& {McNamara}}{{Gitti}
  et~al.}{2012}]{myriam12}
{Gitti} M.,  {Brighenti} F.,   {McNamara} B.~R.,  2012, \mn@doi [Advances in
  Astronomy] {10.1155/2012/950641}, \href
  {https://ui.adsabs.harvard.edu/#abs/2012AdAst2012E...6G} {2012, 950641}

\bibitem[\protect\citeauthoryear{{Green} et~al.,}{{Green}
  et~al.}{2016}]{green16}
{Green} T.~S.,  et~al., 2016, \mn@doi [\mnras] {10.1093/mnras/stw1338}, \href
  {http://adsabs.harvard.edu/abs/2016MNRAS.461..560G} {461, 560}

\bibitem[\protect\citeauthoryear{{Gupta} et~al.,}{{Gupta}
  et~al.}{2017}]{gupta17}
{Gupta} N.,  et~al., 2017, \mn@doi [\mnras] {10.1093/mnras/stx095}, \href
  {http://adsabs.harvard.edu/abs/2017MNRAS.467.3737G} {467, 3737}

\bibitem[\protect\citeauthoryear{{Hennig} et~al.,}{{Hennig}
  et~al.}{2017}]{hennig17}
{Hennig} C.,  et~al., 2017, \mn@doi [\mnras] {10.1093/mnras/stx175}, \href
  {http://adsabs.harvard.edu/abs/2017MNRAS.467.4015H} {467, 4015}

\bibitem[\protect\citeauthoryear{{Hlavacek-Larrondo}, {Fabian}, {Edge},
  {Ebeling}, {Sanders}, {Hogan}  \& {Taylor}}{{Hlavacek-Larrondo}
  et~al.}{2012}]{larrondo12}
{Hlavacek-Larrondo} J.,  {Fabian} A.~C.,  {Edge} A.~C.,  {Ebeling} H.,
  {Sanders} J.~S.,  {Hogan} M.~T.,   {Taylor} G.~B.,  2012, \mn@doi [\mnras]
  {10.1111/j.1365-2966.2011.20405.x}, \href
  {http://adsabs.harvard.edu/abs/2012MNRAS.421.1360H} {421, 1360}

\bibitem[\protect\citeauthoryear{{Hlavacek-Larrondo}, {Fabian}, {Edge},
  {Ebeling}, {Allen}, {Sanders}  \& {Taylor}}{{Hlavacek-Larrondo}
  et~al.}{2013}]{larrondo13}
{Hlavacek-Larrondo} J.,  {Fabian} A.~C.,  {Edge} A.~C.,  {Ebeling} H.,  {Allen}
  S.~W.,  {Sanders} J.~S.,   {Taylor} G.~B.,  2013, \mn@doi [\mnras]
  {10.1093/mnras/stt283}, \href
  {https://ui.adsabs.harvard.edu/#abs/2013MNRAS.431.1638H} {431, 1638}

\bibitem[\protect\citeauthoryear{{Hlavacek-Larrondo}
  et~al.,}{{Hlavacek-Larrondo} et~al.}{2015}]{larrondo15}
{Hlavacek-Larrondo} J.,  et~al., 2015, \mn@doi [\apj]
  {10.1088/0004-637X/805/1/35}, \href
  {http://adsabs.harvard.edu/abs/2015ApJ...805...35H} {805, 35}

\bibitem[\protect\citeauthoryear{{Hogan} et~al.,}{{Hogan}
  et~al.}{2015}]{hogan15}
{Hogan} M.~T.,  et~al., 2015, \mn@doi [\mnras] {10.1093/mnras/stv1517}, \href
  {http://adsabs.harvard.edu/abs/2015MNRAS.453.1201H} {453, 1201}

\bibitem[\protect\citeauthoryear{{Janssen}, {R{\"o}ttgering}, {Best}  \&
  {Brinchmann}}{{Janssen} et~al.}{2012}]{janssen12}
{Janssen} R.~M.~J.,  {R{\"o}ttgering} H.~J.~A.,  {Best} P.~N.,   {Brinchmann}
  J.,  2012, \mn@doi [\aap] {10.1051/0004-6361/201219052}, \href
  {http://adsabs.harvard.edu/abs/2012A%26A...541A..62J} {541, A62}

\bibitem[\protect\citeauthoryear{{Kauffmann}, {Heckman}  \& {Best}}{{Kauffmann}
  et~al.}{2008}]{kauffmann08}
{Kauffmann} G.,  {Heckman} T.~M.,   {Best} P.~N.,  2008, \mn@doi [\mnras]
  {10.1111/j.1365-2966.2007.12752.x}, \href
  {http://adsabs.harvard.edu/abs/2008MNRAS.384..953K} {384, 953}

\bibitem[\protect\citeauthoryear{{Klein} et~al.,}{{Klein}
  et~al.}{2018}]{klein18}
{Klein} M.,  et~al., 2018, \mn@doi [\mnras] {10.1093/mnras/stx2929}, \href
  {http://adsabs.harvard.edu/abs/2018MNRAS.474.3324K} {474, 3324}

\bibitem[\protect\citeauthoryear{{Klein} et~al.,}{{Klein}
  et~al.}{2019}]{klein19}
{Klein} M.,  et~al., 2019, \mn@doi [\mnras] {10.1093/mnras/stz1463}, \href
  {https://ui.adsabs.harvard.edu/abs/2019MNRAS.tmp.1397K} {p.~1397}

\bibitem[\protect\citeauthoryear{{Kravtsov}, {Vikhlinin}  \&
  {Meshcheryakov}}{{Kravtsov} et~al.}{2018}]{kravtsov18}
{Kravtsov} A.~V.,  {Vikhlinin} A.~A.,   {Meshcheryakov} A.~V.,  2018, \mn@doi
  [Astronomy Letters] {10.1134/S1063773717120015}, \href
  {https://ui.adsabs.harvard.edu/abs/2018AstL...44....8K} {44, 8}

\bibitem[\protect\citeauthoryear{{Kriek}, {van Dokkum}, {Labb{\'e}}, {Franx},
  {Illingworth}, {Marchesini}  \& {Quadri}}{{Kriek} et~al.}{2009}]{fast}
{Kriek} M.,  {van Dokkum} P.~G.,  {Labb{\'e}} I.,  {Franx} M.,  {Illingworth}
  G.~D.,  {Marchesini} D.,   {Quadri} R.~F.,  2009, \mn@doi [\apj]
  {10.1088/0004-637X/700/1/221}, \href
  {http://esoads.eso.org/abs/2009ApJ...700..221K} {700, 221}

\bibitem[\protect\citeauthoryear{{Lidman} et~al.,}{{Lidman}
  et~al.}{2012}]{lidman12}
{Lidman} C.,  et~al., 2012, \mn@doi [\mnras]
  {10.1111/j.1365-2966.2012.21984.x}, \href
  {http://adsabs.harvard.edu/abs/2012MNRAS.427..550L} {427, 550}

\bibitem[\protect\citeauthoryear{{Lidman} et~al.,}{{Lidman}
  et~al.}{2013}]{lidman13}
{Lidman} C.,  et~al., 2013, \mn@doi [\mnras] {10.1093/mnras/stt777}, \href
  {https://ui.adsabs.harvard.edu/\#abs/2013MNRAS.433..825L} {433, 825}

\bibitem[\protect\citeauthoryear{{Lin} \& {Mohr}}{{Lin} \&
  {Mohr}}{2004}]{lin04b}
{Lin} Y.-T.,  {Mohr} J.~J.,  2004, \mn@doi [\apj] {10.1086/425412}, \href
  {http://adsabs.harvard.edu/abs/2004ApJ...617..879L} {617, 879}

\bibitem[\protect\citeauthoryear{{Lin} \& {Mohr}}{{Lin} \&
  {Mohr}}{2007}]{lin07}
{Lin} Y.-T.,  {Mohr} J.~J.,  2007, \mn@doi [\apjs] {10.1086/513565}, \href
  {http://adsabs.harvard.edu/abs/2007ApJS..170...71L} {170, 71}

\bibitem[\protect\citeauthoryear{{Lin}, {Mohr}  \& {Stanford}}{{Lin}
  et~al.}{2004}]{lin04a}
{Lin} Y.,  {Mohr} J.~J.,   {Stanford} S.~A.,  2004, \apj, \href
  {http://adsabs.harvard.edu/cgi-bin/nph-bib_query?bibcode=2004ApJ...610..745L&amp;db_key=AST}
  {610, 745}

\bibitem[\protect\citeauthoryear{{Lin}, {Partridge}, {Pober}, {Bouchefry},
  {Burke}, {Klein}, {Coish}  \& {Huffenberger}}{{Lin} et~al.}{2009}]{lin09}
{Lin} Y.,  {Partridge} B.,  {Pober} J.~C.,  {Bouchefry} K.~E.,  {Burke} S.,
  {Klein} J.~N.,  {Coish} J.~W.,   {Huffenberger} K.~M.,  2009, \mn@doi [\apj]
  {10.1088/0004-637X/694/2/992}, \href
  {http://adsabs.harvard.edu/abs/2009ApJ...694..992L} {694, 992}

\bibitem[\protect\citeauthoryear{{Lin}, {Shen}, {Strauss}, {Richards}  \&
  {Lunnan}}{{Lin} et~al.}{2010}]{lin10}
{Lin} Y.-T.,  {Shen} Y.,  {Strauss} M.~A.,  {Richards} G.~T.,   {Lunnan} R.,
  2010, \mn@doi [\apj] {10.1088/0004-637X/723/2/1119}, \href
  {http://adsabs.harvard.edu/abs/2010ApJ...723.1119L} {723, 1119}

\bibitem[\protect\citeauthoryear{{Lin}, {McDonald}, {Benson}  \&
  {Miller}}{{Lin} et~al.}{2015}]{linhenry15}
{Lin} H.~W.,  {McDonald} M.,  {Benson} B.,   {Miller} E.,  2015, \mn@doi [\apj]
  {10.1088/0004-637X/802/1/34}, \href
  {http://adsabs.harvard.edu/abs/2015ApJ...802...34L} {802, 34}

\bibitem[\protect\citeauthoryear{{Lin} et~al.,}{{Lin} et~al.}{2017}]{lin17}
{Lin} Y.-T.,  et~al., 2017, \mn@doi [\apj] {10.3847/1538-4357/aa9bf5}, \href
  {http://adsabs.harvard.edu/abs/2017ApJ...851..139L} {851, 139}

\bibitem[\protect\citeauthoryear{{Lin}, {Huang}  \& {Chen}}{{Lin}
  et~al.}{2018}]{lin18}
{Lin} Y.-T.,  {Huang} H.-J.,   {Chen} Y.-C.,  2018, \mn@doi [\aj]
  {10.3847/1538-3881/aab5b4}, \href
  {http://adsabs.harvard.edu/abs/2018AJ....155..188L} {155, 188}

\bibitem[\protect\citeauthoryear{{Machalski} \& {Godlowski}}{{Machalski} \&
  {Godlowski}}{2000}]{malchalski00}
{Machalski} J.,  {Godlowski} W.,  2000, \aap, \href
  {http://adsabs.harvard.edu/abs/2000A%26A...360..463M} {360, 463}

\bibitem[\protect\citeauthoryear{{Main}, {McNamara}, {Nulsen}, {Russell}  \&
  {Vantyghem}}{{Main} et~al.}{2017}]{main17}
{Main} R.~A.,  {McNamara} B.~R.,  {Nulsen} P.~E.~J.,  {Russell} H.~R.,
  {Vantyghem} A.~N.,  2017, \mn@doi [\mnras] {10.1093/mnras/stw2644}, \href
  {https://ui.adsabs.harvard.edu/abs/2017MNRAS.464.4360M} {464, 4360}

\bibitem[\protect\citeauthoryear{{Mamon}, {Biviano}  \& {Bou{\'e}}}{{Mamon}
  et~al.}{2013}]{mamon13}
{Mamon} G.~A.,  {Biviano} A.,   {Bou{\'e}} G.,  2013, \mn@doi [\mnras]
  {10.1093/mnras/sts565}, \href
  {http://adsabs.harvard.edu/abs/2013MNRAS.429.3079M} {429, 3079}

\bibitem[\protect\citeauthoryear{{Mauch}, {Murphy}, {Buttery}, {Curran},
  {Hunstead}, {Piestrzynski}, {Robertson}  \& {Sadler}}{{Mauch}
  et~al.}{2003}]{mauch03}
{Mauch} T.,  {Murphy} T.,  {Buttery} H.~J.,  {Curran} J.,  {Hunstead} R.~W.,
  {Piestrzynski} B.,  {Robertson} J.~G.,   {Sadler} E.~M.,  2003, \mnras, 342,
  1117

\bibitem[\protect\citeauthoryear{{McAlpine}, {Jarvis}  \&
  {Bonfield}}{{McAlpine} et~al.}{2013}]{mcalpine13}
{McAlpine} K.,  {Jarvis} M.~J.,   {Bonfield} D.~G.,  2013, \mn@doi [\mnras]
  {10.1093/mnras/stt1638}, \href
  {http://adsabs.harvard.edu/abs/2013MNRAS.436.1084M} {436, 1084}

\bibitem[\protect\citeauthoryear{{McClintock} et~al.,}{{McClintock}
  et~al.}{2019}]{mcclintock19}
{McClintock} T.,  et~al., 2019, \mn@doi [\mnras] {10.1093/mnras/sty2711}, \href
  {https://ui.adsabs.harvard.edu/abs/2019MNRAS.482.1352M} {482, 1352}

\bibitem[\protect\citeauthoryear{{McDonald} et~al.,}{{McDonald}
  et~al.}{2012}]{mcdonald12}
{McDonald} M.,  et~al., 2012, \mn@doi [\nat] {10.1038/nature11379}, \href
  {http://adsabs.harvard.edu/abs/2012arXiv1208.2962M} {488, 349}

\bibitem[\protect\citeauthoryear{{McDonald} et~al.,}{{McDonald}
  et~al.}{2013}]{mcdonald13}
{McDonald} M.,  et~al., 2013, \mn@doi [\apj] {10.1088/0004-637X/774/1/23},
  \href {http://adsabs.harvard.edu/abs/2013ApJ...774...23M} {774, 23}

\bibitem[\protect\citeauthoryear{{McNamara}, {Nulsen}, {Wise}, {Rafferty},
  {Carilli}, {Sarazin}  \& {Blanton}}{{McNamara} et~al.}{2005}]{mcnamara05}
{McNamara} B.~R.,  {Nulsen} P.~E.~J.,  {Wise} M.~W.,  {Rafferty} D.~A.,
  {Carilli} C.,  {Sarazin} C.~L.,   {Blanton} E.~L.,  2005, \mn@doi [\nat]
  {10.1038/nature03202}, \href
  {http://adsabs.harvard.edu/cgi-bin/nph-bib_query?bibcode=2005Natur.433...45M&db_key=AST}
  {433, 45}

\bibitem[\protect\citeauthoryear{{McNamara} et~al.,}{{McNamara}
  et~al.}{2014}]{mcnamara14}
{McNamara} B.~R.,  et~al., 2014, \mn@doi [\apj] {10.1088/0004-637X/785/1/44},
  \href {https://ui.adsabs.harvard.edu/\#abs/2014ApJ...785...44M} {785, 44}

\bibitem[\protect\citeauthoryear{{Merloni} et~al.,}{{Merloni}
  et~al.}{2012}]{merloni12}
{Merloni} A.,  et~al., 2012, preprint, \href
  {http://adsabs.harvard.edu/abs/2012arXiv1209.3114M} {} (\mn@eprint {arXiv}
  {1209.3114})

\bibitem[\protect\citeauthoryear{{Mills}}{{Mills}}{1981}]{mills81}
{Mills} B.~Y.,  1981, Proceedings of the Astronomical Society of Australia,
  \href {http://adsabs.harvard.edu/abs/1981PASAu...4..156M} {4, 156}

\bibitem[\protect\citeauthoryear{{Mohr} \& {Evrard}}{{Mohr} \&
  {Evrard}}{1997}]{mohr97}
{Mohr} J.~J.,  {Evrard} A.~E.,  1997, \mn@doi [\apj] {10.1086/304957}, \href
  {https://ui.adsabs.harvard.edu/abs/1997ApJ...491...38M} {491, 38}

\bibitem[\protect\citeauthoryear{{Mohr}, {Mathiesen}  \& {Evrard}}{{Mohr}
  et~al.}{1999}]{mohr99}
{Mohr} J.~J.,  {Mathiesen} B.,   {Evrard} A.~E.,  1999, \mn@doi [\apj]
  {10.1086/307227}, \href
  {https://ui.adsabs.harvard.edu/abs/1999ApJ...517..627M} {517, 627}

\bibitem[\protect\citeauthoryear{{Morganson} et~al.,}{{Morganson}
  et~al.}{2018}]{morganson18}
{Morganson} E.,  et~al., 2018, \mn@doi [\pasp] {10.1088/1538-3873/aab4ef},
  \href {http://adsabs.harvard.edu/abs/2018PASP..130g4501M} {130, 074501}

\bibitem[\protect\citeauthoryear{{Mortonson}, {Hu}  \& {Huterer}}{{Mortonson}
  et~al.}{2011}]{mortonson11}
{Mortonson} M.~J.,  {Hu} W.,   {Huterer} D.,  2011, \mn@doi [\prd]
  {10.1103/PhysRevD.83.023015}, \href
  {http://adsabs.harvard.edu/abs/2011PhRvD..83b3015M} {83, 023015}

\bibitem[\protect\citeauthoryear{{Murphy}, {Mauch}, {Green}, {Hunstead},
  {Piestrzynska}, {Kels}  \& {Sztajer}}{{Murphy} et~al.}{2007}]{murphy07}
{Murphy} T.,  {Mauch} T.,  {Green} A.,  {Hunstead} R.~W.,  {Piestrzynska} B.,
  {Kels} A.~P.,   {Sztajer} P.,  2007, \mn@doi [\mnras]
  {10.1111/j.1365-2966.2007.12379.x}, \href
  {http://adsabs.harvard.edu/abs/2007MNRAS.382..382M} {382, 382}

\bibitem[\protect\citeauthoryear{{Navarro}, {Frenk}  \& {White}}{{Navarro}
  et~al.}{1997}]{navarro97}
{Navarro} J.~F.,  {Frenk} C.~S.,   {White} S.~D.~M.,  1997, \mn@doi [\apj]
  {10.1086/304888}, \href {http://adsabs.harvard.edu/abs/1997ApJ...490..493N}
  {490, 493}

\bibitem[\protect\citeauthoryear{{Ogrean}, {Hatch}, {Simionescu},
  {B{\"o}hringer}, {Br{\"u}ggen}, {Fabian}  \& {Werner}}{{Ogrean}
  et~al.}{2010}]{ogrean10}
{Ogrean} G.~A.,  {Hatch} N.~A.,  {Simionescu} A.,  {B{\"o}hringer} H.,
  {Br{\"u}ggen} M.,  {Fabian} A.~C.,   {Werner} N.,  2010, \mn@doi [\mnras]
  {10.1111/j.1365-2966.2010.16718.x}, \href
  {https://ui.adsabs.harvard.edu/#abs/2010MNRAS.406..354O} {406, 354}

\bibitem[\protect\citeauthoryear{{Panagoulia}, {Fabian}  \&
  {Sanders}}{{Panagoulia} et~al.}{2014}]{panagoulia14}
{Panagoulia} E.~K.,  {Fabian} A.~C.,   {Sanders} J.~S.,  2014, \mn@doi [\mnras]
  {10.1093/mnras/stt2349}, \href
  {https://ui.adsabs.harvard.edu/abs/2014MNRAS.438.2341P} {438, 2341}

\bibitem[\protect\citeauthoryear{{Pracy} et~al.,}{{Pracy}
  et~al.}{2016}]{pracy16}
{Pracy} M.~B.,  et~al., 2016, \mn@doi [\mnras] {10.1093/mnras/stw910}, \href
  {http://adsabs.harvard.edu/abs/2016MNRAS.460....2P} {460, 2}

\bibitem[\protect\citeauthoryear{{Predehl} et~al.,}{{Predehl}
  et~al.}{2010}]{predehl10}
{Predehl} P.,  et~al., 2010, in Society of Photo-Optical Instrumentation
  Engineers (SPIE) Conference Series.  (\mn@eprint {arXiv} {1001.2502}),
  \mn@doi{10.1117/12.856577}

\bibitem[\protect\citeauthoryear{{Rafferty}, {McNamara}, {Nulsen}  \&
  {Wise}}{{Rafferty} et~al.}{2006}]{rafferty06}
{Rafferty} D.~A.,  {McNamara} B.~R.,  {Nulsen} P.~E.~J.,   {Wise} M.~W.,  2006,
  \mn@doi [\apj] {10.1086/507672}, \href
  {http://adsabs.harvard.edu/abs/2006ApJ...652..216R} {652, 216}

\bibitem[\protect\citeauthoryear{{Robertson}}{{Robertson}}{1991}]{robertson91}
{Robertson} J.~G.,  1991, Australian Journal of Physics, \href
  {http://adsabs.harvard.edu/abs/1991AuJPh..44..729R} {44, 729}

\bibitem[\protect\citeauthoryear{{Rose} et~al.,}{{Rose} et~al.}{2019}]{rose19}
{Rose} T.,  et~al., 2019, \mn@doi [\mnras] {10.1093/mnras/stz406}, \href
  {https://ui.adsabs.harvard.edu/\#abs/2019MNRAS.485..229R} {485, 229}

\bibitem[\protect\citeauthoryear{{Rozo} \& {Rykoff}}{{Rozo} \&
  {Rykoff}}{2014}]{rozo_rykoff14}
{Rozo} E.,  {Rykoff} E.~S.,  2014, \mn@doi [\apj] {10.1088/0004-637X/783/2/80},
  \href {http://adsabs.harvard.edu/abs/2014ApJ...783...80R} {783, 80}

\bibitem[\protect\citeauthoryear{{Rozo}, {Rykoff}, {Bartlett}  \&
  {Evrard}}{{Rozo} et~al.}{2014a}]{rozo14a}
{Rozo} E.,  {Rykoff} E.~S.,  {Bartlett} J.~G.,   {Evrard} A.,  2014a, \mn@doi
  [\mnras] {10.1093/mnras/stt2091}, \href
  {http://adsabs.harvard.edu/abs/2014MNRAS.438...49R} {438, 49}

\bibitem[\protect\citeauthoryear{{Rozo}, {Evrard}, {Rykoff}  \&
  {Bartlett}}{{Rozo} et~al.}{2014b}]{rozo14b}
{Rozo} E.,  {Evrard} A.~E.,  {Rykoff} E.~S.,   {Bartlett} J.~G.,  2014b,
  \mn@doi [\mnras] {10.1093/mnras/stt2160}, \href
  {http://adsabs.harvard.edu/abs/2014MNRAS.438...62R} {438, 62}

\bibitem[\protect\citeauthoryear{{Rykoff} et~al.,}{{Rykoff}
  et~al.}{2012}]{rykoff12}
{Rykoff} E.~S.,  et~al., 2012, \mn@doi [\apj] {10.1088/0004-637X/746/2/178},
  \href {http://adsabs.harvard.edu/abs/2012ApJ...746..178R} {746, 178}

\bibitem[\protect\citeauthoryear{{Rykoff} et~al.,}{{Rykoff}
  et~al.}{2014}]{rykoff14}
{Rykoff} E.~S.,  et~al., 2014, \mn@doi [\apj] {10.1088/0004-637X/785/2/104},
  \href {http://adsabs.harvard.edu/abs/2014ApJ...785..104R} {785, 104}

\bibitem[\protect\citeauthoryear{{Rykoff} et~al.,}{{Rykoff}
  et~al.}{2016}]{rykoff16}
{Rykoff} E.~S.,  et~al., 2016, \mn@doi [\apjs] {10.3847/0067-0049/224/1/1},
  \href {http://adsabs.harvard.edu/abs/2016ApJS..224....1R} {224, 1}

\bibitem[\protect\citeauthoryear{{Salpeter}}{{Salpeter}}{1955}]{salpeter55}
{Salpeter} E.~E.,  1955, \mn@doi [\apj] {10.1086/145971}, \href
  {http://adsabs.harvard.edu/abs/1955ApJ...121..161S} {121, 161}

\bibitem[\protect\citeauthoryear{{Saro} et~al.,}{{Saro} et~al.}{2015}]{saro15}
{Saro} A.,  et~al., 2015, \mn@doi [\mnras] {10.1093/mnras/stv2141}, \href
  {http://adsabs.harvard.edu/abs/2015MNRAS.454.2305S} {454, 2305}

\bibitem[\protect\citeauthoryear{{Sehgal}, {Bode}, {Das},
  {Hernandez-Monteagudo}, {Huffenberger}, {Lin}, {Ostriker}  \&
  {Trac}}{{Sehgal} et~al.}{2010}]{sehgal10}
{Sehgal} N.,  {Bode} P.,  {Das} S.,  {Hernandez-Monteagudo} C.,  {Huffenberger}
  K.,  {Lin} Y.,  {Ostriker} J.~P.,   {Trac} H.,  2010, \mn@doi [\apj]
  {10.1088/0004-637X/709/2/920}, \href
  {http://adsabs.harvard.edu/abs/2010ApJ...709..920S} {709, 920}

\bibitem[\protect\citeauthoryear{{Semler} et~al.,}{{Semler}
  et~al.}{2012}]{semler12}
{Semler} D.~R.,  et~al., 2012, \mn@doi [\apj] {10.1088/0004-637X/761/2/183},
  \href {http://adsabs.harvard.edu/abs/2012ApJ...761..183S} {761, 183}

\bibitem[\protect\citeauthoryear{{Simpson}, {Westoby}, {Arumugam}, {Ivison},
  {Hartley}  \& {Almaini}}{{Simpson} et~al.}{2013}]{simpson13}
{Simpson} C.,  {Westoby} P.,  {Arumugam} V.,  {Ivison} R.,  {Hartley} W.,
  {Almaini} O.,  2013, \mn@doi [\mnras] {10.1093/mnras/stt940}, \href
  {http://adsabs.harvard.edu/abs/2013MNRAS.433.2647S} {433, 2647}

\bibitem[\protect\citeauthoryear{{Smol{\v c}i{\'c}} et~al.,}{{Smol{\v c}i{\'c}}
  et~al.}{2009}]{smolcic09}
{Smol{\v c}i{\'c}} V.,  et~al., 2009, \mn@doi [\apj]
  {10.1088/0004-637X/696/1/24}, \href
  {http://esoads.eso.org/abs/2009ApJ...696...24S} {696, 24}

\bibitem[\protect\citeauthoryear{{Smol{\v c}i{\'c}} et~al.,}{{Smol{\v c}i{\'c}}
  et~al.}{2017}]{smolcic17}
{Smol{\v c}i{\'c}} V.,  et~al., 2017, \mn@doi [\aap]
  {10.1051/0004-6361/201730685}, \href
  {https://ui.adsabs.harvard.edu/abs/2017A%26A...602A...6S} {602, A6}

\bibitem[\protect\citeauthoryear{{Soergel} et~al.,}{{Soergel}
  et~al.}{2016}]{soergel16}
{Soergel} B.,  et~al., 2016, \mn@doi [\mnras] {10.1093/mnras/stw1455}, \href
  {http://adsabs.harvard.edu/abs/2016MNRAS.461.3172S} {461, 3172}

\bibitem[\protect\citeauthoryear{{Sommer}, {Basu}, {Pacaud}, {Bertoldi}  \&
  {Andernach}}{{Sommer} et~al.}{2011}]{sommer11}
{Sommer} M.~W.,  {Basu} K.,  {Pacaud} F.,  {Bertoldi} F.,   {Andernach} H.,
  2011, \mn@doi [\aap] {10.1051/0004-6361/201016150}, \href
  {http://adsabs.harvard.edu/abs/2011A%26A...529A.124S} {529, A124}

\bibitem[\protect\citeauthoryear{{Stern} et~al.,}{{Stern}
  et~al.}{2019}]{stern19}
{Stern} C.,  et~al., 2019, \mn@doi [\mnras] {10.1093/mnras/stz234}, \href
  {http://adsabs.harvard.edu/abs/2019MNRAS.485...69S} {485, 69}

\bibitem[\protect\citeauthoryear{{Stott} et~al.,}{{Stott}
  et~al.}{2012}]{stott12}
{Stott} J.~P.,  et~al., 2012, \mn@doi [\mnras]
  {10.1111/j.1365-2966.2012.20764.x}, \href
  {http://adsabs.harvard.edu/abs/2012MNRAS.422.2213S} {422, 2213}

\bibitem[\protect\citeauthoryear{{Strazzullo}, {Pannella}, {Owen}, {Bender},
  {Morrison}, {Wang}  \& {Shupe}}{{Strazzullo} et~al.}{2010}]{strazzullo10}
{Strazzullo} V.,  {Pannella} M.,  {Owen} F.~N.,  {Bender} R.,  {Morrison}
  G.~E.,  {Wang} W.-H.,   {Shupe} D.~L.,  2010, \mn@doi [\apj]
  {10.1088/0004-637X/714/2/1305}, \href
  {http://adsabs.harvard.edu/abs/2010ApJ...714.1305S} {714, 1305}

\bibitem[\protect\citeauthoryear{{Sunyaev} \& {Zel'dovich}}{{Sunyaev} \&
  {Zel'dovich}}{1972}]{sunyaev72}
{Sunyaev} R.~A.,  {Zel'dovich} Y.~B.,  1972, Comments on Astrophysics and Space
  Physics, \href
  {http://adsabs.harvard.edu/cgi-bin/nph-bib_query?bibcode=1972CoASP...4..173S&amp;db_key=AST}
  {4, 173}

\bibitem[\protect\citeauthoryear{{Temi}, {Amblard}, {Gitti}, {Brighenti},
  {Gaspari}, {Mathews}  \& {David}}{{Temi} et~al.}{2018}]{temi18}
{Temi} P.,  {Amblard} A.,  {Gitti} M.,  {Brighenti} F.,  {Gaspari} M.,
  {Mathews} W.~G.,   {David} L.,  2018, \mn@doi [\apj]
  {10.3847/1538-4357/aab9b0}, \href
  {https://ui.adsabs.harvard.edu/\#abs/2018ApJ...858...17T} {858, 17}

\bibitem[\protect\citeauthoryear{{Tremblay} et~al.,}{{Tremblay}
  et~al.}{2016}]{tremblay16}
{Tremblay} G.~R.,  et~al., 2016, \mn@doi [\nat] {10.1038/nature17969}, \href
  {http://adsabs.harvard.edu/abs/2016Natur.534..218T} {534, 218}

\bibitem[\protect\citeauthoryear{{Vantyghem} et~al.,}{{Vantyghem}
  et~al.}{2019}]{vantyghem19}
{Vantyghem} A.~N.,  et~al., 2019, \mn@doi [\apj] {10.3847/1538-4357/aaf1b4},
  \href {https://ui.adsabs.harvard.edu/\#abs/2019ApJ...870...57V} {870, 57}

\bibitem[\protect\citeauthoryear{{Voit}, {Donahue}, {Bryan}  \&
  {McDonald}}{{Voit} et~al.}{2015}]{voit16}
{Voit} G.~M.,  {Donahue} M.,  {Bryan} G.~L.,   {McDonald} M.,  2015, \mn@doi
  [\nat] {10.1038/nature14167}, \href
  {http://adsabs.harvard.edu/abs/2015Natur.519..203V} {519, 203}

\bibitem[\protect\citeauthoryear{{Von Der Linden}, {Best}, {Kauffmann}  \&
  {White}}{{Von Der Linden} et~al.}{2007}]{linden07}
{Von Der Linden} A.,  {Best} P.~N.,  {Kauffmann} G.,   {White} S.~D.~M.,  2007,
  \mn@doi [\mnras] {10.1111/j.1365-2966.2007.11940.x}, \href
  {https://ui.adsabs.harvard.edu/abs/2007MNRAS.379..867V} {379, 867}

\bibitem[\protect\citeauthoryear{{Webb} et~al.,}{{Webb} et~al.}{2017}]{webb17}
{Webb} T. M.~A.,  et~al., 2017, \mn@doi [\apj] {10.3847/2041-8213/aa7749},
  \href {https://ui.adsabs.harvard.edu/\#abs/2017ApJ...844L..17W} {844, L17}

\bibitem[\protect\citeauthoryear{{Yang}, {Tozzi}, {Yu}, {Lusso}, {Gaspari},
  {Gilli}, {Nardini}  \& {Risaliti}}{{Yang} et~al.}{2018}]{yang18}
{Yang} L.,  {Tozzi} P.,  {Yu} H.,  {Lusso} E.,  {Gaspari} M.,  {Gilli} R.,
  {Nardini} E.,   {Risaliti} G.,  2018, \mn@doi [\apj]
  {10.3847/1538-4357/aabfd7}, \href
  {https://ui.adsabs.harvard.edu/abs/2018ApJ...859...65Y} {859, 65}

\bibitem[\protect\citeauthoryear{{Zhang} et~al.,}{{Zhang}
  et~al.}{2016}]{zhang16}
{Zhang} Y.,  et~al., 2016, \mn@doi [\apj] {10.3847/0004-637X/816/2/98}, \href
  {https://ui.adsabs.harvard.edu/abs/2016ApJ...816...98Z} {816, 98}

\bibitem[\protect\citeauthoryear{{de Haan} et~al.,}{{de Haan}
  et~al.}{2016}]{dehaan16}
{de Haan} T.,  et~al., 2016, \mn@doi [\apj] {10.3847/0004-637X/832/1/95}, \href
  {http://adsabs.harvard.edu/abs/2016ApJ...832...95D} {832, 95}

\bibitem[\protect\citeauthoryear{{de Zotti}, {Ricci}, {Mesa}, {Silva},
  {Mazzotta}, {Toffolatti}  \& {Gonz{\'a}lez-Nuevo}}{{de Zotti}
  et~al.}{2005}]{dezotti05}
{de Zotti} G.,  {Ricci} R.,  {Mesa} D.,  {Silva} L.,  {Mazzotta} P.,
  {Toffolatti} L.,   {Gonz{\'a}lez-Nuevo} J.,  2005, \mn@doi [\aap]
  {10.1051/0004-6361:20042108}, \href
  {http://adsabs.harvard.edu/abs/2005A%26A...431..893D} {431, 893}

\makeatother
\end{thebibliography}

\section*{Author Affiliations}

\Melbourne School of Physics, University of Melbourne, Parkville, VIC 3010, Australia \\
\Munich Faculty of Physics, Ludwig-Maximilians-Universit\"{a}t, Scheinerstr.\ 1, 81679 Munich, Germany \\
\ExcellenceCluster Excellence Cluster Origins, Boltzmannstr.\ 2, 85748 Garching, Germany \\
\MPE Max Planck Institute for Extraterrestrial Physics, Giessenbachstr.\ 85748 Garching, Germany \\
\STANFORDkavli Kavli Institute for Particle Astrophysics \& Cosmology, P. O. Box 2450, Stanford University, Stanford, CA 94305, USA\\
\SLAC SLAC National Accelerator Laboratory, Menlo Park, CA 94025, USA\\
\Fermilab Fermi National Accelerator Laboratory, P. O. Box 500, Batavia, IL 60510, USA\\
\Madrid Instituto de Fisica Teorica UAM/CSIC, Universidad Autonoma de Madrid, 28049 Madrid, Spain\\
\UCL Department of Physics \& Astronomy, University College London, Gower Street, London, WC1E 6BT, UK\\
\HARVARD Center for Astrophysics $\vert$ Harvard \& Smithsonian, 60 Garden Street, Cambridge, MA 02138, USA\\
\MIT Kavli Institute for Astrophysics and Space Research, Massachusetts Institute of Technology, 77 Massachusetts Avenue, Cambridge, MA 02139\\
\CIEMAT Centro de Investigaciones Energ\'eticas, Medioambientales y Tecnol\'ogicas (CIEMAT), Madrid, Spain\\
\RIOLAB Laborat\'orio Interinstitucional de e-Astronomia - LIneA, Rua Gal. Jos\'e Cristino 77, Rio de Janeiro, RJ - 20921-400, Brazil\\
\ILLINOIS Department of Astronomy, University of Illinois at Urbana-Champaign, 1002 W. Green Street, Urbana, IL 61801, USA\\
\Urbana National Center for Supercomputing Applications, 1205 West Clark St., Urbana, IL 61801, USA\\
\BARCELONA Institut de F\'{\i}sica d'Altes Energies (IFAE), The Barcelona Institute of Science and Technology, Campus UAB, 08193 Bellaterra (Barcelona) Spain\\
\ASIAA Academia Sinica Institute of Astronomy and Astrophysics, 11F of AS/NTU Astronomy-Mathematics Building, No.1, Sec. 4, Roosevelt Rd, Taipei 10617, Taiwan\\
\RIOOBS Observat\'orio Nacional, Rua Gal. Jos\'e Cristino 77, Rio de Janeiro, RJ - 20921-400, Brazil\\
\INDIA Department of Physics, IIT Hyderabad, Kandi, Telangana 502285, India\\
\SANTACRUZ Santa Cruz Institute for Particle Physics, Santa Cruz, CA 95064, USA\\
\MICHIGANA Department of Astronomy, University of Michigan, Ann Arbor, MI 48109, USA\\
\MICHIGANP Department of Physics, University of Michigan, Ann Arbor, MI 48109, USA\\
\IEEC Institut d'Estudis Espacials de Catalunya (IEEC), 08034 Barcelona, Spain\\
\ICE Institute of Space Sciences (ICE, CSIC),  Campus UAB, Carrer de Can Magrans, s/n,  08193 Barcelona, Spain\\
\STANFORD Department of Physics, Stanford University, 382 Via Pueblo Mall, Stanford, CA 94305, USA\\
\Ohio Department of Physics, The Ohio State University, Columbus, OH 43210, USA\\
\Macquarie Australian Astronomical Optics, Macquarie University, North Ryde, NSW 2113, Australia\\
\ANU The Research School of Astronomy and Astrophysics, Australian National University, ACT 2601, Australia\\
\PAULO Departamento de F\'isica Matem\'atica, Instituto de F\'isica, Universidade de S\~ao Paulo, CP 66318, S\~ao Paulo, SP, 05314-970, Brazil\\
\AandM George P. and Cynthia Woods Mitchell Institute for Fundamental Physics and Astronomy, and Department of Physics and Astronomy, Texas A\&M University, College Station, TX 77843,  USA\\
\CATALAN Instituci\'o Catalana de Recerca i Estudis Avan\c{c}ats, E-08010 Barcelona, Spain\\
\PRINCETON Department of Astrophysical Sciences, Princeton University, Peyton Hall, Princeton, NJ 08544, USA\\
\PORTO Instituto de F\'\i sica, UFRGS, Caixa Postal 15051, Porto Alegre, RS - 91501-970, Brazil\\
\TriesteA INAF-Osservatorio Astronomico di Trieste, via Tiepolo 11, I-34143 Trieste, Italy\\
\TriesteB IFPU - Institute for Fundamental Physics of the Universe, Via Beirut 2, 34014 Trieste, Italy\\
\TriesteC INAF-Osservatorio Astronomico di Trieste, via G. B. Tiepolo 11, I-34143 Trieste, Italy\\
\SOUTHAMPTON School of Physics and Astronomy, University of Southampton,  Southampton, SO17 1BJ, UK\\
\Lancaster Department of Physics, Lancaster University, Lancaster, LA1 4YB, UK\\
\OAK Computer Science and Mathematics Division, Oak Ridge National Laboratory, Oak Ridge, TN 37831\\
\ARGONNE Argonne National Laboratory, 9700 South Cass Avenue, Lemont, IL 60439, USA\\
\LaSerena Cerro Tololo Inter-American Observatory, Casilla 603, La Serena, Chile\\

\end{document}